\documentclass[%
aip,
 amsmath,amssymb,
reprint,%
nofootinbib,
nobibnotes,
floatfix,
]{revtex4-1}

\usepackage{graphicx}
\usepackage{dcolumn}
\usepackage{bm}
\usepackage{subfigure}
\usepackage{xcolor}
\usepackage{microtype}

\usepackage[utf8]{inputenc}
\usepackage[T1]{fontenc}
\usepackage{mathptmx}

\begin{document}

\preprint{AIP/123-QED}

\title[PB-2b CHWP]{A cryogenic continuously rotating half-wave plate for the POLARBEAR-2b cosmic microwave background receiver}

\author{C. A. Hill}
 \email{chill@lbl.gov.}
 \affiliation{Department of Physics, University of California, Berkeley, CA 94720, USA}
 \affiliation{Physics Division, Lawrence Berkeley National Laboratory, Berkeley, CA 94720, USA}

\author{A. Kusaka}
 \affiliation{Physics Division, Lawrence Berkeley National Laboratory, Berkeley, CA 94720, USA}
 \affiliation{Department of Physics, University of Tokyo, Bunkyo-ku, Tokyo 113-0033, Japan}
 \affiliation{Kavli Institute for the Physics and Mathematics of the Universe (WPI), The University of Tokyo Institutes for Advanced Study, The University of Tokyo, Kashiwa, Chiba 277-8583, Japan}
 \affiliation{Research Center for the Early Universe, University of Tokyo, Tokyo 113-0033, Japan}

\author{P. Ashton}
 \affiliation{Department of Physics, University of California, Berkeley, CA 94720, USA}
 \affiliation{Physics Division, Lawrence Berkeley National Laboratory, Berkeley, CA 94720, USA}
 \affiliation{Kavli Institute for the Physics and Mathematics of the Universe (WPI), The University of Tokyo Institutes for Advanced Study, The University of Tokyo, Kashiwa, Chiba 277-8583, Japan}
 
\author{P. Barton}
 \affiliation{Nuclear Science Division, Lawrence Berkeley National Laboratory, Berkeley, CA 94720, USA}
 
\author{T. Adkins}
 \affiliation{Department of Physics, University of California, Berkeley, CA 94720, USA}
 
\author{K. Arnold}
 \affiliation{Department of Physics, University of California, San Diego, La Jolla, CA 92037, USA}
 
\author{B. Bixler}
 \affiliation{Department of Physics, University of California, San Diego, La Jolla, CA 92037, USA}
 
\author{S. Ganjam}
 \altaffiliation{Currently at Yale University.}
  \affiliation{Department of Physics, University of California, Berkeley, CA 94720, USA}
 \affiliation{Physics Division, Lawrence Berkeley National Laboratory, Berkeley, CA 94720, USA}
 
\author{A. T. Lee}
 \affiliation{Department of Physics, University of California, Berkeley, CA 94720, USA}
 \affiliation{Physics Division, Lawrence Berkeley National Laboratory, Berkeley, CA 94720, USA}
 
\author{F. Matsuda}
 \affiliation{Kavli Institute for the Physics and Mathematics of the Universe (WPI), The University of Tokyo Institutes for Advanced Study, The University of Tokyo, Kashiwa, Chiba 277-8583, Japan}
 
\author{T. Matsumura}
 \affiliation{Kavli Institute for the Physics and Mathematics of the Universe (WPI), The University of Tokyo Institutes for Advanced Study, The University of Tokyo, Kashiwa, Chiba 277-8583, Japan}

\author{Y. Sakurai}
 \affiliation{Kavli Institute for the Physics and Mathematics of the Universe (WPI), The University of Tokyo Institutes for Advanced Study, The University of Tokyo, Kashiwa, Chiba 277-8583, Japan}
 
\author{R. Tat}
 \altaffiliation{Currently at the California Institute of Technology.}
  \affiliation{Department of Physics, University of California, Berkeley, CA 94720, USA}
 \affiliation{Physics Division, Lawrence Berkeley National Laboratory, Berkeley, CA 94720, USA}
 
\author{Y. Zhou}
 \affiliation{Department of Physics, University of California, Berkeley, CA 94720, USA}

\date{\today}

\begin{abstract}
We present the design and laboratory evaluation of a cryogenic continuously rotating half-wave plate (CHWP) for the POLARBEAR-2b (PB-2b) cosmic microwave background (CMB) receiver, the second installment of the Simons Array. PB-2b will observe at 5,200~m elevation in the Atacama Desert of Chile in two frequency bands centered at 90 and 150~GHz. In order to suppress atmospheric 1/f noise and mitigate systematic effects that arise when differencing orthogonal detectors, PB-2b modulates linear sky polarization using a CHWP rotating at 2~Hz. The CHWP has a 440~mm clear aperture diameter and is cooled to $\approx$~50~K in the PB-2b receiver cryostat. It consists of a low-friction superconducting magnetic bearing (SMB) and a low-torque synchronous electromagnetic motor, which together dissipate <~2~W. During cooldown, a grip-and-release mechanism centers the rotor to <~0.5~mm, and during continuous rotation, an incremental optical encoder measures the rotor angle with a noise level of 0.1~$\mathrm{\mu rad / \sqrt{Hz}}$. We discuss the experimental requirements for the PB-2b CHWP, the designs of its various subsystems, and the results of its evaluation in the laboratory. The presented CHWP has been deployed to Chile and is expected to see first light on PB-2b in 2020 or 2021.
\end{abstract}

\maketitle


\section{Introduction}
\label{sec:into}

Cosmic microwave background (CMB) polarization anisotropies have emerged as a powerful and versatile probe of cosmology. CMB polarization is decomposed into parity-even ``E-modes,'' which are sourced by primordial density fluctuations, and parity-odd ``B-modes,'' which are not.\cite{seljak_signature_1997} Primordial B-modes are, however, thought to be generated by inflation, a theorized epoch of rapid expansion $\sim$~$10^{-36}$~s after the Big Bang.\cite{guth_inflationary_1981,seljak_direct_1999,zaldarriaga_all-sky_1997} A detection of primordial B-modes would therefore confirm inflation and illuminate physics on grand unified energy scales.

In principle, B-modes offer a clean probe of inflation, but in reality, gravitational lensing\cite{kamionkowski_probe_1997,zaldarriaga_all-sky_1997,zaldarriaga_gravitational_1998} and polarized galactic emission\cite{planck_collaboration_planck_2018} produce B-mode foregrounds. Lensing B-mode power peaks on arcminute scales and is sensitive to cosmological constituents such as neutrinos.\cite{hu_mass_2002,lewis_weak_2006} Galactic B-mode power is predominantly generated by synchrotron and thermal dust emission, each of which has a distinct frequency spectrum.\cite{bennett_nine-year_2013,planck_collaboration_planck_2018} Lensing and galactic foregrounds obfuscate the inflationary signal, and as degree-scale polarization measurements improve,\cite{the_bicep-2_and_keck_collaborations_constraints_2018} B-mode foreground removal is becoming increasingly important. Delensing, or the process of removing lensing B-modes to search for primordial B-modes, requires exquisitely sensitive CMB maps with arcminute resolution and wide sky coverage.\cite{knox_limit_2002,kesden_separation_2002,seljak_gravitational_2004,sherwin_delensing_2015,wu_measurement_2019}  Equivalently, subtracting galactic B-modes requires multiple observation frequencies to leverage the unique spectral signatures of synchrotron and thermal dust emission.\cite{planck_collaboration_planck_2019} As a result, modern CMB experiments strive to cover wider ranges of angular scales and microwave frequencies.

The Simons Array (SA) is a CMB observatory located on Cerro Toco in the Atacama Desert of Chile at an elevation of 5,200~m. It aims to achieve state-of-the-art large-angular-scale sensitivity using multi-chroic, large-aperture telescopes, enabling both delensing and galactic foreground subtraction in the hunt for primordial B-modes. SA consists of three telescopes in total, each housing a dichroic POLARBEAR-2 (PB-2) receiver cryostat: PB-2a and PB-2b observe at 90 and 150~GHz, while PB-2c observes at 220 and 270~GHz. Each telescope has a 2.5~m primary mirror in an off-axis Gregorian configuration\cite{tran_comparison_2008} and generates 5.2, 3.5, 2.7, and 2.2 arcmin beams at 90, 150, 220, and 270~GHz, respectively.  Each receiver cryostat has a 0.5~m vacuum window aperture, three 4~K alumina reimaging lenses, a 4~K Lyot stop, and a 365~mm diameter, 0.3~K focal plane of 7,588 transition-edge sensor (TES) bolometers. PB-2a achieved first light in January 2019,\cite{kaneko_deployment_2020} PB-2b\cite{howe_design_2018} has been deployed to Chile, and PB-2c is under development.

PB-2b is the first SA receiver to employ a cryogenic half-wave plate (CHWP) polarization modulator, which is the subject of this paper. The following sections present the requirements, design, and evaluation of the PB-2b CHWP rotation system, leaving a discussion of the optical performance for a future publication. Sec.~\ref{sec:hwp_modulation} overviews HWP polarization modulation, Sec.~\ref{sec:requirements} presents the CHWP requirements, Sec.~\ref{sec:chwp_design} presents the design of all CHWP subsystems, Sec.~\ref{sec:lab_testing} presents the results of CHWP laboratory testing, and Sec.~\ref{sec:conclusion} discusses the implications of the presented research.


\section{Half-wave plate modulation}
\label{sec:hwp_modulation}

Precise characterization of celestial microwave polarization requires careful control of systematic effects. One such systematic effect for ground-based telescopes is atmospheric noise. Even in the high and dry conditions of Cerro Toco, the atmosphere is much brighter than the CMB and fluctuates both spatially and temporally due to clouds, turbulence, and weather systems.\cite{errard_modeling_2015} These fluctuations manifest as 1/f noise in the detector data that degrades sensitivity on large angular scales. Another such systematic effect is intensity-to-polarization leakage (I-to-P), which often arises from optical effects at non-normal incidence and from mismatched response between orthogonal polarimeters.\cite{shimon_cmb_2008} This I-to-P signal fluctuates with unpolarized intensity and is difficult to null when differencing detectors to extract linear sky polarization.

\begin{figure*}
    \includegraphics[width=0.98\linewidth, trim=1.2cm 7cm 1.2cm 4cm, clip]{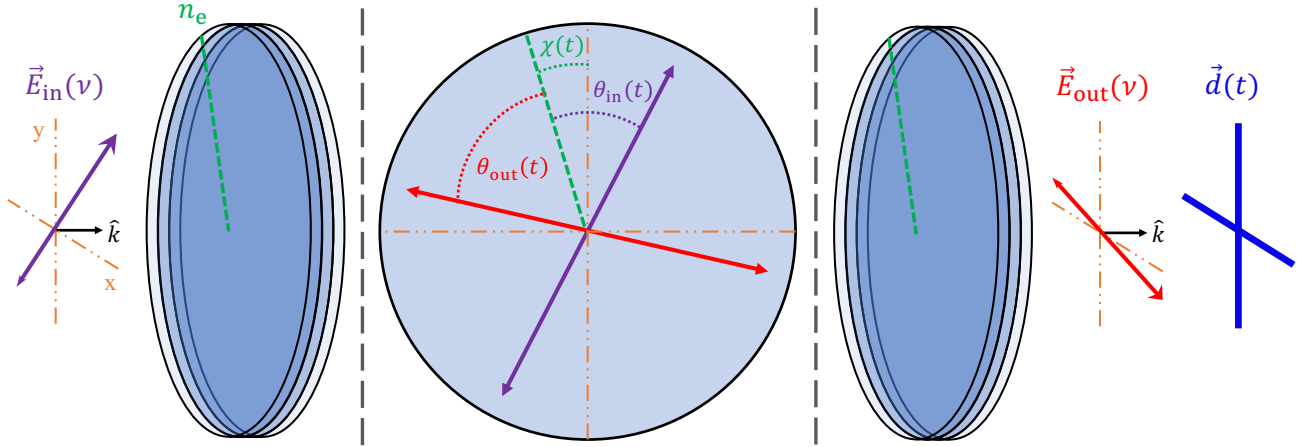}
    \caption{A cartoon of the sapphire achromatic HWP. Linearly polarized input light, with wave vector $\hat{k}$ and field vector $\vec{E}_{\mathrm{in}}(\nu)$, is rotated by twice its polarization angle $\theta_{\mathrm{in}}(t)$ with respect to the HWP axis---which coincides with the first sapphire piece's extraordinary crystal axis $n_{\mathrm{e}}$---plus a frequency-dependent phase $2 \phi(\nu)$. A continuously rotating HWP spins at a constant velocity $\mathrm{d} \chi / \mathrm{d} t = 2 \pi f_{\mathrm{HWP}}$, modulating the output field vector $E_{\mathrm{out}}(t)$ at $2 f_{\mathrm{HWP}}$ and the detected polarimeter power $\vec{d}(t)$ at $4 f_{\mathrm{HWP}}$.}
    \label{fig:hwp_cartoon}
\end{figure*}

Continuous polarization modulation is a common technique to mitigate 1/f noise and I-to-P leakage, and rotating half-wave plates (HWPs) are used widely in the millimeter and submillimeter community.\cite{johnson_maxipol_2007,klein_cryogenic_2011,kusaka_modulation_2014,moncelsi_empirical_2014,bryan_cryogenic_2016,ashton_interstellar_2017,takakura_performance_2017,hill_design_2016,ritacco_polarimetry_2017,harper_hawc_2018,bryan_optical_2018} An HWP is a birefringent medium geometrically tuned to generate a $\pi$ phase delay between light polarized along its ordinary and extraordinary axes. As depicted in Fig.~\ref{fig:hwp_cartoon}, a HWP rotates input linear polarization by
\begin{equation}
    \Delta \theta = -2 \left[ \theta_{\mathrm{in}} - \phi(\nu) \right] \, ,
\end{equation}
\noindent
where $\nu$ is the microwave frequency, $\theta_{\mathrm{in}}$ is the input polarization angle with respect to the HWP's extraordinary axis, and $\phi(\nu)$ is a phase factor that, in general, depends on frequency. When rotated, the HWP modulates the input linear polarization and distinguishes it from both intensity and downstream I-to-P. When rotated rapidly, the HWP up-converts the input signal above 1/f intensity fluctuations, in turn suppressing 1/f noise in the demodulated data. The modulated detector output is
\begin{equation}
    d_{\mathrm{m}}(t) = I_{\mathrm{in}}(t) + \epsilon \mathrm{Re} \{ [Q_{\mathrm{in}}(t) + i U_{\mathrm{in}}(t)] m(\chi(t)) \} \, ,
    \label{eq:det_mod}    
\end{equation}
\noindent
where $t$ is time, $I_{\mathrm{in}}(t)$, $Q_{\mathrm{in}}(t)$, and $U_{\mathrm{in}}(t)$ are the input Stokes vectors, $\epsilon$ is the HWP's linear polarization modulation efficiency, $\chi$ is the HWP's rotation angle, and 
\begin{equation}
    m(\chi) = \exp[-i 4 \chi]
\end{equation}
\noindent
is the modulation function. In this scheme, the modulation frequency is four times the HWP's rotation frequency
\begin{equation}
    f_{\mathrm{m}} = 4 f_{\mathrm{HWP}} \, ,
    \label{eq:carrier_freq}
\end{equation}
where $2 \pi f_{\mathrm{HWP}} = \mathrm{d} \chi / \mathrm{d} t$.

Naturally birefringent media, such as single-crystal sapphire, are inherently narrow-band: they preserve linear polarization for a single frequency but increasingly convert to elliptical polarization with increasing detection bandwidth. While wide-band HWPs can be engineered using capacitive grids\cite{pisano_achromatic_2006} or metamaterial layers,\cite{coughlin_metamaterial_2018} a common solution is the Pancharatnam sapphire achromatic HWP (AHWP).\cite{pancharatnam_achromatic_1955} In a sapphire AHWP, multiple $\alpha$-cut sapphire disks with tuned thicknesses are stacked such that their relative crystal-axis orientation is optimized for broadband linear polarization modulation efficiency.\cite{matsumura_performance_2009,komatsu_demonstration_2019} Sapphire HWPs have been demonstrated on both small- and large-aperture CMB instruments,\cite{essinger-hileman_systematic_2016,kusaka_results_2018,takakura_performance_2017,ThePOLARBEARCollaboration2020} making them an appealing technology for SA.


\subsection{The PB-2b cryogenic HWP}
\label{sec:pb2_chwp}

PB-2b adopts an achromatic sapphire continuously rotating \textit{cryogenic} HWP (CHWP) and builds on the experience of its predecessor experiments POLARBEAR (PB-1) and PB-2a. PB-1 uses an ambient-temperature sapphire HWP at its telescope's prime focus,\cite{takakura_performance_2017,ThePOLARBEARCollaboration2020} while PB-2a uses an ambient-temperature sapphire AHWP at its vacuum window.\cite{hill_design_2016} While an ambient HWP is practical to implement, its thermal emission substantially increases detected photon noise. Therefore, PB-2b cools its HWP to cryogenic temperatures to minimize parasitic optical power and maximize detector sensitivity.

While stepped CHWPs have deployed on a variety of mm and sub-mm experiments,\cite{moncelsi_empirical_2014,bryan_cryogenic_2016,ashton_interstellar_2017,harper_hawc_2018,bryan_optical_2018} the continuously rotating CHWP is an emerging technology.\cite{klein_cryogenic_2011,johnson_large-diameter_2017,hill_large-diameter_2018,Sakurai2018a} Operating a 500~mm-diameter, tens-of-kilogram, spinning instrument in a cryogenic vacuum space poses a menagerie of challenges, including differential thermal contraction, frictional dissipation, and cryo-mechanical durability. The PB-2b CHWP design addresses these challenges with several hardware advancements.

\begin{figure*}
\begin{subfigure}
    \centering
    \includegraphics[width=0.98\linewidth, trim=1.5cm 5.5cm 1.5cm 5cm, clip]{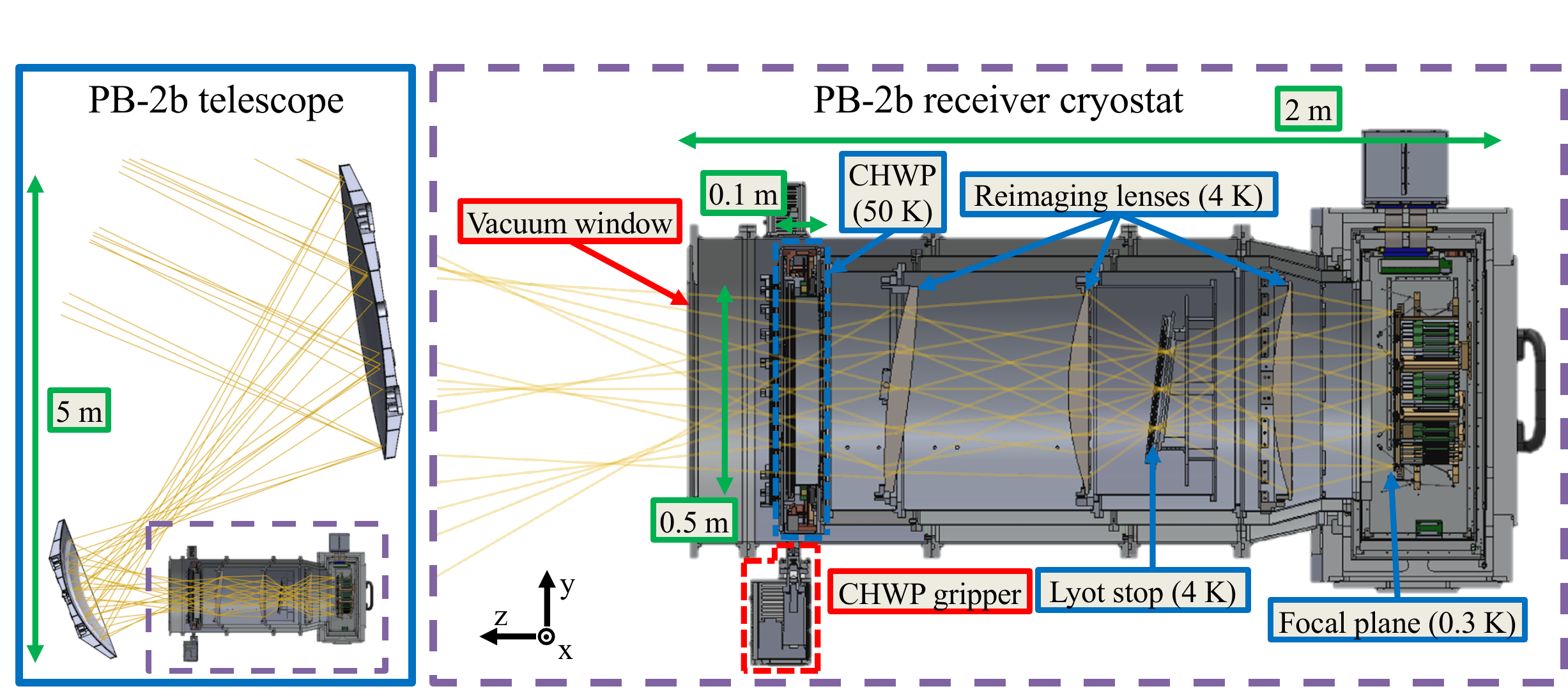}
    \caption{An optical ray trace of the PB-2b telescope (left) and receiver cryostat (right), superposed onto computer-aided-design (CAD) cross-sectional views. Red boxes mark warm components, and blue boxes mark cold components. The primary and secondary mirrors form an off-axis Gregorian configuration and create a sky image at the skyward-most field lens through a 0.5~m vacuum window. Three 4~K alumina lenses reimage onto a telecentric, 0.3~K focal plane, and a 4~K Lyot stop defines the primary mirror illumination. The CHWP is located between the secondary mirror and the field lens and has a 440~mm clear aperture diameter.}
    \label{fig:pb2_ray_trace}
\end{subfigure}
\begin{subfigure}
    \centering
    \includegraphics[width=0.98\linewidth, trim=1.5cm 4.7cm 1.5cm 3cm, clip]{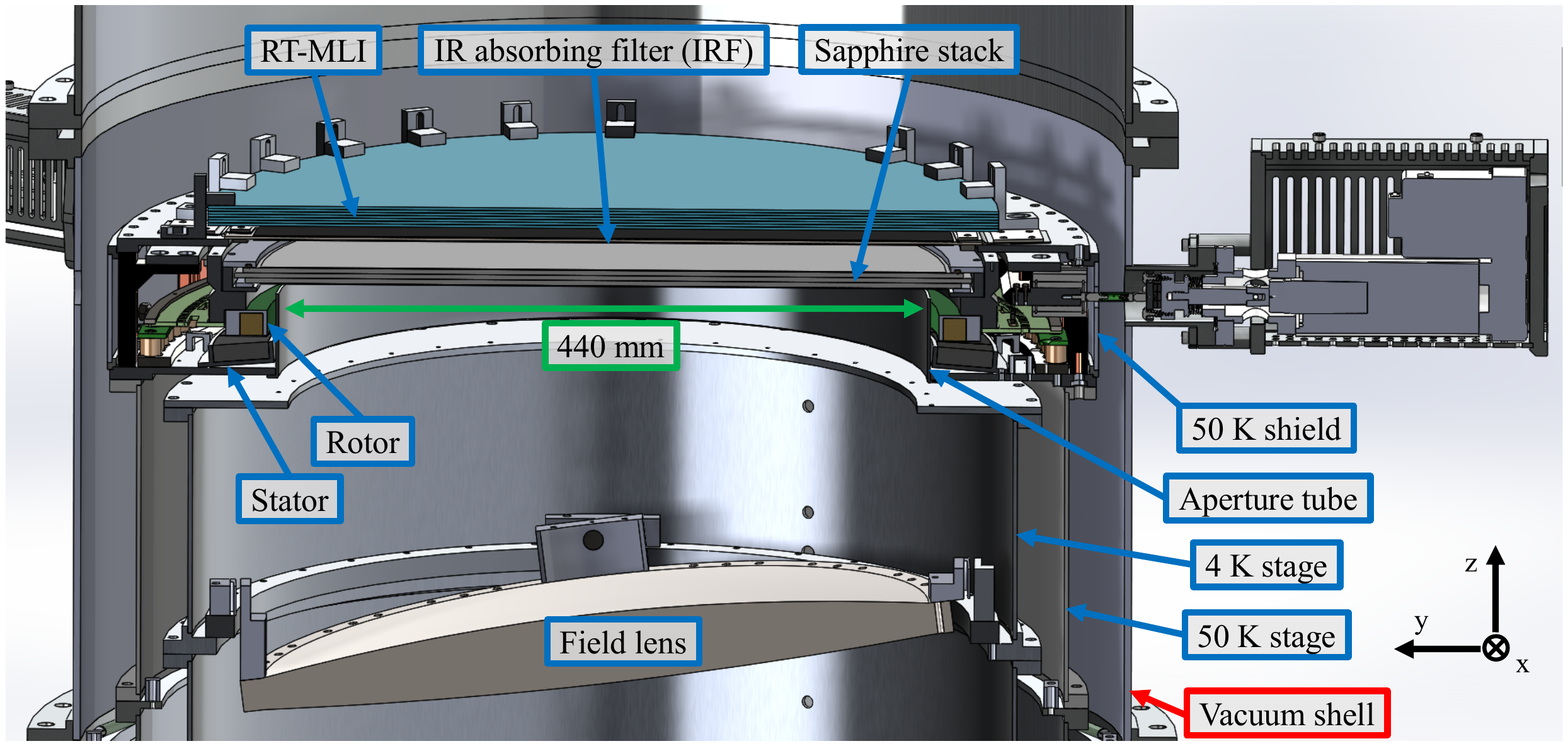}
    \caption{A zoomed CAD cross section of the CHWP integrated into the PB-2b optics tube, highlighting CHWP-relevant optical and thermal components. Red boxes mark warm parts, and blue boxes mark cold parts. The CHWP's clear aperture is defined by a stationary, HR-10-lined aperture tube (absorber not shown) that hides non-optical rotating components from the telescope's beam.}
    \label{fig:chwp_render}
\end{subfigure}
\end{figure*}

As shown in Figs.~\ref{fig:pb2_ray_trace}~and~\ref{fig:chwp_render}, the PB-2b CHWP is located within the receiver cryostat's optics tube, which has a 0.8~m diameter vacuum shell and two cryogenic stages cooled by a pulse tube refrigerator\footnote{Cryomech PT415: https://www.cryomech.com/products/pt415/} (PTR) to $\approx$~50~K and $\approx$~4~K. The CHWP operates on the 50~K stage near the telescope's Gregorian focus in front of three reimaging lenses and behind a vacuum window, radio-transmissive multi-layer insulation (RT-MLI),\cite{choi_radio-transparent_2013} and an alumina infrared absorbing filter (IRF).\cite{inoue_cryogenic_2014} Its drive system consists of a superconducting magnetic bearing and a synchronous electromagnetic motor, and its rotation angle is monitored by a slot-chopped optical encoder. In the following sections, we discuss the requirements and design of the CHWP, as well as its performance during laboratory evaluation.

\section{CHWP requirements}
\label{sec:requirements}

The CHWP must meet several optical, thermal, noise, and operational requirements in order to effectively mitigate 1/f noise and I-to-P leakage while not introducing its own systematic effects. These requirements are central to the CHWP design and depend on the specifics of PB-2b's telescope, cryogenic, and detector systems. 

When it was conceived, PB-2b was virtually a copy of PB-2a,\cite{matsumura_polarbear-2_2012} which does not use a CHWP. As a result, the CHWP is retrofitted to the original PB-2b design and centers around maintaining ``heritage'' system performance. Table~\ref{tab:requirements} summarizes the PB-2b CHWP's numerical requirements and achieved values. In the following subsections, we detail the CHWP's design drivers and how they flow into its hardware specifications. 

\begin{table}
\caption{The CHWP's numerical requirements and achieved values.}
\label{tab:requirements}
\begin{ruledtabular}
\begin{tabular}{l l l}
\textbf{Parameter} & \textbf{Requirement} & \textbf{Achieved} \\
\hline
Assembly outer diameter & $\leq$ 700~mm & 690~mm \\
\hline
Assembly height & $\leq$ 100~mm & 95~mm \\
\hline
Clear aperture diameter & $\geq$ 430~mm & 440~mm \\
\hline
Rotor temperature $T_{\mathrm{rotor}}$ & $\leq$ 55~K & < 53~K \\
\hline
Thermal dissipation $P_{\mathrm{stator}}$ & $\leq$ 2~W & < 1.3 W \\
\hline
Rotor thermalization time & $\leq$ 36~hr & 5~hr \\
\hline
Rotation frequency $f_{\mathrm{HWP}}$ & $\geq$ 2~Hz & $\leq$ 2.8~Hz \\
\hline
Angle encoder noise $\sigma_{\chi}$ & $\ll$ 3 $\mathrm{\mu rad / \sqrt{Hz}}$ & 0.1~$\mathrm{\mu rad / \sqrt{Hz}}$ \\
\hline
B-field interference $\sigma_{\mathrm{B}}$ @ 4$f_{\mathrm{HWP}}$ & $\ll$~20~$\mathrm{\mu G / \sqrt{Hz}}$ & <~10~$\mathrm{\mu G / \sqrt{Hz}}$\\
\hline
$T_{\mathrm{rotor}}$ stability $\sigma_{T_{\mathrm{rotor}}}$ @ 1~mHz & $\ll$ 1~$\mathrm{mK / \sqrt{Hz}}$ & < 0.3~$\mathrm{mK / \sqrt{Hz}}$ \\
\end{tabular}
\end{ruledtabular}
\end{table}


\subsection{Spatial and optical requirements}
\label{sec:spatial_requirements}

The initial set of requirements are on the CHWP's physical dimensions. As shown in Figs.~\ref{fig:pb2_ray_trace} and~\ref{fig:chwp_render}, the assembly must fit within the confines of the optics tube's vacuum shell, limiting its cryogenic diameter to $\leq$~700~mm. Additionally, its height, or dimension along the z-axis, must be $\leq$~100~mm so that the window aperture does not extend too far toward the secondary mirror, where the telescope's beam~\footnote{The telescope's beam is defined to be the collection of reverse-time-sense rays from all detectors on the focal plane.} diverges more rapidly than within the receiver. These dimensional constraints influence much of how the CHWP subsystems are built and arranged, as well as their clearances and alignment tolerances.

\begin{figure}[!t]
    \centering
    \includegraphics[width=0.98\linewidth, trim=0cm 4cm 0cm 5.7cm, clip]{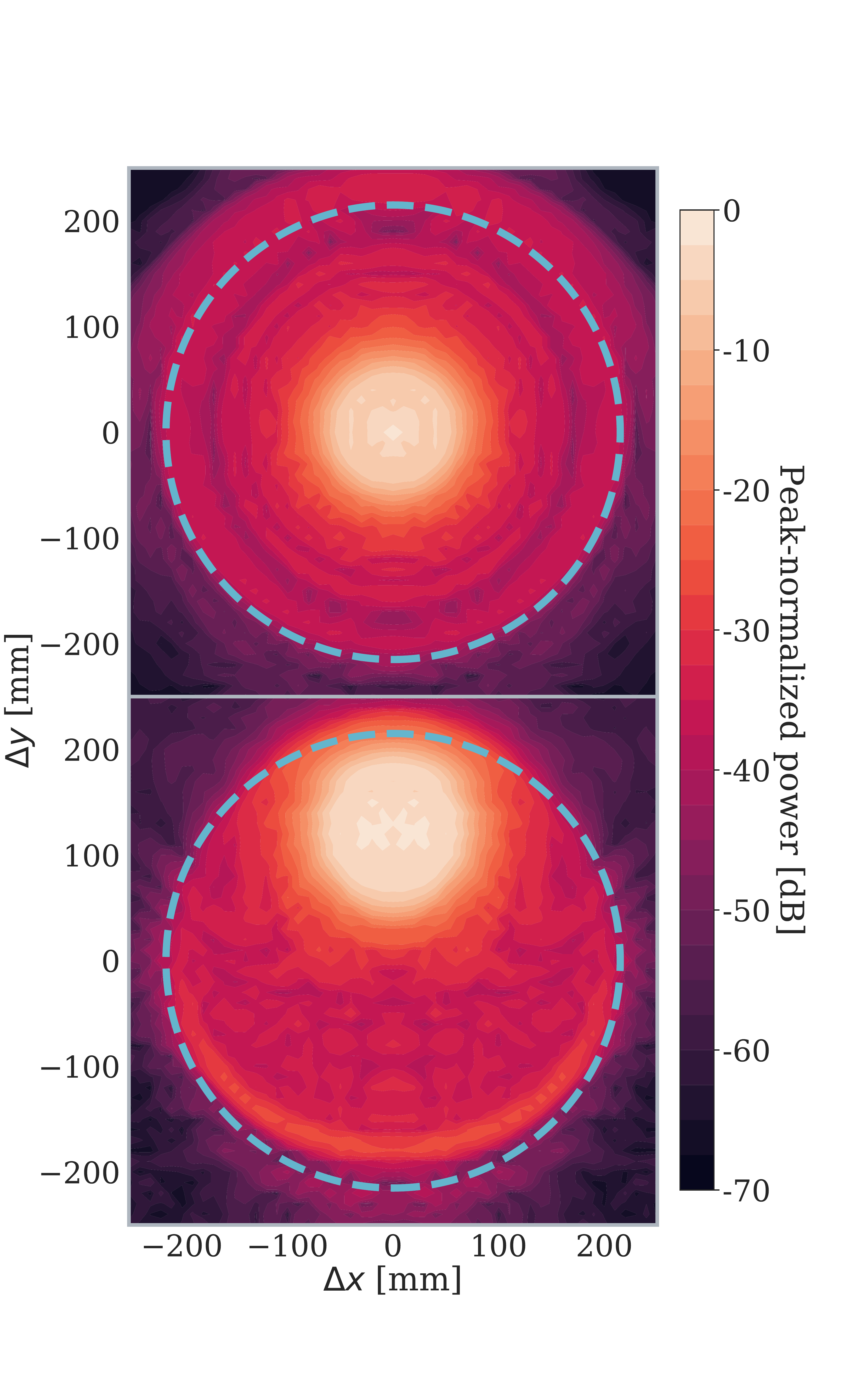}
    \caption{The reverse-time-sense, peak-normalized, 90~GHz, $x$-polarization illumination pattern onto the CHWP aperture plane from detector pixels at the center (top panel) and $-y$ edge (bottom panel) of the focal plane. The 430~mm aperture requirement is marked with a dashed cyan circle about the telescope's chief ray, and the total spillover for the center (edge) pixel is 0.2\% (0.1\%).}
    \label{fig:chwp_aperture}
\end{figure}

While a presentation of sapphire-stack requirements is beyond the scope of this paper, we do discuss the clear aperture diameter, which is central to the cryo-mechanical design. The CHWP is located sky-side of the field lens (see Fig.~\ref{fig:pb2_ray_trace}), and its clear aperture must be large enough to admit the beam between the secondary mirror and the Gregorian focus. We use GRASP\footnote{GRASP: https://www.ticra.com/software/grasp/} to simulate the polarized, 90~GHz response of detector pixels across the focal plane, and Fig.~\ref{fig:chwp_aperture} shows the $x$-polarization illumination of the central and $-y$ edge pixels at the CHWP aperture. To set a requirement, we vary the CHWP aperture diameter and compare the telescope's far-field beam to that with the CHWP aperture removed. We evaluate a collection of merit figures, including angular resolution, spill over the primary mirror, side-lobe amplitude, and differential pointing and ellipticity.\cite{shimon_cmb_2008} We find that a 430~mm aperture has no impact on the central pixel and has edge-pixel impacts that are subdominant to preexisting systematic effects in the telescope optics. Therefore, we require that the CHWP aperture encapsulate a 430~mm diameter circle about the telescope's chief ray, including a margin for manufacturing and alignment tolerances.

Additionally, we require all rotating, non-optical CHWP components to be hidden from the beam in order to minimize rotation-synchronous signals. To satisfy this requirement, the bearing and sapphire must be sufficiently oversized to encapsulate the optical baffling that defines the CHWP aperture, as shown in Fig.~\ref{fig:chwp_render}. 


\subsection{Thermal requirements}
\label{sec:thermal_requirements}

\begin{figure}[!t]
    \includegraphics[width=0.98\linewidth, trim=0cm 1cm 3cm 3cm, clip]{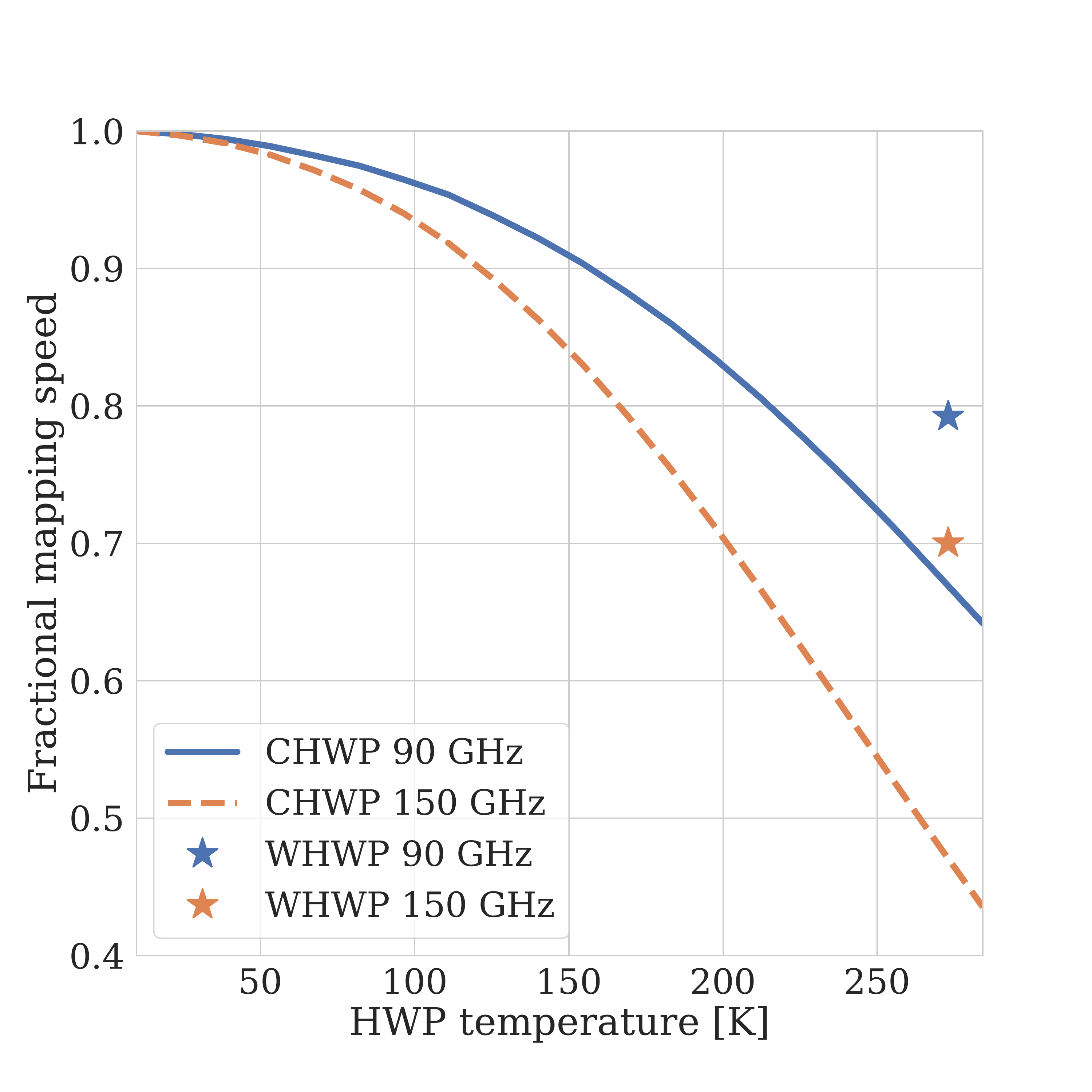}
    \caption{The fractional impact of rotor temperature on PB-2b mapping speed with respect to that of a 4~K HWP, compared to that of a PB-2a-style warm HWP (WHWP) at 273~K.\cite{hill_design_2016} Note that the mapping speed gain of a 4~K CHWP over a 50~K CHWP is $\approx$ 1\%.}
\label{fig:mapping_speed} 
\end{figure}

The next set of requirements are on the CHWP's thermal impact. Because the PB-2b detectors are expected to be photon-noise limited, CHWP-induced optical power can dramatically degrade experiment sensitivity. Fig.~\ref{fig:mapping_speed} shows the simulated\cite{hill_bolocalc_2018} fractional impact of rotor temperature on PB-2b mapping speed, compared to that of a PB-2a-style warm HWP (WHWP).\cite{hill_design_2016} Mapping speed quantifies the number of detector-hours needed to reach a specified CMB map depth, and therefore fractional mapping speed is analogous to detector yield or observation efficiency. In this simulation, the CHWP is assumed to have an epoxy-based anti-reflection (AR) coating,\cite{rosen_epoxy-based_2013} and sensitivity degradation with increasing CHWP temperature is primarily due to increasing mm-wave emission, which in turn increases detected photon noise. As shown by the WHWP points, some of the sensitivity loss at $\approx$~270~K can be reclaimed by optimizing the AR coating for lower ambient emissivity, but the biggest sensitivity gain comes from cooling the HWP to cryogenic temperatures. Mapping speed retention is >~97\% when the CHWP is at <~50~K, and the sensitivity loss vs. CHWP temperature is gentle <~100~K. Therefore, the dominant driver of the rotor's temperature is not the CHWP's mm-wave emission but instead its IR emission onto the field lens. 

In the heritage PB-2b design without the CHWP, the 4~K field lens is radiatively loaded by the IRF (see Fig.~\ref{fig:chwp_render}). Therefore, if the CHWP becomes warmer than the IRF or if its dissipation onto the 50~K stage is too large, the CHWP-induced 4~K load will exceed that of the heritage configuration, which may in turn increase lens and Lyot temperatures and even degrade the performance of the sub-kelvin stage. Sec.~\ref{sec:thermal_design} and App.~\ref{app:thermal_sim} discuss the CHWP thermal system in detail, and Fig.~\ref{fig:rotor_temp} shows the relationship between rotor temperature and field-lens load. We require that the 4~K load not exceed that of the heritage configuration at 85\% confidence, in turn requiring the rotor temperature to be
\begin{equation}
    T_{\mathrm{rotor}} \leq 55 \; \mathrm{K} \, .
    \label{eq:rotor_temp}
\end{equation}

It is unavoidable that dissipation from the CHWP motor and bearing will warm the 50~K stage, raising IRF and rotor temperatures. The goal, however, is to keep this temperature rise small enough to not warm 4~K components. Utilizing load-curve measurements of the PB-2b optics tube,\cite{howe_polarbear-2_2019} we mandate that the CHWP warm the 50~K stage <~1~K with respect to the heritage configuration and therefore set the CHWP's 50~K dissipation requirement to be
\begin{equation}
    P_{\mathrm{stator}} \leq 2 \; \mathrm{W} \, .
    \label{eq:stator_power}
\end{equation}

We next require that the CHWP not add to the PB-2b cooldown time. This mandate is important both for restarting observations following a disruption in the field and for rapid-turnaround receiver testing in the lab. The heritage PB-2b system takes about five days to cool, and that duration is limited by the cooldown of the focal plane, which reaches 4~K about 36~hours after the optics tube.\cite{howe_polarbear-2_2019} The CHWP's cooldown time is limited by that of the rotor, which has a large thermal mass and is heatsinked by unfastened interfaces (see Sec.~\ref{sec:gripper_design}). Therefore, we require that the rotor reach its base temperature before the focal plane has thermalized, or within 36~hours of the 50~K stage.

Finally, we require that CHWP operation not vibrate the focal plane structure. CHWP-induced vibrations can generate microphonic heating that may modulate the focal plane's temperature and degrade detector gain stability.\cite{howe_polarbear-2_2019} To avoid such issues, we require no measurable difference in focal-plane temperature between when the rotor is spinning and when it is not.


\subsection{Noise requirements}
\label{sec:noise_requirements}

In order for the CHWP to effectively suppress 1/f noise, $f_{\mathrm{m}} = 4 f_{\mathrm{HWP}}$ must be large enough such that all modulation sidebands, which are contained within $4 f_{\mathrm{HWP}} \pm \Delta f$, are above the Atacama's 1$\sim$2~Hz knee frequency,\cite{takakura_performance_2017,kusaka_modulation_2014} or the frequency at which 1/f noise power and white noise power are equal. Assuming a $0.4^{\circ}$/s telescope scan speed,\cite{the_polarbear_collaboration_measurement_2020,ThePOLARBEARCollaboration2020} the sky is fully resolved by $\approx$~4~Hz of temporal bandwidth.\cite{takakura_performance_2017} Therefore, we require
\begin{equation}
    f_{\mathrm{HWP}} \geq 2 \; \mathrm{Hz}
    \label{eq:rot_freq_requirement} \, ,
\end{equation}
\noindent
which places the demodulation band $4f_{\mathrm{HWP}} \pm \Delta f = 8 \pm 4$~Hz comfortably clear of atmospheric fluctuations. PB-1 and the Atacama B-mode Search (ABS) also rotate their HWPs at $\approx$~2~Hz in Chile and achieve excellent 1/f suppression in their demodulated detector data.\cite{takakura_performance_2017,kusaka_modulation_2014}

While rejecting optical 1/f noise, the CHWP introduces other noise sources that can degrade data quality. These CHWP-induced noise sources are largely common-mode across the focal plane and hence do not average down during detector coaddition. Therefore, we require the noise equivalent CMB temperature ($\mathrm{NET_{CMB}}$) of each CHWP noise source to be much less than the forecasted PB-2b array-averaged $\mathrm{NET_{CMB}}$\cite{suzuki_polarbear-2_2016}
\begin{equation}
    \mathrm{NET_{CMB}^{HWP}} \ll \mathrm{NET_{CMB}^{arr}} = 5.8 \; \mathrm{\mu K_{CMB} / \sqrt{Hz}} \, .
    \label{eq:arr_noise}
\end{equation}
\noindent
We use this bound to set requirements on three primary CHWP contaminants: encoder angle jitter, magnetic interference, and rotor temperature stability.

Firstly, encoder angle noise injects noise into the demodulated detector data. The CHWP is located behind the telescope's off-axis mirrors, which induce I-to-P along the $y$ direction (see Fig.~\ref{fig:pb2_ray_trace}). This I-to-P is modulated at 4$f_{\mathrm{HWP}}$, and an imperfect rotor angle measurement will imperfectly subtract the telescope signal during demodulation. PB-1 measures up to $\approx$~180~$\mathrm{mK_{\mathrm{RJ}}}$ of I-to-P at the prime focus of an SA-style telescope, largely due to emission from the primary reflector.\cite{takakura_performance_2017} Because the PB-2b CHWP is located behind the primary \textit{and} secondary mirrors, we conservatively\footnote{In reality, we expect the I-to-P from the secondary reflector to be less than that of the primary due to its being ellipsoidal with a smaller incident angle. However, the secondary mirror's emission has not been studied in detail, and so we assert that each mirror generates similar I-to-P leakage in order to set a conservative angle jitter requirement.} assert an I-to-P amplitude of
\begin{equation}
    A_{4}^{0} \sim 400 \; \mathrm{mK_{\mathrm{RJ}}} \, .
    \label{eq:4f_amp}
\end{equation}
\noindent
This mirror polarization is modulated at 4$f_{\mathrm{HWP}}$, and a mismeasurement $\Delta \chi$ of the CHWP angle $\chi$ modifies the 4$f_{\mathrm{HWP}}$ amplitude $A_{4}(\chi)$ as
\begin{eqnarray}
    A_{4}(\chi) &&= A_{4}^{0} \mathrm{cos}(4 \left[ \chi + \Delta \chi \right] ) \nonumber \\ &&\approx A_{4}^{0} \mathrm{cos}(4 \chi) - 4 \, A_{4}^{0} \mathrm{sin}(4 \chi) \Delta \chi \, ,
    \label{eq:freq_mod}
\end{eqnarray}
\noindent
such that
\begin{equation}
    \Delta A_{4} = -4 \, A_{4}^{0} \mathrm{sin}(4 \chi) \Delta \chi \, .
    \label{eq:delta_chi}
\end{equation}
Assuming that the angle jitter is random, its average effect on the detected noise spectrum is
\begin{equation}
    \mathrm{NET_{CMB}^{HWP}} = 2 \sqrt{2} \,  \frac{\mathrm{d} T_{\mathrm{CMB}}}{\mathrm{d} T_{\mathrm{RJ}}} \, A_{4}^{0} \, \sigma_{\chi} \, ,
    \label{eq:encoder_noise}
\end{equation}
\noindent
where $\sigma_{\chi}$ is the root mean square (RMS) of $\Delta \chi$ and $\mathrm{d} T_{\mathrm{CMB}} / \mathrm{d} T_{\mathrm{RJ}} = 1.7$ is evaluated at 150~GHz. Mandating that $\mathrm{NET_{CMB}^{HWP}} \ll \mathrm{NET_{CMB}^{arr}}$ (see Eq.~\ref{eq:arr_noise}), the angle noise requirement is
\begin{equation}
    \sigma_{\chi} \ll 3 \, \mathrm{\mu rad / \sqrt{Hz}} \, .
\end{equation}

Secondly, CHWP-induced magnetic interference can also inject noise into the demodulated data. The PB-2b detector array consists of aluminum manganese (AlMn) transition-edge sensors (TESs) amplified by superconducting quantum interference devices (SQUIDs). Both AlMn TESs and SQUIDs are sensitive to ambient magnetic fields, and because the PB-2b CHWP consists of a magnetic bearing and an electromagnetic motor (see Sec.~\ref{sec:chwp_design}), magnetic interference at the focal plane must be carefully controlled.

PB-2b uses gradiometric Series SQUID Array Amplifiers (SSAAs)\cite{stiehl_time-division_2011} fed back using Digital Active Nulling,\cite{haan_improved_2012} stabilized by a low-frequency flux-locked loop,\cite{Bender2014} and shielded via a combination of $\mu$-metal and superconducting niobium.\cite{xu_combined_1987}  This particular SQUID configuration is robust to time-varying magnetic fields, especially those which are uniform over each SQUID's $\sim$~10~$\mathrm{mm^{2}}$ area. Therefore, the more considerable CHWP-specific concern is magnetic interference in the detectors themselves.\footnote{Private communication with Prof. Tijmen de Haan (KEK High Energy Research Organization) and Prof. Darcy Barron (University of New Mexico), May 2020.}

Magnetic pickup at 4$f_{\mathrm{HWP}}$ mimics modulated sky polarization, especially if the magnetic signal drifts over the course of an observation. The superconducting transition temperature $T_{\mathrm{c}}$ of AlMn varies slightly with ambient magnetic field $B$,\cite{deiker_superconducting_2004} and a change in $T_{\mathrm{c}}$ changes the bolometer's saturation power $P_{\mathrm{sat}}$, which in turn mimics a change in optical power $P_{\mathrm{opt}}$. We can convert the CHWP-induced magnetic jitter $\sigma_{\mathrm{B}}$ to $\mathrm{NET_{CMB}}$ as
\begin{equation}
    \mathrm{NET_{CMB}^{HWP}} = \frac{\mathrm{d} T_{\mathrm{CMB}}}{\mathrm{d} P_{\mathrm{opt}}} \frac{\mathrm{d} P_{\mathrm{sat}}}{\mathrm{d} T_{\mathrm{c}}} \frac{\mathrm{d} T_{\mathrm{c}}}{\mathrm{d} B} \sigma_{B} \, .
    \label{eq:pbias_bfield}
\end{equation}
\noindent
Measurements of the PB-2b AlMn TESs give $\mathrm{d} T_{\mathrm{c}} / \mathrm{d} B = 0.3$~mK/G\cite{vavagiakis_magnetic_2018} and $\mathrm{d} P_{\mathrm{sat}} / \mathrm{d} T_{\mathrm{c}} = 0.1$~pW/mK,\cite{westbrook_polarbear-2_2018} while simulations of the PB-2b optics\cite{hill_bolocalc_2018} give $\mathrm{d} T_{\mathrm{CMB}} / \mathrm{d} P_{\mathrm{opt}} = 12$~$\mathrm{\mu K_{CMB} / aW}$. We again require that $\mathrm{NET_{CMB}^{HWP}} \ll \mathrm{NET_{CMB}^{arr}}$ such that 
\begin{equation}
    \sigma_{\mathrm{B}} \; @ \; 4f_{\mathrm{HWP}} \ll 20 \; \mathrm{\mu G / \sqrt{Hz}} \, .
    \label{eq:mag_requirement}
\end{equation}
\noindent

The final noise requirement is on the CHWP rotor temperature stability.  Thermal emission from the CHWP can be modulated at 4$f_{\mathrm{HWP}}$ due to non-uniformities in the sapphire stack.\cite{salatino_studies_2018} In turn, variations in CHWP temperature will modulate the amplitude of this thermal signal, mimicking polarized sky fluctuations in the demodulated detector data. At 50~K, the thermal 4$f_{\mathrm{HWP}}$ amplitude is expected to arise predominantly due to thickness and index variations in the AR coating, as sapphire is very transparent at mm wavelengths and low temperatures.\cite{parshin_silicon_1995,braginsky_experimental_1987} While a detailed study of AR coating uniformity is beyond the scope of this paper, we can set an empirical limit by once again invoking the PB-1 HWP's 4$f_{\mathrm{HWP}}$ amplitude (see Eq.~\ref{eq:4f_amp}), which is $\sim$~10\% of its thermal emission.\cite{takakura_performance_2017} We assume that the PB-2b CHWP's 4$f_{\mathrm{HWP}}$ thermal signal will also be 10\% of its thermal emission to set a conservative\footnote{While a substantial portion of the PB-1 4$f_{\mathrm{HWP}}$ signal is due to polarized emission from the primary mirror (as discussed earlier in Sec.~\ref{sec:requirements}), we assume here that it is entirely due to HWP non-uniformity in order to set a conservative requirement on CHWP temperature stability.} requirement on its temperature stability.

Rotor temperature fluctuations $\sigma_{T_{\mathrm{rotor}}}$ can be converted to $\mathrm{NET_{CMB}}$ as
\begin{equation}
     \mathrm{NET_{CMB}^{HWP}} = \frac{0.1 \epsilon}{\eta} \frac{\mathrm{d} T_{\mathrm{CMB}}}{\mathrm{d} T_{\mathrm{RJ}}} \sigma_{T_{\mathrm{rotor}}} \, ,
\end{equation}
\noindent
where $\epsilon = 0.02$ is the assumed CHWP emissivity,~\cite{rosen_epoxy-based_2013} $\eta = 0.9$ is the optical efficiency of the telescope + atmosphere between the CHWP and the CMB, and $\mathrm{d} T_{\mathrm{CMB}} / \mathrm{d} T_{\mathrm{RJ}} = 1.7$ is evaluated at 150~GHz. Noting that rotor temperature fluctuations follow a 1/f spectrum, we require $\mathrm{NET_{CMB}^{HWP}} \ll \mathrm{NET_{CMB}^{arr}}$ on 1,000~s timescales---long enough to encapsulate many long-baseline telescope traversals---which imposes a rotor temperature stability requirement of 
\begin{equation}
    \sigma_{T_{\mathrm{rotor}}} \; @ \; 1 \; \mathrm{mHz} \ll 1~\mathrm{mK / \sqrt{Hz}} \, .
    \label{eq:chwp_temp_requirement}
\end{equation}


\subsection{Operational requirements}
\label{sec:operational_requirements}

As the only moving component inside of the receiver cryostat, the CHWP poses unique risks to experiment operations. PB-2b intends to observe for at least three years in Chile, and the CHWP must operate robustly for as long as it is needed. This situational constraint gives rise to several operational requirements, which we highlight below.

First, the CHWP must be robust against component degradation. Because warming, opening, and cooling the receiver cryostat cost weeks of observation time, the rotor should be able to undergo >~100,000,000 revolutions without being serviced. This robustness requirement drives many aspects of the CHWP design, such as a zero-contact drive and multiple hardware redundancies. Second, standard CHWP operation must require no human intervention. Critical monitoring data, such as encoder data packet drops, rotational velocity, temperatures, and gripper status, are automatically logged. Additionally, the CHWP control software comprises a system of classes, configuration files, and command-line executables designed to interface with the observatory control program in order to fully automate routine operations. Rotor temperature monitoring is particularly important and is discussed in Sec.~\ref{sec:rotor_thermometry}.

Third, we mandate an automated shutdown procedure to keep the rotor centered if site power is lost. Weather events are not uncommon on Cerro Toco and can cause sudden generator failures. If PTR cooling terminates and site network access is unavailable, the CHWP must automatically stop and stow such that it can reliably restart with minimal performance degradation. Shutdown and recovery testing was central to system evaluation and is discussed in Sec.~\ref{sec:shut_down}. Fourth, in the event of critical equipment failures, an issue with the sapphire stack, or an otherwise unforeseen problem, the CHWP must be simple to service on the telescope. This requirement motivates the CHWP's location near the vacuum window where it can be accessed with minimal disassembly, as discussed in Sec.~\ref{sec:assembly_procedure}.


\section{CHWP design}
\label{sec:chwp_design}

\begin{figure}[!t]
    \begin{subfigure}
        \centering
        \includegraphics[width=0.98\linewidth, trim={3.5cm, 7.5cm, 4.5cm, 6.5cm}, clip]{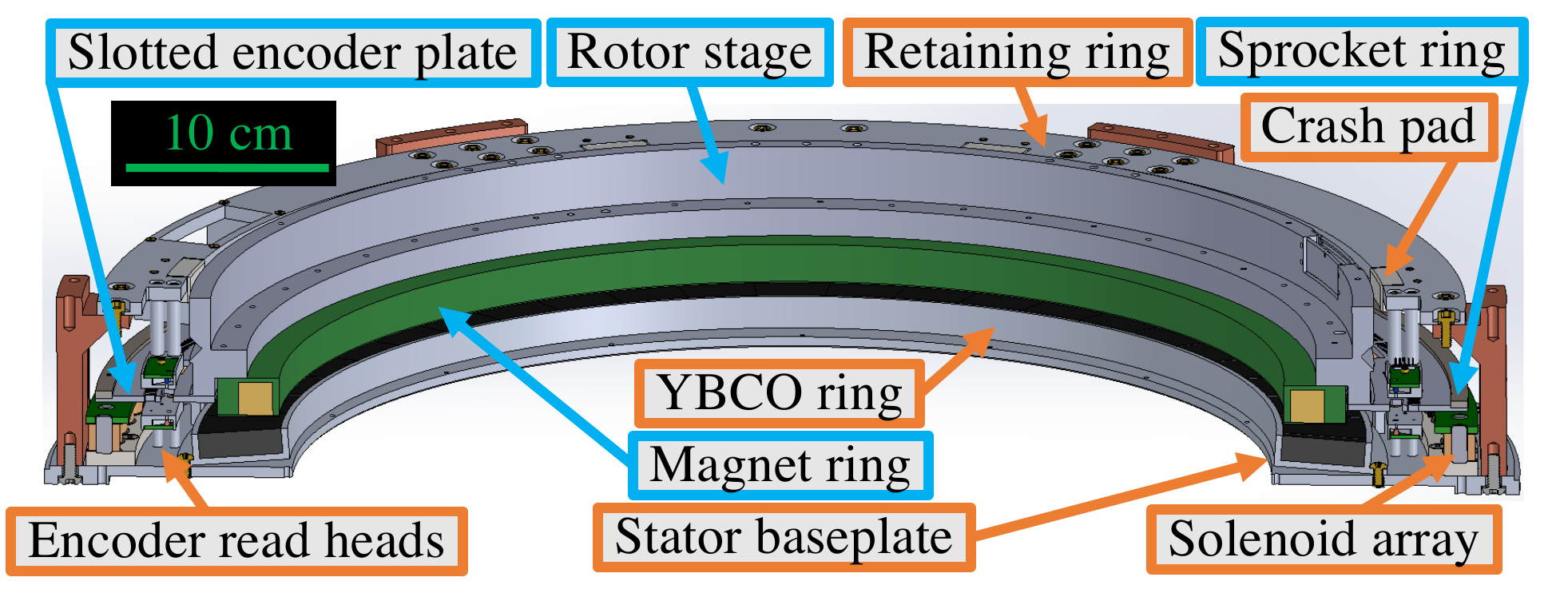}
    \end{subfigure}
    \begin{subfigure}
        \centering
        \includegraphics[width=0.98\linewidth, trim={3.5cm, 4.0cm, 5.5cm, 3.5cm}, clip]{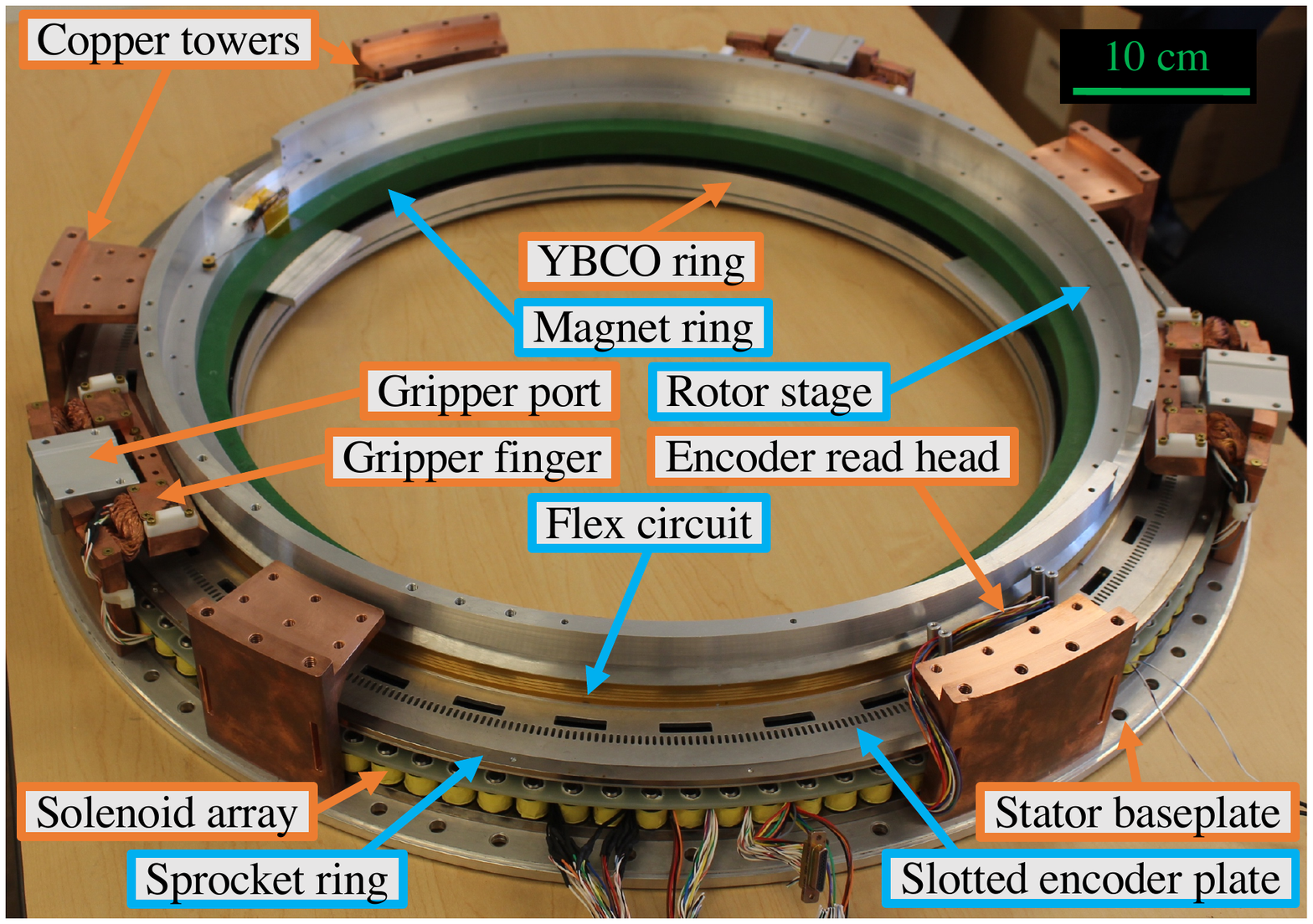}
    \end{subfigure}
    \caption{The CHWP rotation mechanism without the sapphire stack. Rotating components are labeled with cyan boxes, and non-rotating components are labeled with orange boxes. \textbf{Top panel:} a CAD cross section of the rotation mechanism with the rotor floating 5~mm above the stator. The YBCO ring is buckled due to residual thermal stress between the YBCO tiles and their aluminum fixture. \textbf{Bottom panel:} a photograph of the rotation mechanism with the retaining ring removed and the gripper ports, gripper fingers, and flex circuit added. The magnet and YBCO rings are separated by three 4~mm shims. All internal wiring, including that of the solenoids, photodiodes, LEDs, and thermometers, exit the assembly via four 25-pin Micro-D connectors at the bottom of the photo.}
    \label{fig:chwp_tabletop}
\end{figure}

Given the requirements presented in Sec.~\ref{sec:requirements}, we now discuss the design of the CHWP mechanism, shown in Figs.~\ref{fig:chwp_render} and~\ref{fig:chwp_tabletop}. The CHWP consists of a rotor, stator, motor, encoder, and gripper, all integrated into a compact assembly on the 50~K stage. The following sections detail the designs of each subsystem.


\subsection{Location}
\label{sec:location}

The CHWP is located on the 50~K stage, which has several advantages over possible 4~K locations, such as at the Lyot stop. First, the CHWP is skyward of all reimaging optics, mitigating the impact of Fresnel I-to-P generated at the lens surfaces. Second, at 50~K, the sapphire stack's mm-wave emission is already deeply subdominant to that of other optical elements, limiting the sensitivity gain of cooling further (see Sec.~\ref{sec:thermal_requirements}). Third, the PTR's first stage has substantially more cooling power ($\sim$~10~W/K) than its second ($\sim$~1~W/K), relaxing CHWP dissipation requirements. Fourth, the focal plane structure is anchored to the 4~K stage but not the 50~K, easing CHWP vibrational requirements. Fifth, the 50~K location necessitates minimal optics-tube assembly/disassembly to install/access the CHWP, making it easy to both integrate into the receiver and service on the telescope, if necessary. Finally, the 50~K CHWP is modular, with its addition to the PB-2b instrument requiring no adjustment to the heritage lens shapes or positions, thermal filtering, or optics tube assembly procedure. This final feature was particularly powerful for PB-2b integration and testing, allowing the CHWP and optics tube to be validated in parallel prior to full system integration, hence accelerating the receiver commissioning process.

\subsection{Optical design}
\label{sec:optical_design}

As shown in Fig.~\ref{fig:chwp_render}, the CHWP sapphire stack consists of three $\approx$~505~mm diameter, 3.8~mm-thick sapphire windows\footnote{Guizhou Haotian Optoelectronics Technologies: http://www.ghtot.com/} with a $\approx$~0.7~mm thick dual-layer AR coating on each of its outermost surfaces. The stack is held in an aluminum cradle, which uses tubular springs\footnote{Spira Manufacturing Corporation: https://www.spira-emi.com/} to absorb differential thermal contraction and keep the sapphire from shifting during cooldown and rotation. The cradle has a 490~mm clear aperture diameter, is 20~mm thick, and weighs $\approx$~10~kg including the sapphire stack.

The CHWP's clear aperture is defined by a stationary 440~mm diameter tube lined with HR-10 absorber.\footnote{Eccosorb: https://www.laird.com/rfmicrowave-absorbers-dielectrics} The aperture tube is 60~mm tall and forms a 5~mm gap with the back face of the sapphire stack, ``hiding'' non-optical components from the telescope's beam and restricting the propagation of any CHWP-induced stray light. The CHWP's aperture diameter is limited by bearing manufacturing constraints (see Sec.~\ref{sec:bearing}) yet meets the 430~mm requirement while providing 5~mm of radial alignment tolerance. 


\subsection{Bearing}
\label{sec:bearing}

The CHWP bearing needs to be low-friction, have a large bore diameter, and be mechanically robust at low temperatures. Cryogenic ball bearings with a $\sim$~500~mm bore are not commercially available and are challenging to develop due to thermal contraction, vibration, and durability issues. Therefore, the PB-2b CHWP employs a superconducting magnetic bearing (SMB), as shown in Fig.~\ref{fig:chwp_tabletop}. The SMB operates by flux pinning in an azimuthally symmetric geometry. When suspending a uniformly magnetized ring (the rotor) above a type-II superconducting ring (the stator) and subsequently cooling the superconductor below its transition temperature, the rotor's permanent magnetic field becomes trapped in the stator's superconducting bulk, constraining the rotor in the axial and radial directions while allowing it to rotate in azimuth. SMBs are implemented in other CHWP systems for CMB observation\cite{matsumura_cosmic_2006,klein_cryogenic_2011,sakurai_estimation_2017,Sakurai2018a,Sakurai2020} and their cryogenic robustness is well demonstrated.

The PB-2b SMB is manufactured by Adelwitz Technologiezentrum GmbH.\footnote{ATZ: http://www.atz-gmbh.com/} The magnet ring consists of 16 $22.5^{\circ}$, 97~mm~$\times$~16~mm annular segments of Neodymium (NdFeB), glued contiguously into an encapsulating G10 fixture to form a highly uniform, $\approx$~5,000~G surface field. The superconducting ring consists of 46 $7.8^{\circ}$, 35~mm~$\times$~13~mm annular segments of yttrium barium copper oxide (YBCO), which has a $T_{\mathrm{c}} \approx 90$~K, glued into an encapsulating aluminum fixture to form a contiguous type-II superconductor. Both the rotor and stator have a 470~mm inner diameter, which is large enough to fit around the aperture tube with a 5~mm radial clearance (see Fig.~\ref{fig:chwp_render}).

The SMB's effectiveness relies on a small separation between the permanent magnet and YBCO. SMB stiffness is a steep function of rotor-stator separation,\cite{hull_effect_2000} and therefore controlling the rotor's axial position near the YBCO's transition temperature is critical to controlling the SMB's spring constant. The nominal rotor-stator separation in the PB-2b CHWP system is 5~mm, for which we measure a bearing spring constant of $\approx$~300~N/mm. In addition, SMB friction must be small enough both to achieve the required rotation speed and to mitigate heat dissipation. Eddy losses are induced by eddy currents in nearby metal and scale as $\Delta B^{2}$,\cite{bean_magnetization_1964} where $\Delta B$ is the rotor's peak-to-peak magnetic field variation. Hysteresis losses arise due to paramagnetic hysteresis in the YBCO and scale as $\Delta B^{3}$.\cite{zeisberger_losses_1998} Therefore, the uniformity of the magnet ring is critical to minimizing dissipation during continuous rotation. A measurement of rotor friction is presented in Sec.~\ref{sec:thermal_performance}.


\subsection{Gripper}
\label{sec:gripper_design}

Magnetic levitation is a cryogenic phenomenon, and the SMB only engages below the YBCO's $\approx$~90~K superconducting transition temperature. Above this temperature, the YBCO does not flux pin, and the rotor is effectively decoupled from the stator. Therefore, the CHWP employs a ``gripper'' to support the rotor during cooldown and keep it aligned until levitation initiates. The gripper consists of three subassemblies azimuthally distributed about the rotor (see Fig.~\ref{fig:chwp_in_pb2b}): one along $-y$, and the others $\pm$~40~deg\footnote{The opening angle of the top two subassemblies is less than $120^{\circ}$ to avoid mechanical interference with the optics tube's PTR attachment.} about $+y$. Each subassembly has a linearly actuating vacuum feedthrough at 300~K, a flexible, thermally isolating ``arm'' between 300 and 50~K, and a 50~K ``finger'' that engages the rotor stage.  When the gripper fingers are extended, the rotor is gripped, and when they are retracted, the rotor is released.

\begin{figure}[!t]
    \begin{subfigure}
        \centering
        \includegraphics[width=0.98\linewidth, trim=2cm 4cm 1.4cm 4.2cm, clip]{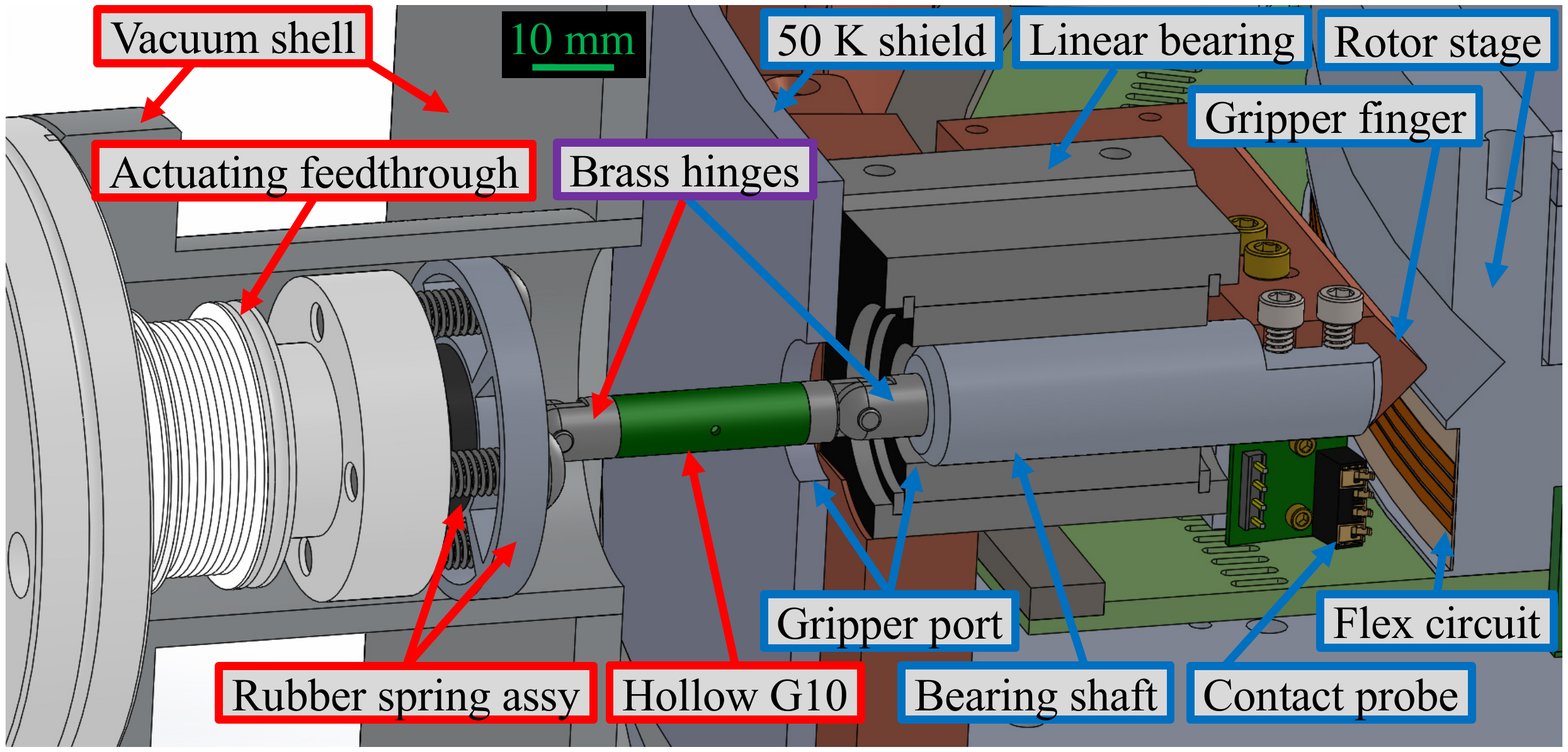}
    \end{subfigure}
    \begin{subfigure}
        \centering
        \includegraphics[width=0.98\linewidth, trim=4.5cm 6cm 4.5cm 5cm, clip]{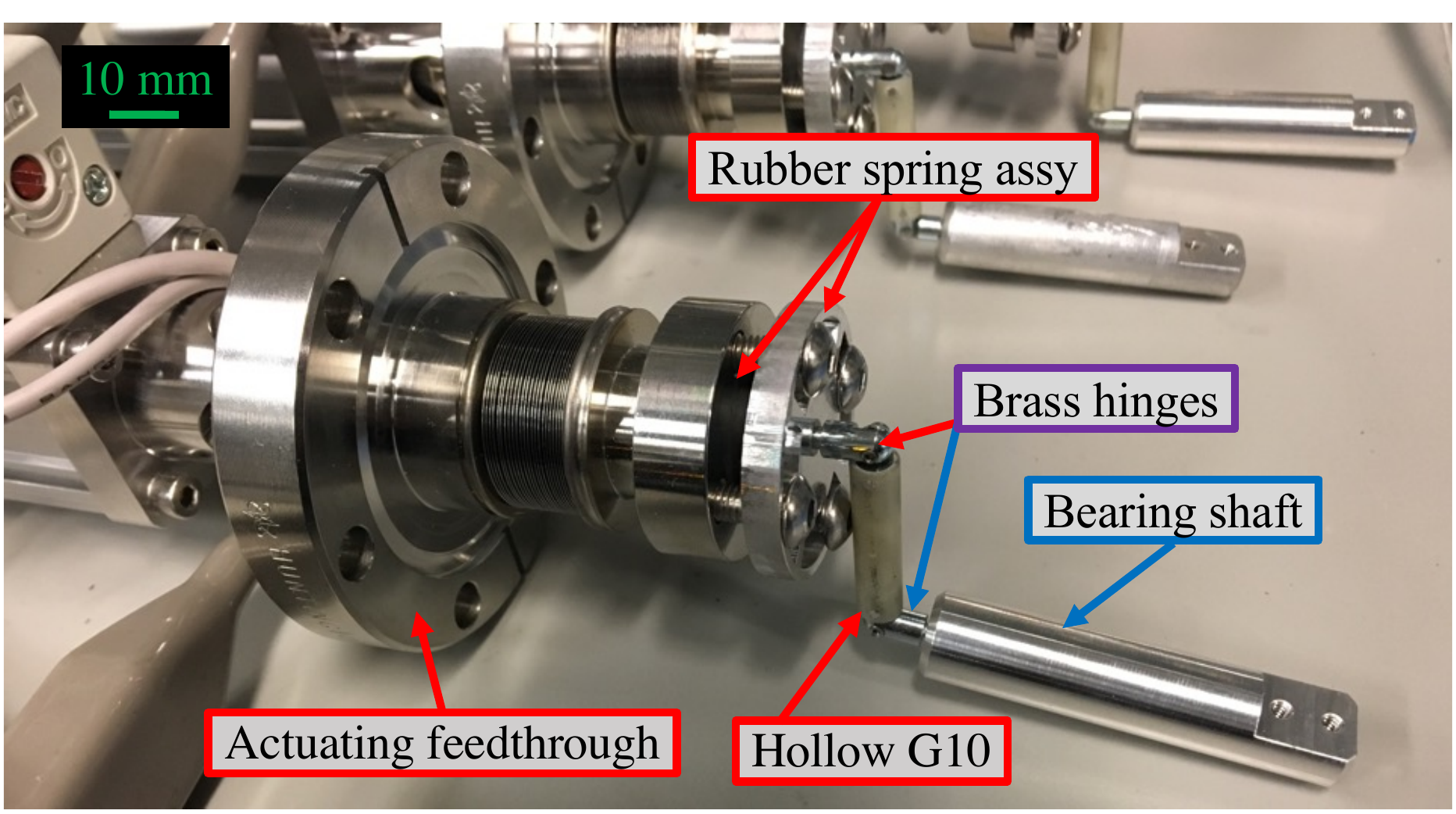}
    \end{subfigure}
    \begin{subfigure}
        \centering
        \includegraphics[width=0.98\linewidth, trim=5cm 5cm 5.8cm 3.5cm, clip]{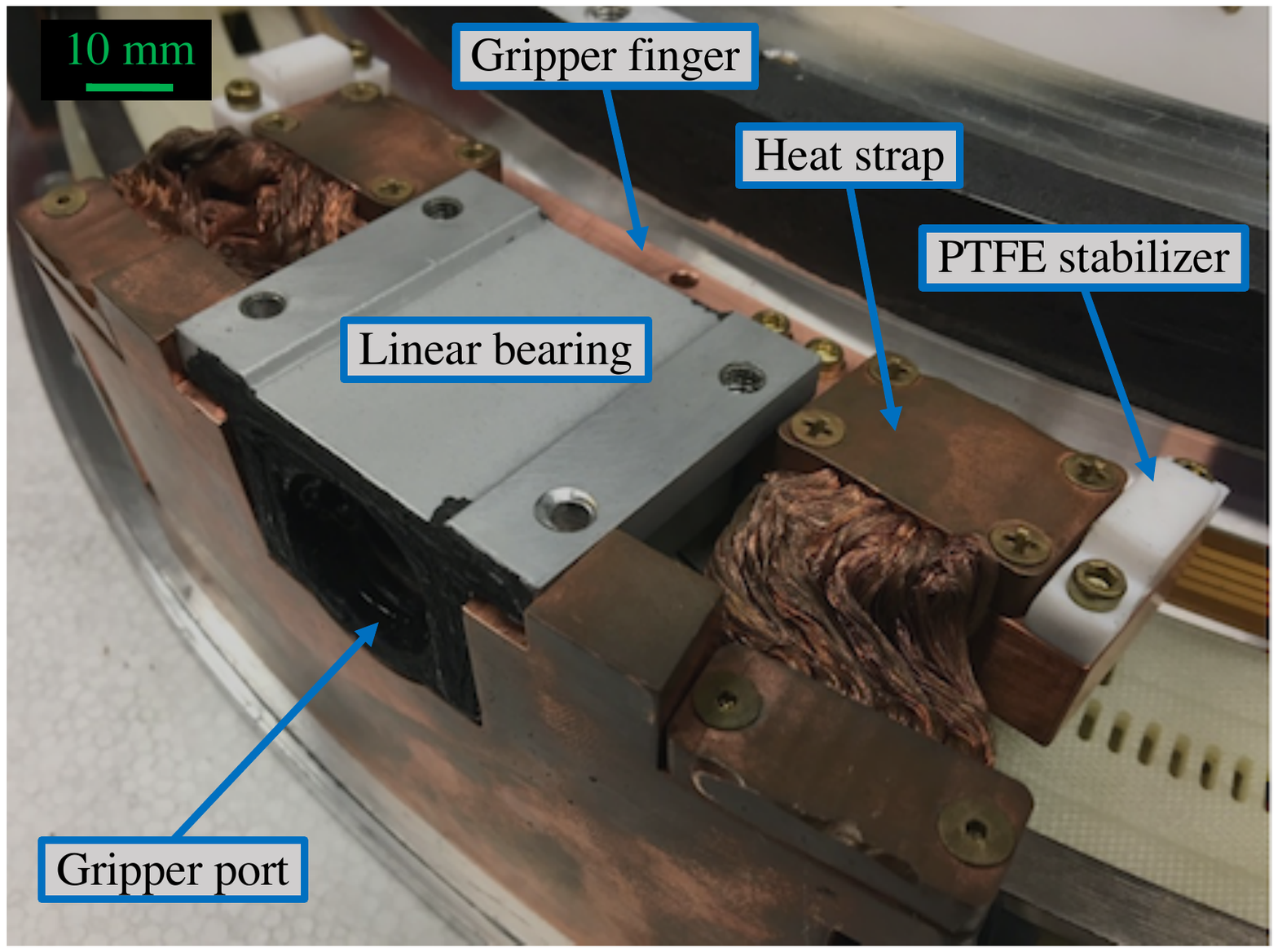}
    \end{subfigure}
    \caption{A detailed view of a single gripper subassembly with the retaining ring removed for visual clarity. Red boxes mark warm components, and blue boxes mark cold components. \textbf{Top panel:} a CAD cross section of a gripper subassembly with the rotor un-gripped and floating.  \textbf{Middle panel:} a photograph of a feedthrough actuator attached to a gripper arm. \textbf{Bottom panel:} a photograph of the linear bearing and the copper gripper finger without the gripper arm.}
    \label{fig:gripper_assy}
\end{figure}

Fig.~\ref{fig:gripper_assy} shows a CAD rendering and photographs of a single gripper subassembly. The feedthrough\footnote{Huntington Mechanical Laboratories: https://huntvac.com/} consists of a linear actuator\footnote{SMC Corporation: https://www.smcusa.com/} with a vacuum-compatible bellows assembly that mounts to a ConFlat port on the vacuum shell. The feedthrough bolts to a rubber spring assembly, which adds compliance both along and about the radial direction. This spring assembly in turn connects to the gripper arm, which is composed of a hinged thermal isolator joined to a 6~mm diameter, 50~mm long aluminum bearing shaft. The thermal isolator comprises two brass hinges epoxied to either end of a 24~mm long, 0.8~mm-thick-walled hollow G10 tube. The hinges accommodate differential thermal contraction along the receiver cryostat's axial direction and relax alignment tolerances between the 300~K and 50~K stages.

The bearing shaft enters the 50~K assembly through a gripper port, which includes a 10~mm diameter hole in the 50~K shield and a Frelon-lined linear bearing within which the bearing shaft slides. Because this port introduces a potential 300~K light leak, the outer surface of the linear bearing and the inner surface of the 50~K shield are blackened with carbon-loaded Stycast\cite{persky_review_1999} to limit the propagation of any stray light. The innermost end of the bearing shaft bolts to the gripper finger, which is a 6~mm deep, $90^{\circ}$-angled, oxygen-free high-thermal-conductivity (OFHC) copper wedge, attached to the CHWP stator via two flexible OFHC copper-braid heat straps. When the rotor is gripped, the finger fits into an identically shaped, azimuthally symmetric groove on the rotor stage, heat sinking the rotor while constraining its axial position.

The gripper finger also contains a spring-loaded probe that contacts an azimuthally symmetric flex circuit when the rotor is gripped, permitting four-point readout of a silicon diode thermometer\footnote{Lakeshore DT-670: https://www.lakeshore.com} on the sapphire stack's cradle. The probe is a beryllium copper, four-point battery contact, and the Kapton-based flex circuit\footnote{Q-Flex: https://www.qflexinc.com/} has four 2~mm-wide, 2.5~mm-pitch, gold-plated copper traces soldered to the thermometer's leads.

The complete gripper assembly is composed of three subassemblies that actuate simultaneously to grip and release the rotor. The motors are driven by a parallel-output controller,\footnote{SMC JXC-831: https://www.smcpneumatics.com/JXC831.html} which is commanded using a programmable logic controller (PLC). The typical cooldown configuration for the PB-2b receiver is approximately horizontal (see Fig.~\ref{fig:chwp_in_pb2b}), with the bottom gripper finger supporting the rotor's 17~kg mass during cooldown. Therefore, we employ a 450~N-max motor for the bottom subassembly and 200~N-max motors for the top two.\footnote{SMC LEY-32 and LEY-16: https://www.smcusa.com/}

The gripper is designed for reliability. Each gripper finger attaches to its bearing shaft with two titanium bolts to avoid fastener fatigue and has two PTFE stabilizers that slide along the bottom face of the retaining ring (see Fig.~\ref{fig:chwp_tabletop}) to constrain rotation about the gripper-port axis. The G10 tubes and hinges are joined with a step to avoid compression failures, and the hinges are screened to withstand more than four times the rotor's weight. In addition, the rotor's center of mass lies in the plane of its triangular groove, and the bottom finger's copper wedge is slightly V-shaped, forming a $176^{\circ}$ angle about the radial direction. These two features allow the rotor to be constrained by the bottom subassembly alone when the receiver is horizontal, hence providing insurance against possible gripper failure modes such as motor control issues and imperfect subassembly synchronization.


\subsection{Motor}
\label{sec:motor_design} 

The CHWP motor needs to drive stable 2~Hz rotation while being low-dissipation, low-magnetic-interference, and mechanically robust at cryogenic temperatures. While a belt drive with an external motor has been successfully operated during weeks of CMB observation,\cite{klein_cryogenic_2011,the_ebex_collaboration_ebex_2018} the PB-2b CHWP must run for years, motivating a drive system free of mechanical fatigue. Therefore, we utilize a brushless, three-phase, synchronous electromagnetic motor driven by custom electronics. We discuss the motor's cryogenic assembly, driver, and efficiency in the following subsections.


\subsubsection{Motor cryogenic assembly}
\label{sec:motor_mech} 

\begin{figure}[!t]
    \begin{subfigure}
        \centering
        \includegraphics[width=0.98\linewidth, trim=1cm 3cm 1cm 3cm, clip]{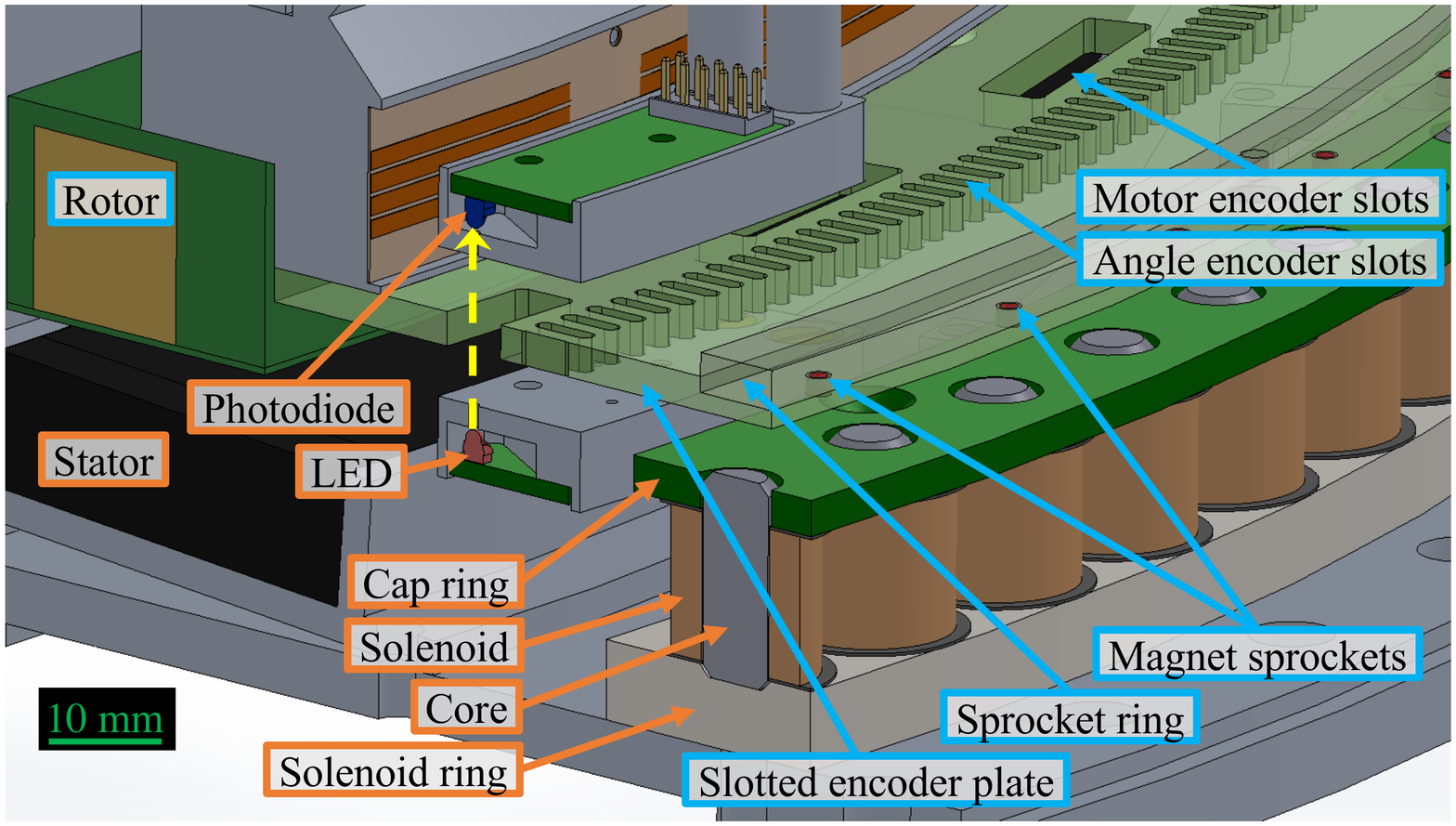}
    \end{subfigure}
    \begin{subfigure}
        \centering
        \includegraphics[trim={4.5cm, 6cm, 4.5cm, 7.5cm}, clip, width=0.98\linewidth]{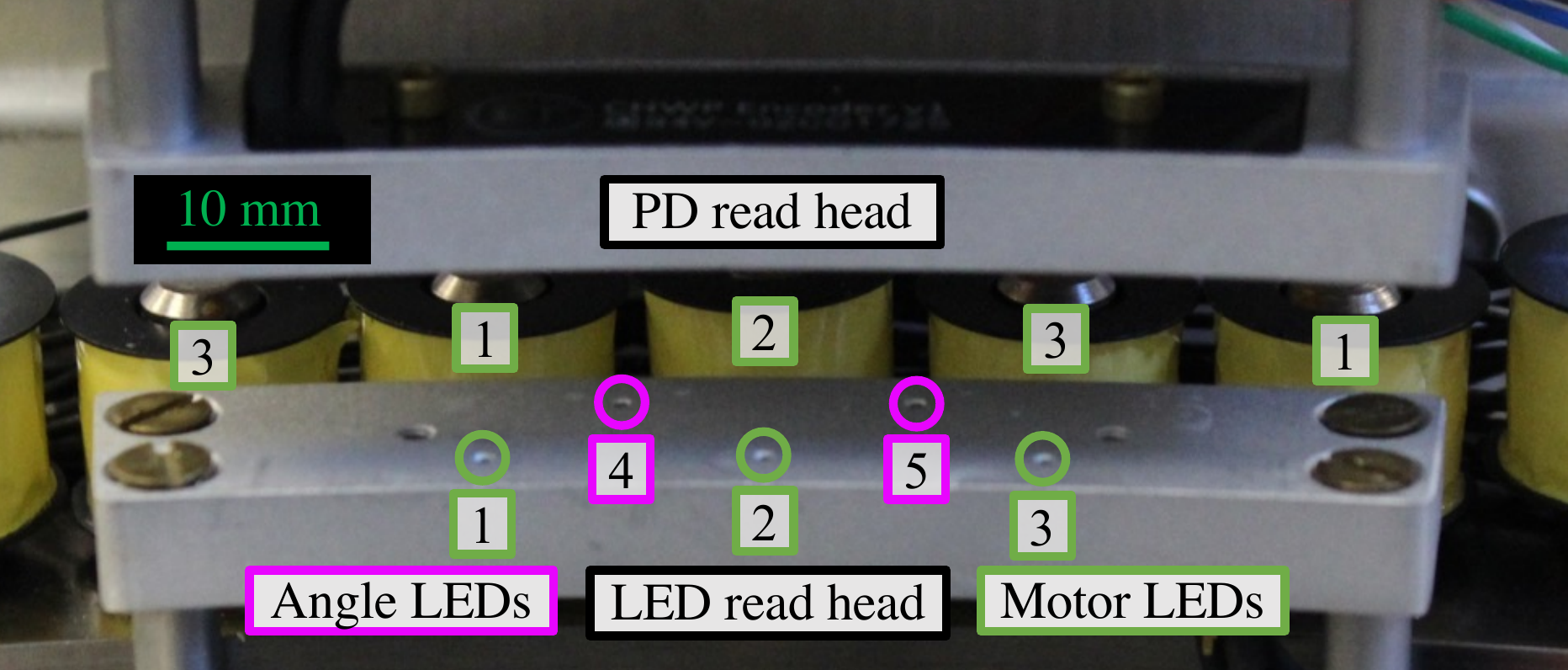}
    \end{subfigure}
    \caption{\textbf{Top panel:} a zoomed CAD cross section of the CHWP motor. Rotating components are labeled with cyan boxes, and stationary components are labeled with orange boxes. The slotted encoder plate and sprocket ring are semi-transparent for visual clarity. \textbf{Bottom panel:} a photograph of the encoder read heads and nearby solenoids with the slotted encoder plate removed. The collimator holes for the motor LEDs are circled in green and labeled by phase, while those of the angle LEDs are circled in magenta. The solenoids are also labeled by phase and are radially aligned with their corresponding motor LED-PD pairs.}
    \label{fig:motor_mech_assy}
\end{figure}

The motor's cryogenic assembly is shown in Fig.~\ref{fig:motor_mech_assy}. Its active component consists of 114 solenoids\footnote{APW Company: https://apwelectromagnets.com/fc-6035.html} with low-carbon-steel magnetic cores glued with equal spacing onto a 650~mm diameter low-carbon-steel ring on the stator. Each coil has an 8.6~mH inductance and a $\approx$ 3~$\Omega$ resistance at 50~K, and the motor's three phases are driven across three regularly interspersed groups of 38 coils. In order to reduce the motor's equivalent resistance and hence reduce the voltage needed to operate it, the solenoid array is further divided into four sections---two with 30 total coils, two with 27 total coils---that are driven in parallel. The motor's passive component consists of 76 1.5~mm diameter $\times$ 1.5~mm tall NdFeB magnet sprockets, which have a 6,700~G surface field and are glued with equal spacing and alternating polarity onto a 650~mm diameter low-carbon-steel sprocket ring on the rotor.

The solenoids are energized in three $120^{\circ}$-separated phases with an alternating drive voltage $\pm \; \mathrm{V_{D}}$, creating an oscillating magnetic field that couples to the rotor's magnet sprockets, driving rotation. The sprockets, solenoids, and cores are carefully chosen to avoid cogging while providing enough torque to attain the required rotational velocity. At 2~Hz rotation, the motor delivers only $\approx$~5~N-mm of torque, making it susceptible to physical touches. Therefore, the CHWP assembly includes a system of wire harnesses to facilitate clean cable management. Additionally, despite the <~2~mm rotor-stator alignment requirement presented in Sec.~\ref{sec:motor_eff}, we provide 5~mm of clearance around the rotor to further limit the possibility of a physical touch impacting rotation.

The coils are driven by custom electronics, described in Sec.~\ref{sec:driver}, whose sensing component is an optical encoder on the 50~K stage shown in Fig.~\ref{fig:motor_mech_assy}. The encoder consists of three 940~nm light-emitting diodes\footnote{Vishay VSMB294008G: \\ https://www.vishay.com/docs/84228/vsmb294008rg.pdf} (LEDs) shining onto three reverse-biased photodiodes\footnote{Vishay TEMD1020: \\ https://www.vishay.com/docs/81564/temd1000.pdf} (PDs) through a slotted encoder plate on the rotor. The LEDs and PDs are soldered to printed wiring boards and are housed in aluminum read heads with 1~mm diameter, 3~mm deep collimation holes. As shown in Figs.~\ref{fig:chwp_tabletop} and~\ref{fig:motor_mech_assy}, the LED read head is mounted to four precision-ground aluminum standoffs on the stator baseplate, while the PD read head is similarly mounted to the retaining ring on the opposite side of the slotted encoder plate. Each motor LED-PD pair is azimuthally aligned with the solenoid array, and the read heads are aligned to each other using dowel pins during assembly. The gallium aluminum arsenide LEDs and the silicon PDs are cryogenically screened via dunk tests in liquid nitrogen and have been robust throughout hundreds of hours of testing. Even so, the CHWP has two identical pairs of read heads, as shown on the right- and left-hand sides of Fig.~\ref{fig:chwp_tabletop}, to provide insurance against LED or PD failure. The slotted encoder plate has 38 5~mm wide motor slots whose edges are azimuthally aligned to the magnet sprockets. During rotation, the slot pattern chops the PD input, and the motor driver converts this photocurrent waveform into the solenoid bias voltage, as described in the following subsection. Both the slotted encoder plate and cap ring are made of G10 to minimize motor-induced eddy currents on both the stator and rotor.


\subsubsection{Motor driver}
\label{sec:driver}

The motor drive electronics must be robust, low-noise, and simple to operate. While commercial drivers for synchronous motors are abundant, most involve auxiliary control software, pulse-width-modulated (PWM) waveforms (which can inject high-frequency noise into the receiver), and awkward interfacing to the PB-2b cryogenic assembly. Therefore, we employ a custom driver printed circuit board (PCB) that both meets our requirements and simplifies CHWP operation. The presented PCB is also used to read out the angle encoder signal, which is discussed in Sec.~\ref{sec:angle_encoder}.

\begin{figure*}
    \includegraphics[width=0.98\linewidth, trim={1.5cm, 3.8cm, 1.5cm, 3.5cm}, clip]{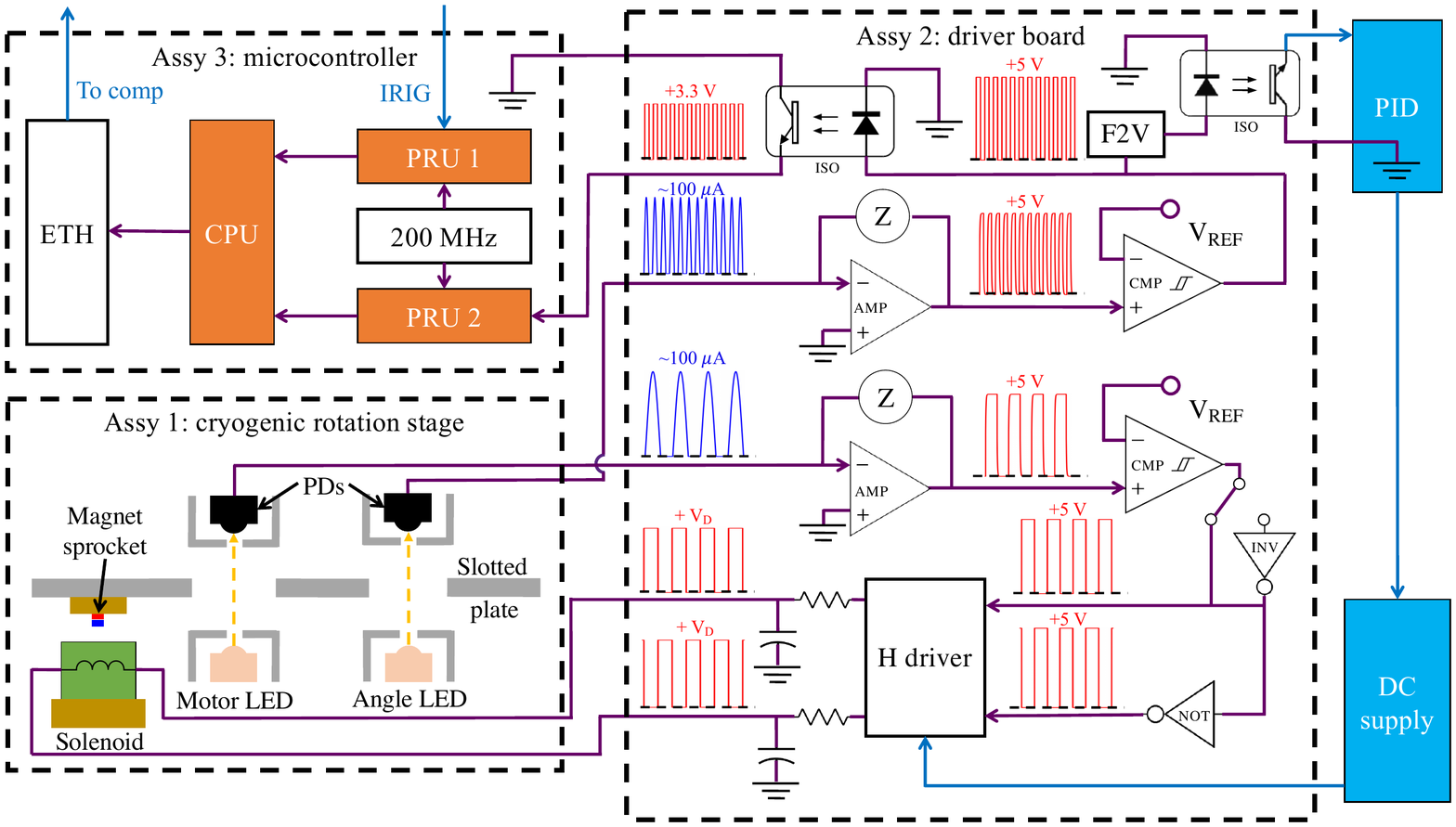}
    \caption{A signal diagram for a single phase of the CHWP motor driver and for a single angle encoder output. On the \textbf{cryogenic rotation stage}, the slotted encoder plate, which has one lane of slots for motor encoding and another for angle encoding, chops the signals of the LED-PD pairs. The \textbf{driver board} converts these photocurrents into voltage signals using transimpedance amplifiers (AMP) with carefully tuned low-pass feedback (Z) and then converts these analog waveforms into 5~V digital signals using comparators (CMP) with potentiometer-tunable reference voltages ($\mathrm{V_{REF}}$). For motor control, both non-inverted and inverted (NOT) signals are passed to an H-bridge driver (H driver), which outputs synchronized, low-pass-filtered, alternating-polarity voltage waveforms $\pm \mathrm{V_{D}}$ to the solenoids. For angle encoder readout, which has a 15$\times$ faster signal than that of the motor, the digital signals are opto-isolated (ISO) and sent to a microcontroller unit for processing. Additionally, the angle encoder digital signal is passed to a frequency-to-voltage converter (F2V) whose opto-isolated output is used by a PID controller to provide feedback to the H-driver voltage and stabilize CHWP rotation. To brake or reverse direction, a switched inverter (INV) applies a global $180^{\circ}$ phase shift to the H-driver input. The \textbf{microcontroller} is a BeagleBone Black, which has two programmable real-time units (PRUs) that share a 200~MHz clock. PRU 1 timestamps and decodes a GPS-synchronized IRIG-B PWM waveform, while PRU 2 timestamps rising and falling edges of the digital angle encoder waveform. These data are written to a shared buffer, which is emptied by the central processing unit (CPU) before being sent to an external computer over Ethernet (ETH).}
    \label{fig:motor_driver_schematic}
\end{figure*}

\begin{figure}[!t]
    \centering
    \includegraphics[trim={3.0cm, 3.5cm, 2.5cm, 3.6cm}, clip, width=0.98\linewidth]{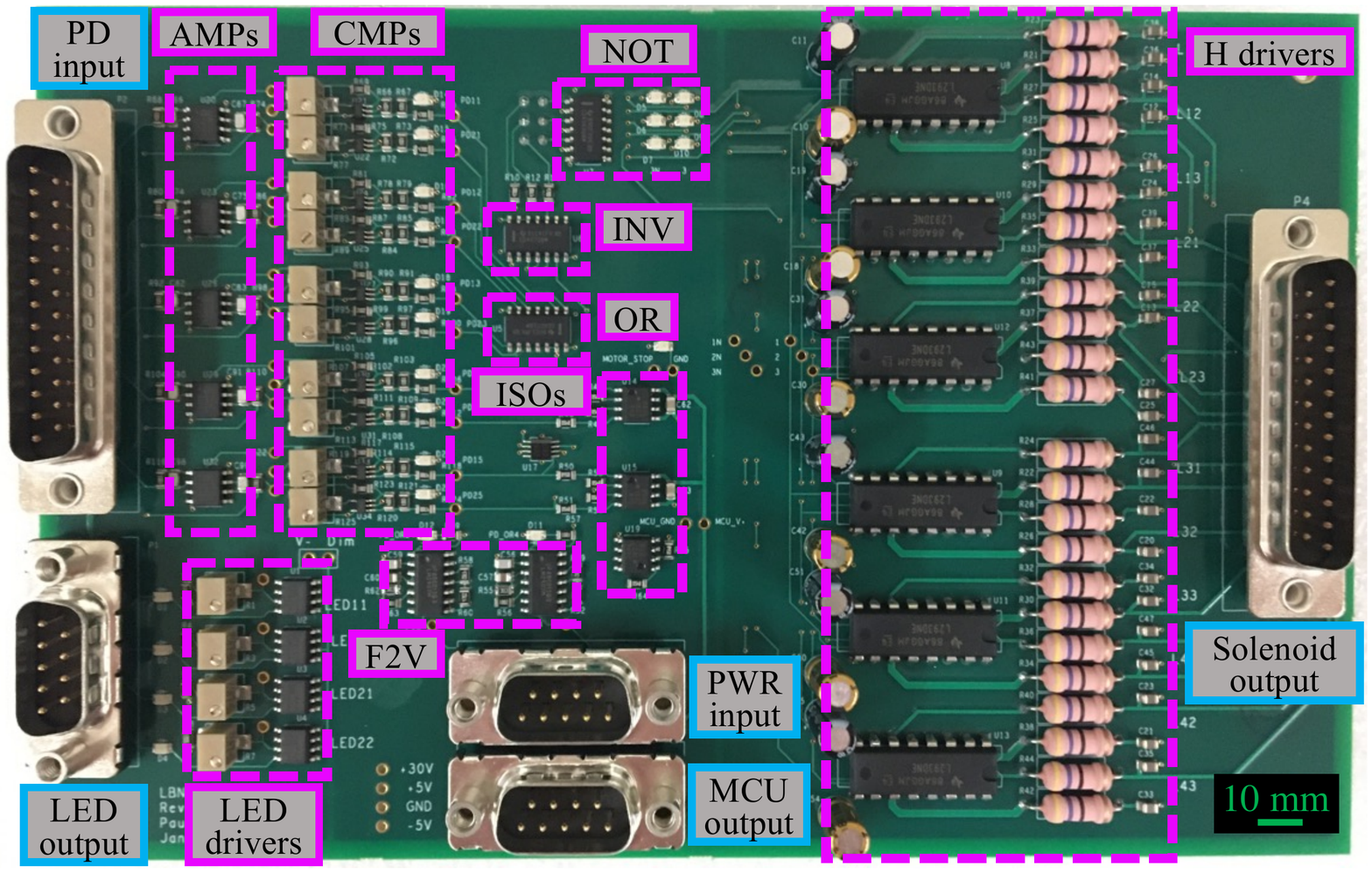}
    \caption{A photograph of the driver PCB, the schematic for which a subset is shown in Fig.~\ref{fig:motor_driver_schematic}. The photocurrents from all ten encoder PDs enter the board through the ``PD input'' connector before being converted to analog signals by the transimpedance amplifiers ``AMPs'' and converted to 5~V digital signals by the comparators ``CMPs.'' At this point, the angle encoder signals are opto-isolated (``ISOs'') and sent to the microcontroller via the ``MCU output'' connector. The motor signals from both PD read heads are ``OR-ed'' before being inverted (``INV'') and input to the ``H drivers,'' which in turn power the solenoids via the ``Solenoid output'' connector. The motor and encoder LEDs are current biased by potentiometer-adjustable ``LED drivers'' via the ``LED output'' connector. DC supply voltages come through the ``PWR input'' connector.}
    \label{fig:motor_driver_photo}
\end{figure}

A single PD amplification chain is shown in Fig.~\ref{fig:motor_driver_schematic}, and the complete driver PCB is shown in Fig.~\ref{fig:motor_driver_photo}. Photocurrent from the PD read head travels through 50~K Manganin ribbon cables,\footnote{Tekdata Interconnect: https://www.tekdata-interconnect.com/} a DB-25 vacuum feedthrough,\footnote{Accu-Glass Products: https://www.accuglassproducts.com/25d2-450} and 300~K, double-shielded, twisted-pair, copper cables\footnote{Alpha Wire 6831: http://www.alphawire.com/} to the driver board where it is converted to an analog voltage by a transimpedance amplifier. The time constant of the amplifier feedback is chosen to suppress high-frequency noise while outputting a symmetrical waveform. This analog signal is then converted to 5~V digital, and the TTL waveforms from each PD read head are OR-ed so that if one LED-PD pair fails, CHWP operation is unaffected. All 12 solenoid chains (three phases in four sections) are energized in parallel by H-bridge drivers whose output is cleaned by a single-pole $\approx$~300~Hz low-pass filter to suppress any high-frequency interference in the cryostat. The PCB's digital logic and H drivers are powered by low-noise, non-switching DC power supplies,\footnote{Kikusui PMX: https://www.kikusui.co.jp/en/} and the PCB layout and ground-plane geometry are specifically designed to avoid contaminating analog signals with digital artifacts.

The CHWP's rotational velocity is naturally steady due to the rotor's large rotational inertia and the motor's small torque. However, the positioning between the motor's solenoids and rotor's magnet sprockets changes slightly when the CHWP's gravitational orientation changes, such as during telescope motion. This modulation in motor coupling slightly modulates the motor's efficiency (see Sec.~\ref{sec:motor_eff}) and causes small, slow drifts in rotational velocity. Therefore, we employ proportional-integral-derivative (PID) feedback\footnote{Omega CNi16D52: https://www.omega.com/en-us/} to the H-driver voltage in order to stabilize the CHWP velocity on long timescales.

Finally, in order to both stop the rotor and spin it in the opposite direction, an inverter switch applies a $180^{\circ}$ phase shift to both H-driver inputs when toggled by an external digital input. Braking is vital during power failure (see Sec.~\ref{sec:shut_down}), and spinning the CHWP in both directions provides a useful data split during analysis.


\subsubsection{Motor efficiency}
\label{sec:motor_eff} 

\begin{figure}[!t]
    \centering
    \includegraphics[width=0.95\linewidth, trim=0.5cm 1cm 2cm 1cm]{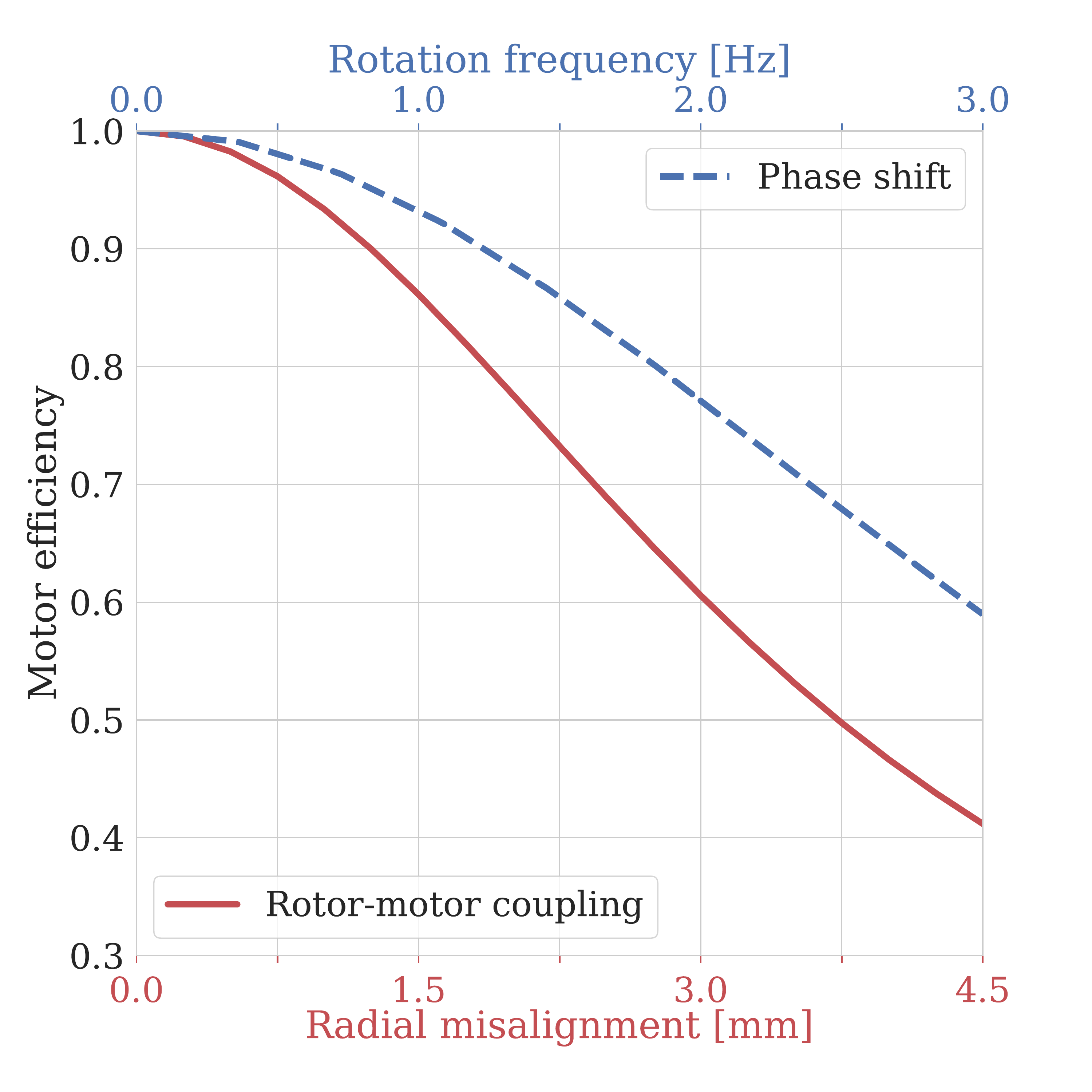}
    \caption{CHWP motor efficiency vs. the rotor's radial misalignment to the motor solenoids and the rotor's rotation frequency. The PB-2b rotor alignment tolerance is <~2~mm, limiting positional efficiency loss to <~20\%. At 2~Hz rotation, the phase-delay efficiency loss is $\approx$~20\%.}
    \label{fig:motor_eff}
\end{figure}

The motor's maximum torque is delivered at start-up when the rotor's sprocket pattern and the solenoid array's rotating magnetic field are in phase. However, at non-zero rotation frequencies, the solenoids' inductance creates a phase shift between the H-driver voltage and the solenoid current, degrading motor efficiency. The impact of this inductive phase shift is shown in Fig.~\ref{fig:motor_eff} and is $\approx$~20\% at 2~Hz rotation.\footnote{In principle, a microcontroller or equivalent could use the F2V output to time-shift H-driver inputs and recover this loss, but as shown in later sections, the existing motor scheme meets the CHWP requirements, and therefore the phase delay does not need to be corrected for PB-2b.} 

Additionally, motor efficiency relies on concentricity between the rotor and stator. When radially misaligned, the coupling between the solenoids and the magnet sprockets weakens, reducing motor torque. Radial misalignment\footnote{Axial displacement is also important for motor coupling but is easily controlled to $\sim$~0.1~mm by the gripper's wedge-and-groove design.} is most likely to occur along the gravitational axis when the cold assembly contracts during cooldown. The simulated impact of rotor radial misalignment on motor efficiency is shown in Fig.~\ref{fig:motor_eff} and assumes a 5~mm rotor-stator axial separation. This simulation motivates a concentricity requirement of $\Delta R <  \: 2$~mm, for which the efficiency loss is <~20\%. 

\subsection{Angle encoder}
\label{sec:angle_encoder}

The rotor angle is measured by an incremental encoder that uses much of the same infrastructure as the motor encoder (see Sec.~\ref{sec:motor_design}). The angle encoder's signal schematic is shown as part of Fig.~\ref{fig:motor_driver_schematic}, and its cryogenic components are shown in Fig.~\ref{fig:motor_mech_assy}. On each encoder read head, two LED-PD pairs peer through $570 - 1 = 569$ slots on the slotted encoder plate with one missing slot to mark the rotor's absolute position. During 2~Hz continuous rotation, the PD photocurrents are chopped at 1.14~kHz, are digitized on the driver board, are opto-isolated to 3.3~V, and are processed by a microcontroller unit (MCU). The two angle encoder LED-PD pairs are offset by half a slot width, enabling quadrature readout to monitor rotation direction.

The MCU is a BeagleBone Black (BBB),\footnote{Beagle Board: https://beagleboard.org/black} which houses two programmable real-time units (PRUs) and an on-board CPU running Linux. The PRU is a lightweight, low-latency processing unit specifically designed to handle single-threaded inputs, making it ideal for angle encoding. One PRU polls the angle encoder signal and uses a 200~MHz clock\footnote{Specifically, both PRUs access the industrial Ethernet protocol (IEP) timer.} to timestamp each rising and falling edge. Simultaneously, a second PRU polls and decodes a GPS-synchronous inter-range instrumentation group B code (IRIG-B)\footnote{Spectrum Instruments TM-4: http://www.spectruminstruments.net/} PWM waveform using the same 200~MHz clock. The encoder and IRIG clock values are written to a shared memory buffer that is periodically emptied by the CPU, which in turn sends IPv4 data packets to an external computer. During post-processing, the rotor angle is reconstructed using the missing reference slot and is interpolated to IRIG time using the MCU clock values. This time-ordered CHWP angle data can then be used to demodulate the detectors, which are also synchronized to IRIG.

While the CHWP angle jitter requirement is $\ll$~$3$~$\mathrm{\mu rad / \sqrt{Hz}}$, the encoder only has a resolution of 5~mrad, necessitating precise interpolation between ticks on the slotted encoder plate. Such an encoding scheme is feasible because the CHWP rotation is very smooth and the encoder is high-signal-to-noise, enabling clean angle reconstruction (see Secs.~\ref{sec:continuous_rotation} and ~\ref{sec:encoder_jitter}).


\subsection{Thermal design}
\label{sec:thermal_design}

The effectiveness of the CHWP system depends centrally on its thermal performance. As shown in Fig.~\ref{fig:chwp_render}, the CHWP is located on the 50~K stage, which is cooled by the first stage of the optics tube's PTR. The rotor is shielded from sky-side radiation by the vacuum window, RT-MLI, and IRF, and it floats between the IRF and the field lens. As discussed in Sec.~\ref{sec:requirements}, the primary objective of the CHWP thermal design is to minimally impact both the 4~K and 50~K stage temperatures with respect to those of the heritage PB-2b configuration. 

When the rotor is stationary, loading on the 4~K stage is lower than in the heritage system. The IRF transmits non-negligibly $\lesssim$~2~THz,\cite{inoue_cryogenic_2014} and when present, the CHWP's sapphire stack absorbs >~90\% of this leaked sky-side power, keeping it from reaching the field lens. Additionally, because the rotor is floating, it acts as multi-layer insulation (MLI), reducing the 4~K load further. When the rotor is spinning, however, the solenoids, LEDs, and rotor generate heat, and if these loads warm the 50~K stage or rotor too much, 4~K improvements could be negated. Therefore, the CHWP design focuses on limiting 50~K dissipation and on keeping the rotor cool. 

\begin{figure}[!t]
    \centering
    \includegraphics[width=0.98\linewidth, trim=0.5cm 5.0cm 2.3cm 6.5cm, clip]{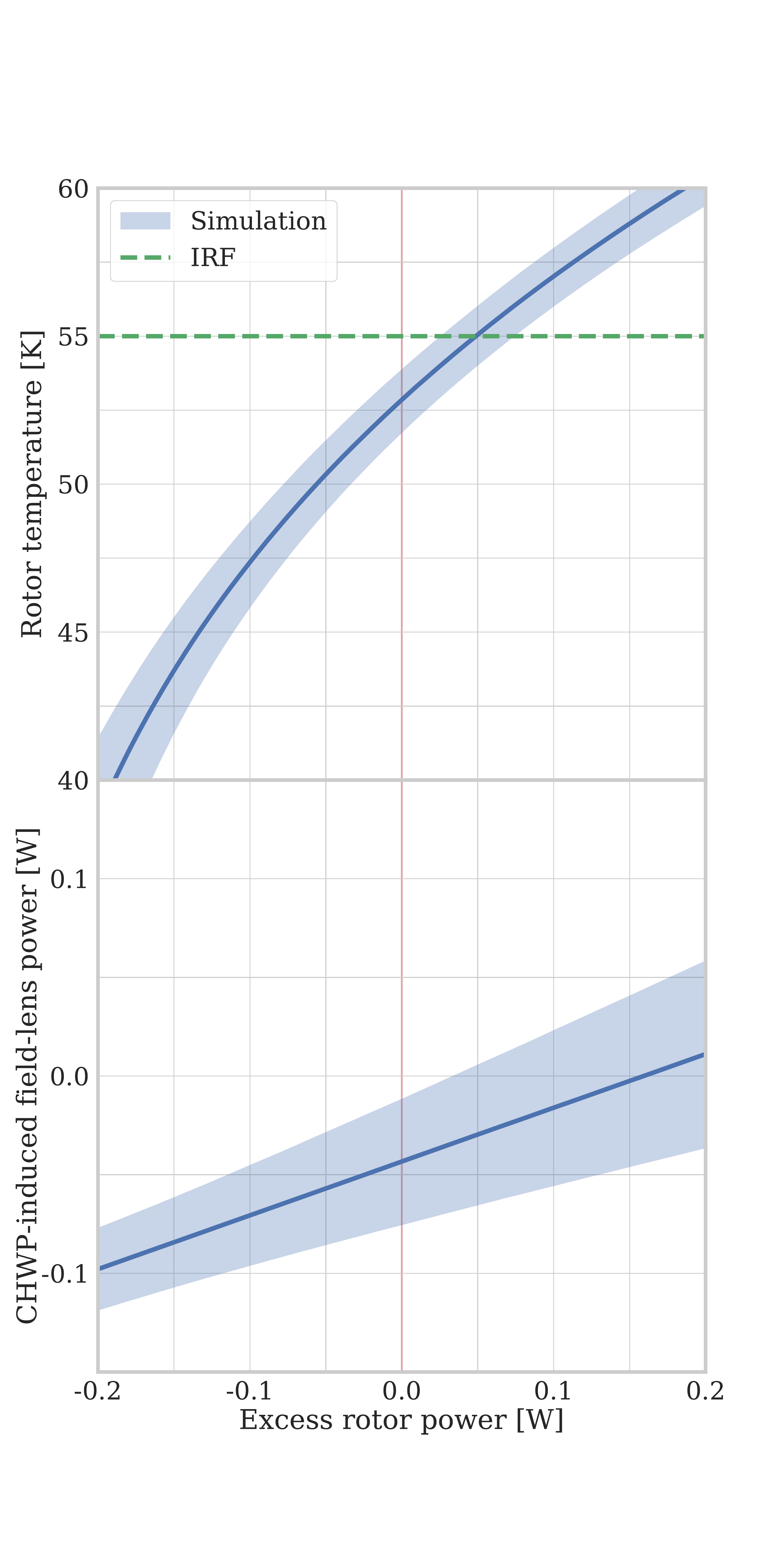}
    \caption{Rotor temperature (top panel) and CHWP-induced power on the field lens (bottom panel) as a function of excess, unmodeled power on the rotor. The solid blue line represents the median, while the shaded regions denote one-sigma uncertainties. Each plot's y-intercept gives the modeled expectation, and the IRF is assumed to be 55~K.}
    \label{fig:rotor_temp}
\end{figure}

We use an analytic simulation to predict rotor temperature and CHWP-induced power on the 50~K and 4~K stages during continuous operation in the field. The details of the model's assumptions, including measurements and calculations of conductivities, radiative couplings, and power transfers, are presented in App.~\ref{app:thermal_sim}. Fig.~\ref{fig:rotor_temp} shows the rotor temperature and the CHWP-induced load on the field lens as a function of excess power on the rotor, where ``excess'' is defined as that which is beyond the model's prediction. The expected rotor temperature is 52~to~54 K and the expected CHWP-induced field lens load is $-70$~to~$-10$~mW with respect to the heritage configuration. CHWP-induced power on the field lens is <~0~mW at 85\% confidence when the rotor is <~55~K, which motivates the rotor temperature requirement presented in Sec.~\ref{sec:thermal_requirements}. CHWP-induced power on the 50~K stage is <~1.3~W at 85\% confidence and is nearly independent of rotor temperature.

\subsection{Rotor thermometry}
\label{sec:rotor_thermometry}

Two modes are used to monitor the rotor's temperature. First, when the rotor is gripped, four spring-loaded contacts on each gripper finger touch four flex-circuit traces on the rotor stage (see Sec.~\ref{sec:gripper_design}). These traces are soldered to a four-wire silicon diode thermometer, which is varnished to the sapphire stack's cradle. While only one contact current biases the diode, all three probe the diode's voltage; therefore this flex-circuit thermometry is also a touch-sensing system used to verify the fit between the gripper's copper wedges and the rotor stage's triangular groove.

Second, during continuous rotation, we use the ring magnet as a thermometer. Neodymium has a temperature-dependent magnetic field\footnote{Neodymium magnetization: \\ http://spontaneousmaterials.com/Papers/TN\_0302.pdf} that generates a $\sim$~1~G/K variation $\approx$~5~mm from the ring magnet's face.\cite{Sakurai2018} We varnish a 1~mm~thick cryogenic Hall sensor\footnote{Lakeshore HGCT-3020: https://shop.lakeshore.com/} to the surface of the YBCO and monitor long-time-scale changes in the ring magnet's average field. Limited by degeneracy with rotor position, the sensitivity of this scheme is a few K, and therefore, Hall-sensor monitoring is intended to flag thermal events rather than subtract rotor temperature drifts from detector data. If the need to re-calibrate the Hall-sensor output arises, the CHWP can be stopped and gripped to cross-check the diode thermometer via the touch probes.


\subsection{Installation}
\label{sec:assembly_procedure}

The CHWP assembly is designed to be modular, allowing for seamless integration into a nearly-fully-assembled PB-2b receiver. As shown in Fig.~\ref{fig:chwp_tabletop}, the CHWP assembles mostly on the benchtop, and when integrated, it is bolted to the optics tube's 50~K stage. At this point, the gripper subassemblies are installed, the retaining ring is added, and the IRF, RT-MLI, and vacuum window close the cryostat. Because CHWP integration requires minimal receiver disassembly, the CHWP and receiver were evaluated in parallel prior to full system validation, accelerating the PB-2b commissioning process. The sapphire stack is also modular, enabling the AR coatings to be easily replaced in the field, if necessary.


\section{Laboratory evaluation}
\label{sec:lab_testing}

\begin{figure}[!t]
    \centering
    \includegraphics[width=0.98\linewidth, trim=4.4cm 4.5cm 4.5cm 3.5cm, clip]{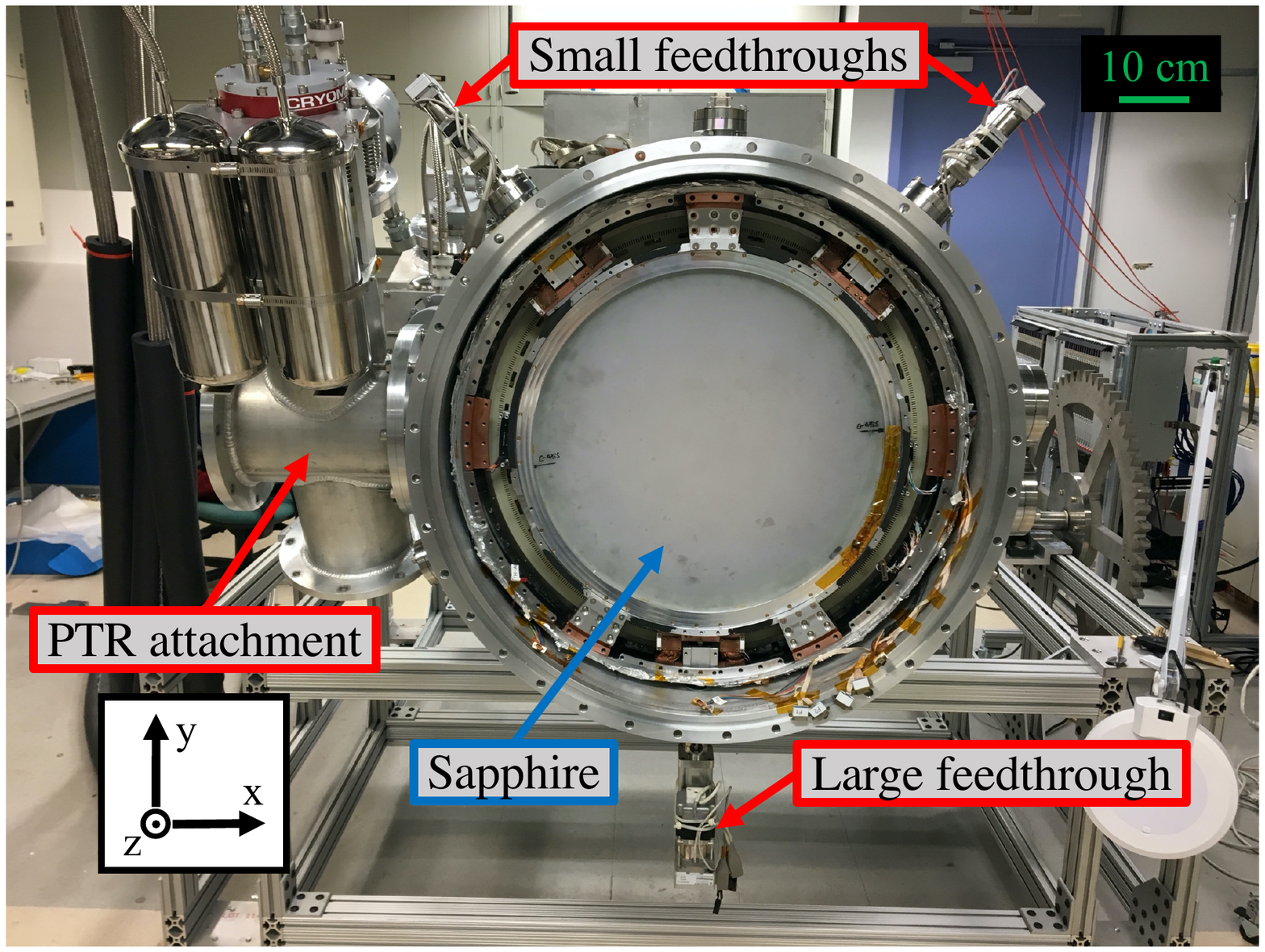}
    \caption{The CHWP mounted in the PB-2b receiver at UCSD. Before the gripper engages, the rotor is held by three installation stanchions that align it to the stator. The top two ``small'' gripper subassemblies are only separated by $80^{\circ}$ to avoid colliding with the PTR attachment. The single sapphire plate has no AR coating, and its frosty appearance is due to its $\sim \; 0.1 \; \mathrm{\mu m}$ RMS surface roughness.}
    \label{fig:chwp_in_pb2b}
\end{figure}

The CHWP is evaluated in two laboratory testing phases. The first phase is conducted in a non-optical setup at Lawrence Berkeley National Laboratory (LBNL). The LBNL cryostat uses a 20~K Gifford-McMahon (GM) refrigerator, and the CHWP's sapphire stack is replaced by a IR-blackened aluminum disk. The second phase integrates the CHWP into a test configuration of the PB-2b receiver at the University of California, San Diego (UCSD), as shown in Fig.~\ref{fig:chwp_in_pb2b}. The UCSD setup employs the full detector array, but the lenses, CHWP, IRF, and vacuum window are not AR coated, and the sapphire stack only includes one 3.8~mm thick sapphire plate. Though optically incomplete, the UCSD configuration allows us to evaluate the cryo-mechanical, thermal, and magnetic interaction between the CHWP and the rest of the experiment.

We divide a discussion of the CHWP evaluation into two subsections. First, we discuss operational testing, including spin-up, continuous rotation, thermal performance, and shutdown. Second, we present the CHWP's data quality and noise impact, including encoder jitter, magnetic interference, and rotor temperature stability.


\subsection{Operational performance}
\label{sec:operational_performance}

In this section, we review CHWP operation, focusing on the gripper, motor, bearing, and thermal performance. We evaluate the interfacing between various subsystems and show that the CHWP executes its essential functions while meeting the requirements shown in Tab.~\ref{tab:requirements}.


\subsubsection{Cooldown}
\label{sec:cooldown}

\begin{figure}[!t]
    \centering
    \includegraphics[width=0.98\linewidth, trim=0.5cm 0.9cm 2.8cm 0.7cm, clip]{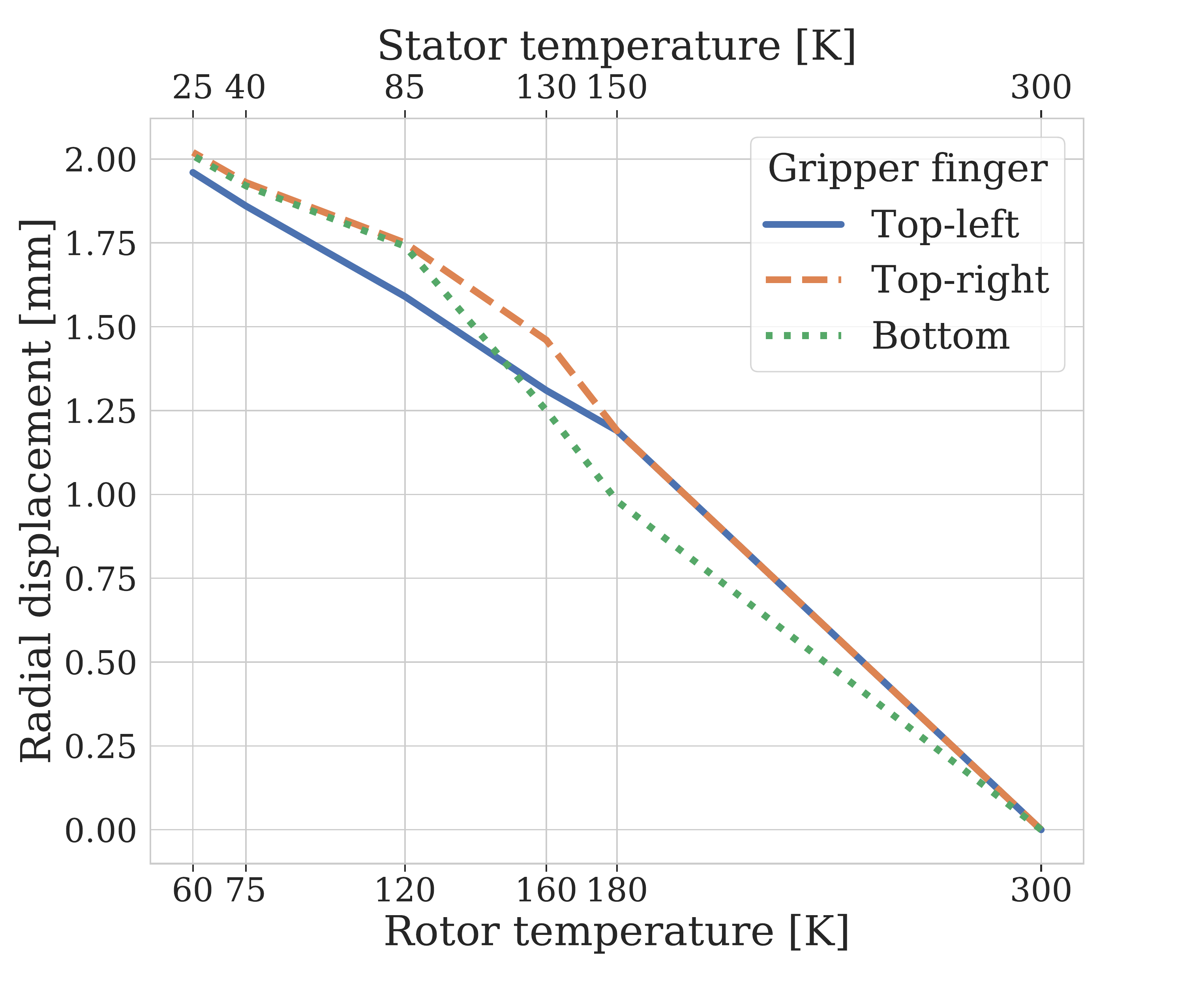}
    \caption{Radial displacement of each gripper finger vs. both rotor and stator temperature during LBNL testing. The gripper-finger labels refer to their locations in Fig.~\ref{fig:chwp_in_pb2b}. The motors are manually commanded at convenient points during the $\approx$~100-hour cooldown, and their relative positions are maintained to $\leq$~0.2~mm throughout.}
    \label{fig:gripper_cooldown}
\end{figure}

When the YBCO is above its $\approx$~90~K transition temperature, the gripper must keep the rotor centered, and between 300 and 50~K, the rotor contracts $\approx$~2~mm radially with respect to the vacuum shell. Therefore during cooldown, the gripper periodically ``re-grips'' the rotor, with each gripper finger inching inwards until the grip force is 50\% of the rotor's weight. This routine both keeps the rotor centered until the YBCO goes superconducting and maintains a conductive cooling path from the rotor to the stator baseplate. Fig.~\ref{fig:gripper_cooldown} shows gripper-finger position vs. both rotor and stator temperature during an LBNL cooldown. In this particular test, the fingers are moved manually at convenient intervals, and their relative positions are maintained to <~0.2~mm, resulting in a rotor-stator coupling efficiency of $\approx$~99\% (see Fig.~\ref{fig:motor_eff}). 

\begin{figure}[!t]
    \centering
    \includegraphics[width=0.98\linewidth, trim=0.5cm 1cm 2.5cm 2.4cm, clip]{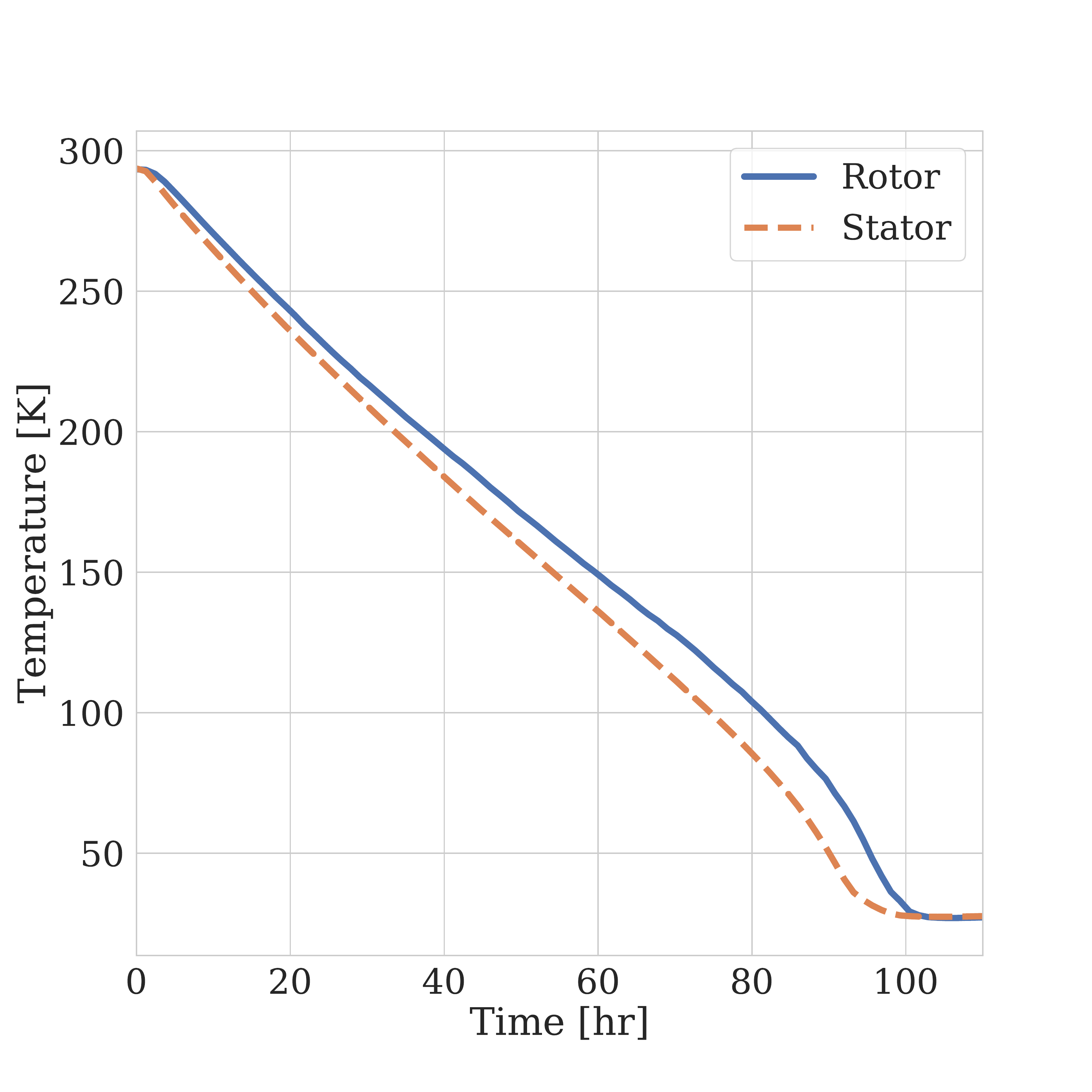}
    \caption{CHWP temperatures during a cooldown to 25 K in the LBNL dark cryostat. While the rotor lags behind the stator by $\approx$~10~K, it thermalizes within five hours of its surroundings.}
    \label{fig:rotor_cooldown}
\end{figure}

Also during cooldown, the rotor's only cooling mechanisms are conduction through the gripper fingers and radiative coupling to the surrounding environment. Because the rotor's heat capacity is $\sim$ 10~kJ/K at 300~K, good rotor-gripper contact conductance is needed for the CHWP to thermalize within 36 hours of the 50~K stage. The rotor stage's triangular groove and gripper fingers' triangular wedges are precision-cut to maximize thermal contact, and the measured rotor-to-gripper conductance is 0.7~W/K at 100~K. Fig.~\ref{fig:rotor_cooldown} shows rotor and stator temperatures during a cooldown in the LBNL cryostat. The rotor lags behind the stator by only five~hours, which is well within the 36-hour requirement. 

\subsubsection{Start-up}
\label{sec:spin_up}

After the YBCO becomes superconducting and the CHWP assembly thermalizes, the rotor is held in place for a few hours to let the bearing ``relax.'' Flux pinning needs time to find its lowest-energy configuration,\cite{postrekhin_dynamics_2001} and if the rotor moves during this relaxation, pinning sites are more likely to escape the superconducting bulk, hence reducing the bearing's spring constant. This flux migration is logarithmic in time, with the vast majority of equilibration occurring within hours of the YBCO transition. After the bearing has relaxed, the gripper fingers retract and the stator supports the rotor's weight. When the receiver is horizontal, as in Fig.~\ref{fig:chwp_in_pb2b}, the rotor is observed to ``sag'' away from its gripped position by 0.5~mm, and because the bearing's spring force is elastic, this displacement decreases with increasing receiver inclination.

\begin{figure}[!t]
    \centering
    \includegraphics[width=0.98\linewidth, trim=0.5cm 1.2cm 2.5cm 3cm, clip]{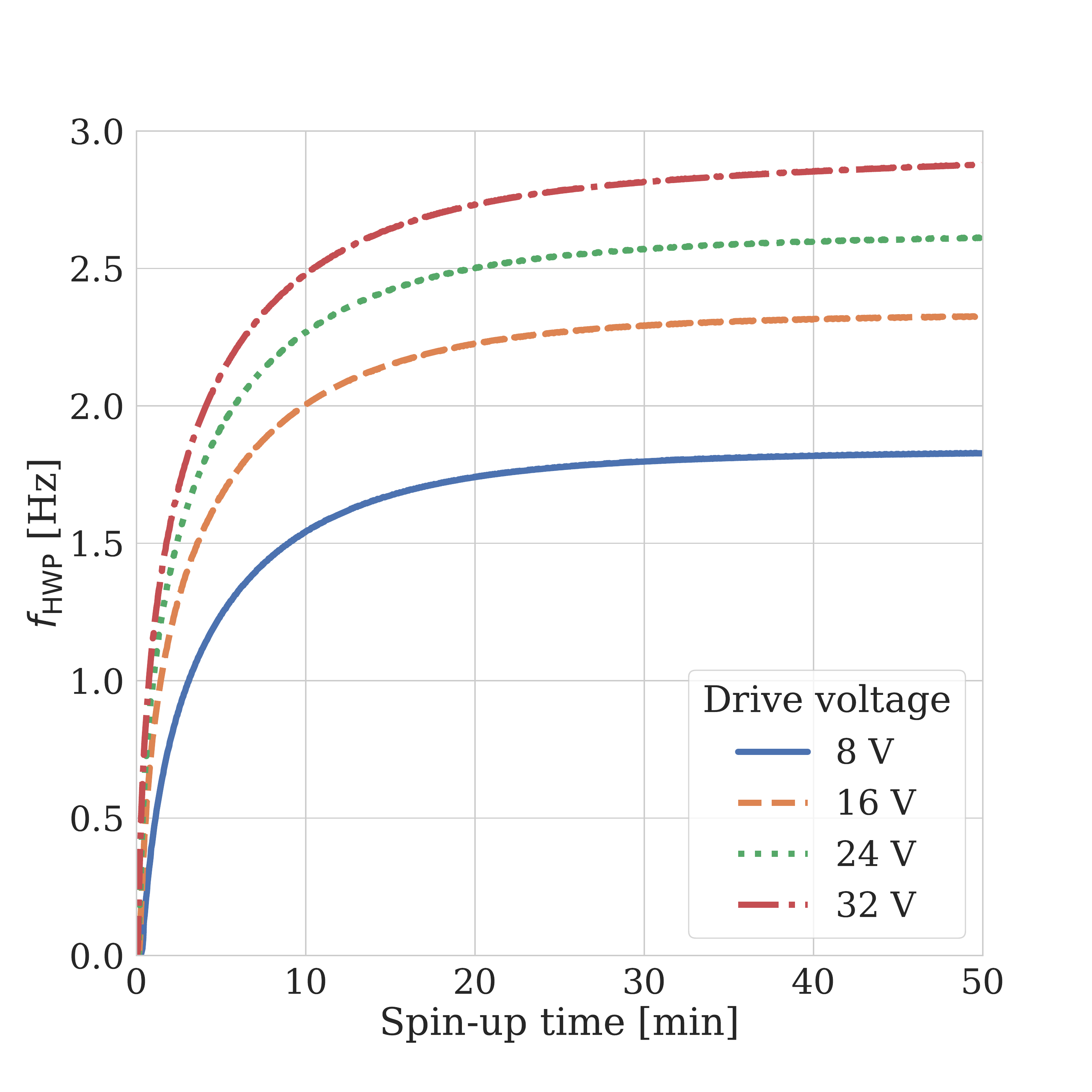}
    \caption{CHWP rotation frequency vs. time during start-up for various H-driver voltages in the LBNL cryostat. The motor attains rotation frequencies up to $\approx$ 2.8 Hz, at which point it becomes limited by motor-phase-delay efficiency (see Fig.~\ref{fig:motor_eff}) and rotor friction (see Fig.~\ref{fig:rot_pow_diss}).}
    \label{fig:spin_up}
\end{figure}

When floating and stationary, forces between the magnet sprockets and the solenoids' ferromagnetic cores keep the rotor azimuthally constrained. In order to overcome this stiction during start-up, the motor energizes with $\approx$~0.4~A (or $\sim$~5$\times$ its current draw during continuous operation). Once rotation commences, cogging quickly diminishes, the rotor transitions into smooth rotation, and the solenoid bias is reduced. Rotation frequency vs. time for various H-driver voltages during spin-up are shown in Fig.~\ref{fig:spin_up}. The achievable rotation frequencies are 1.8-2.8~Hz, and the equilibration time is tens of minutes. While PID control will reduce the start-up time somewhat, the CHWP is intended to run without interruption throughout each day's telescope observations, and therefore this level of latency meets PB-2b's needs.

\subsubsection{Continuous rotation}
\label{sec:continuous_rotation}

Once spun up, the CHWP enters constant velocity mode. To measure the CHWP's open-loop stability, we collect one hour of continuous rotation data in the LBNL test cryostat without PID control. Such a test helps identify any pathologies in the drive system that may manifest as modulations in rotational velocity. While the PB-2b CHWP does not have a velocity stability requirement, steady rotation helps assess encoding accuracy, which is discussed in Sec.~\ref{sec:encoder_jitter}.

\begin{figure}[!t]
    \centering
    \includegraphics[width=0.98\linewidth, trim=0.5cm 1.4cm 2.5cm 3cm, clip]{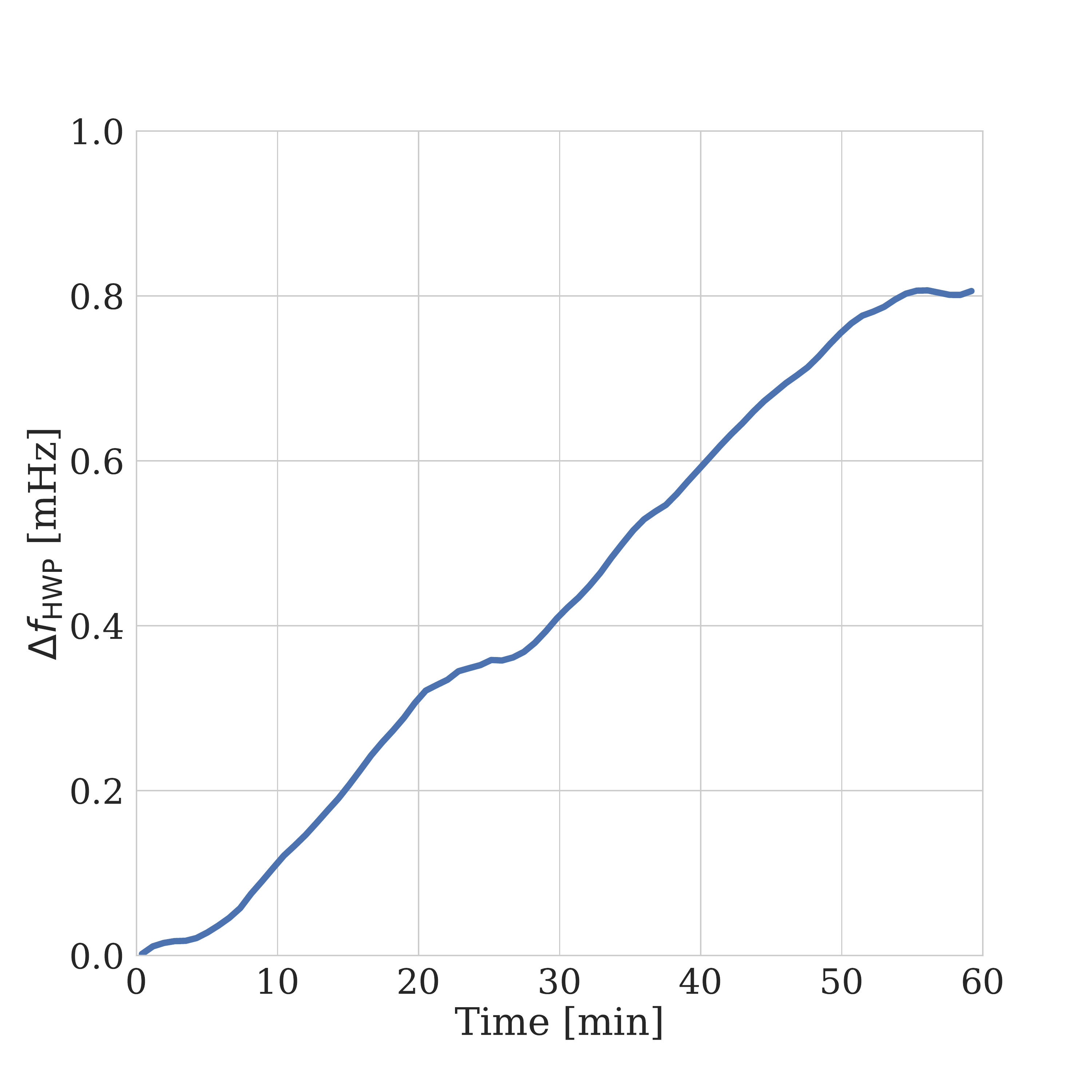}
    \caption{$\Delta f_{\mathrm{HWP}}$ during one hour of continuous rotation in the LBNL cryostat, sampled once per rotation and averaged over one-minute intervals. The mean velocity is $f_{\mathrm{HWP}} = 2.15$~Hz, and therefore the fractional stability is 0.04\%/hr.}
    \label{fig:rot_stability}
\end{figure}

Fig.~\ref{fig:rot_stability} shows velocity drift vs. time at $f_{\mathrm{HWP}} = 2.15 \; \mathrm{Hz}$. The total $\Delta f_{\mathrm{HWP}}= 0.8$~mHz corresponds to a rotational stability of 0.04\%/hr, which is much better than that of ABS\cite{kusaka_modulation_2014} and PB-1. While changes in telescope inclination will slightly modulate rotor-stator concentricity, which in turn modulates motor efficiency (see Sec.~\ref{sec:motor_eff}), we anticipate even better stability when using PID control in the field (see Sec.~\ref{sec:motor_design}).

\subsubsection{Thermal impact}
\label{sec:thermal_performance}

\begin{figure}[!t]
    \centering
    \includegraphics[width=0.98\linewidth, trim=0.4cm 1.2cm 2.3cm 3.2cm, clip]{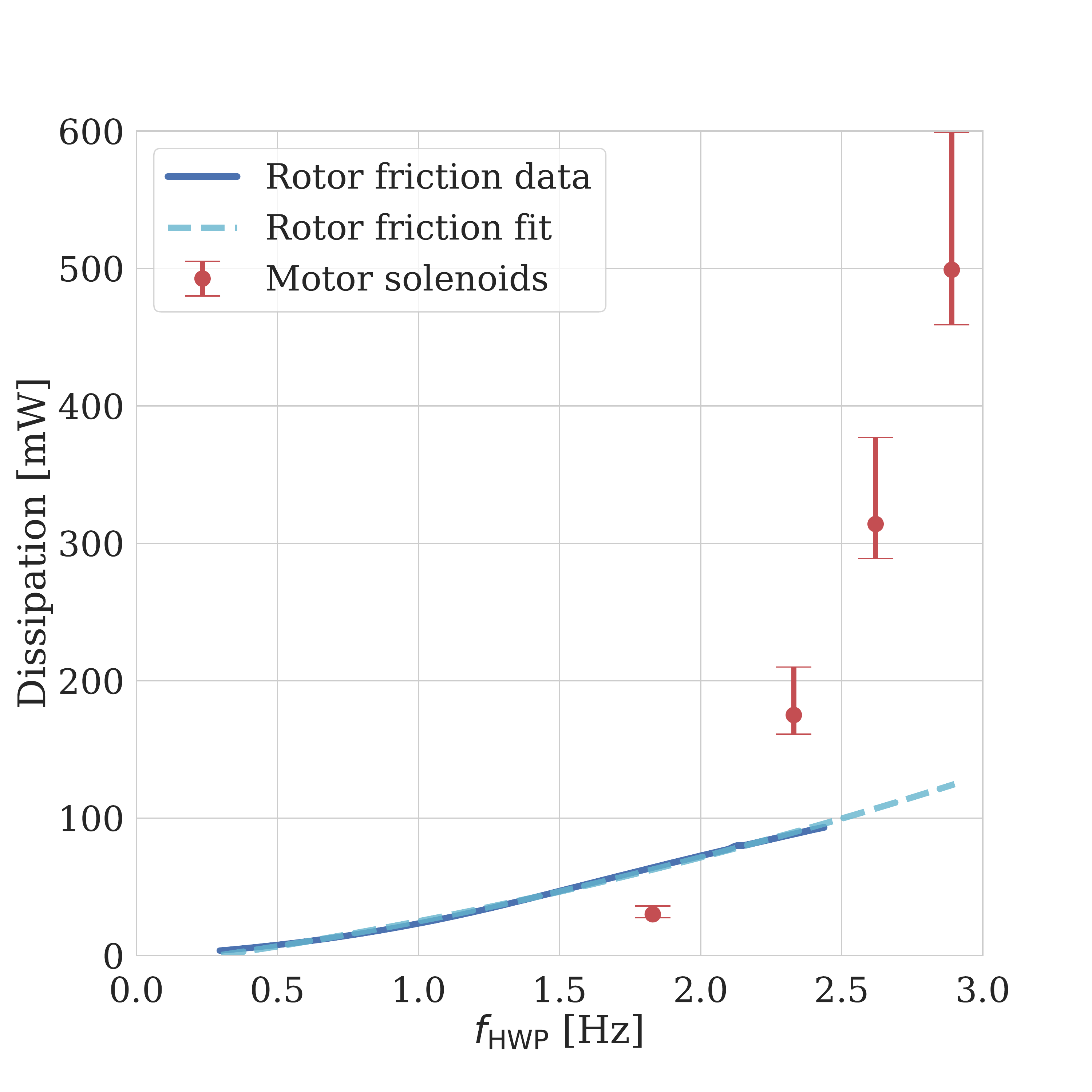}
    \caption{Measured $f_{\mathrm{HWP}}$-dependent dissipation during continuous rotation. As described in Sec.~\ref{sec:thermal_performance}, a polynomial fit shows that rotor friction at 2~Hz is predominantly due to eddy-current losses, while motor dissipation is due to resistive heating in the solenoid array. The motor dissipation's error bars represent possible variations due to rotor positioning, solenoid temperature, and cryostat angle.}
    \label{fig:rot_pow_diss}
\end{figure}

The CHWP's thermal impact is evaluated in three stages. First, we measure rotor friction and motor dissipation in the LBNL cryostat. Rotor friction is determined by measuring deceleration vs. velocity with the motor powered off, and motor dissipation is determined by measuring the current through and resistance of the solenoids. Dissipation as a function of rotation frequency is shown in Fig.~\ref{fig:rot_pow_diss}, and at 2~Hz, rotor and motor heating are 80~mW and <~200~mW, respectively. A polynomial fit to rotor dissipation vs. $f_{\mathrm{HWP}}$ shows that at 2~Hz, most friction is due to eddy currents ($P_{\mathrm{eddy}} \propto f_{\mathrm{HWP}}^{2}$) as opposed to hysteresis loss ($P_{\mathrm{hyst}} \propto f_{\mathrm{HWP}}$). The motor dissipation data points represent measured values with excellent alignment at $T_{\mathrm{rotor}} = 60$~K, and the bands quantify possible variations in motor torque due to variations in rotor-stator concentricity and solenoid temperature. The motor dissipation's steep slope is due to both decreasing motor efficiency and increasing rotor friction with increasing $f_{\mathrm{HWP}}$.

Second, we evaluate cryo-optical performance in the UCSD test setup shown in Fig.~\ref{fig:chwp_in_pb2b}. Even though this configuration has no AR coatings on the IRF, sapphire stack, or lenses, bare alumina and sapphire are largely IR absorptive, allowing us to evaluate the thermal model presented in Sec.~\ref{sec:thermal_design} and App.~\ref{app:thermal_sim}. During UCSD testing, the CHWP's rotor thermometer disconnected, leaving us to infer the rotor's temperature rather than monitor it directly. Using a load curve of the field lens\cite{howe_polarbear-2_2019} and comparing field-lens and 50~K temperatures to a dark run without the CHWP,\cite{howe_polarbear-2_2019} we measure a rotor temperature of <~50~K and a field-lens temperature of 6.4~K, both of which are $\approx$~1$\sigma$ better than the thermal model's expectation after accounting for no AR coatings.\footnote{With one piece of 3.8~mm thick sapphire and with no AR coatings, the rotor temperature and field-lens power in the UCSD test are expected to be lower and higher, respectively, than shown in Fig.~\ref{fig:rotor_temp}.} On the other hand, we measure a 50~K load of $\approx$~2.5~W, which is larger than the 2~W requirement. This test was performed before adopting the solenoid heat sinking scheme described in Sec.~\ref{sec:thermal_design}, which has been shown to reduce solenoid heating by $\sim$ 5x in auxiliary tests. Therefore, we anticipate a stator load consistent with the thermal model's <~1.3~W expectation when the CHWP operates in Chile.

Third, we search for any CHWP-induced heating of the focal plane. We spin the rotor up and down at various drive voltages over several hours in the UCSD setup, and we see no changes in mK temperatures to within a $\approx$~0.3~mK/hr background drift,\footnote{No focal plane temperature regulation was employed during this test, and therefore this background drift is larger than that which is achievable.} implying negligible CHWP-induced focal-plane vibrations. This null result is expected, as the CHWP's resonant frequency $\sqrt{k / m} \approx$~130~Hz is much higher than its $\approx$~2~Hz rotation frequency.


\subsubsection{Shutdown and recovery}
\label{sec:shut_down}

A critical capability for long-term field operation is the clean recovery of the rotor following a power disruption. When the optics tube's PTR shuts off, the CHWP warms with the rest of the experiment, and in the absence of a robust recovery procedure, the rotor may become difficult to recenter without opening the cryostat. Therefore, we employ two redundant procedures to recapture the CHWP in the event of a cryostat warmup.

\begin{figure}[!t]
    \centering
    \includegraphics[width=0.98\linewidth, trim=0.3cm 0.9cm 3.0cm 3.3cm, clip]{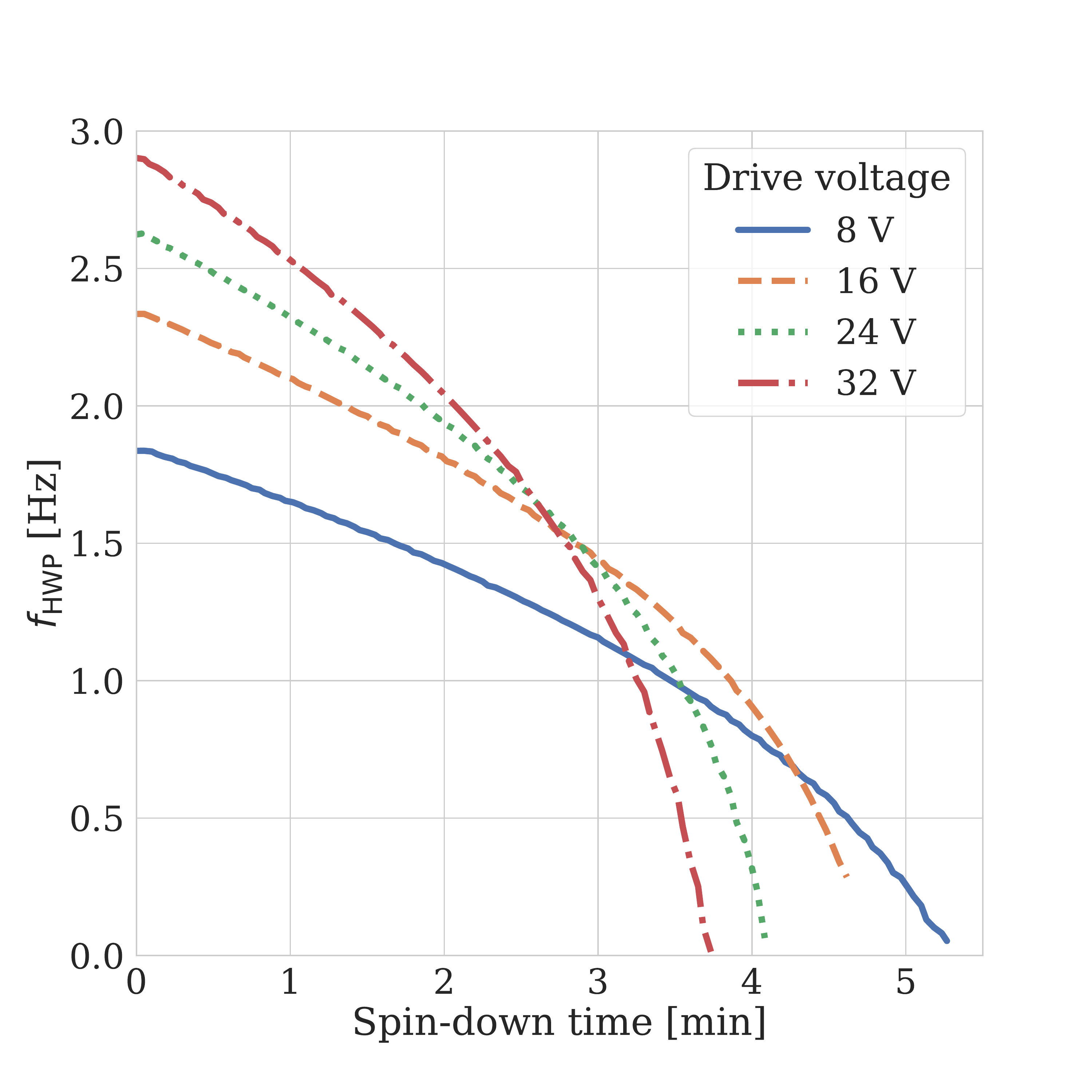}
    \caption{Spin-down tests from various rotation frequencies using various braking voltages. These tests did not employ the PID controller, which will shorten the stopping time for continuous rotation frequencies <~2.8~Hz.}
    \label{fig:spin_down}
\end{figure}

The primary procedure is to stop and grip the rotor before the 50~K stage warms appreciably. The CHWP electronics are powered via an uninterrupted power supply (UPS) that provides a $\sim$~30~min window after site-wide power loss during which the CHWP must be stilled, re-gripped, and stowed. Necessitated by its low friction, the rotor is actively braked by globally inverting the motor's H-driver outputs (see Sec.~\ref{sec:driver}). Fig.~\ref{fig:spin_down} shows rotation frequency vs. time for various braking voltages, and the measured spin-down time is $\approx$~5~min, which is sufficiently shorter than the UPS duration. After rotation stops, the rotor is gripped loosely to provide a margin for thermal expansion, the gripper motors' brakes are applied, and the CHWP electronics are shut down. When site power is later restored, the cooldown procedure in Sec.~\ref{sec:spin_up} commences, and nominal CHWP operation resumes. We have tested this emergency stop and re-grip procedure multiple times in the LBNL cryostat and have shown that it keeps the rotor centered to within 0.5~mm.

In the event of an issue when stopping and re-gripping, we employ a backup procedure to recover the rotor after it has ``fallen.'' As shown in Fig.~\ref{fig:chwp_tabletop}, the retaining ring includes six PTFE crash pads designed to gently ``catch'' the rotor and limit its misalignment after the bearing disengages. Once site power is restored, we orient the receiver horizontally, as shown in Fig.~\ref{fig:chwp_in_pb2b}, lift the rotor along the $+y$ direction using the bottom gripper finger (see Sec.~\ref{sec:gripper_design}), and then grab it using the top two fingers. This after-the-fact technique is less than ideal, as it requires the rotor to be secured and realigned from an indeterminate position. Nonetheless, it provides necessary insurance against opening the cryostat, which is an expensive operation. We have tested this backup procedure in the LBNL and UCSD cryostats and have shown it to recenter the rotor to within 1.5~mm of its original position, which is good enough to reattain 2~Hz rotation.

\subsection{Data quality}
\label{sec:data_quality}

In this section, we review the impact of the CHWP on experiment data quality within the context of the noise requirements presented in Sec.~\ref{sec:noise_requirements}. Specifically, we discuss encoder jitter, magnetic interference, and rotor temperature stability, highlighting the achieved values in Tab.~\ref{tab:requirements}.


\subsubsection{Angle encoder performance}
\label{sec:encoder_jitter}

We use the same hour-long data set presented in Sec.~\ref{sec:continuous_rotation} to evaluate the angle encoder described in Sec.~\ref{sec:angle_encoder}. For purely historical reasons, this test is performed using an Arduino MCU,~\footnote{Arduino Leonardo: https://store.arduino.cc/usa/arduino-leonardo-eth} which employs two separate 16~MHz clocks, instead of the BBB, which employs one shared 200~MHz clock. Additionally, this test did not utilize PID control and therefore relies on open-loop motor stability to maintain steady rotation. Our primary goal in this section is to measure the angle encoder's white noise level and compare it to the $\ll$~$3$~$\mathrm{\mu rad / \sqrt{Hz}}$ requirement in Tab.~\ref{tab:requirements}.

As described in Sec.~\ref{sec:angle_encoder}, the rotor angle $\chi(t)$ is reconstructed by linearly interpolating angle encoder ticks to IRIG time $t$ as
\begin{equation}
    \chi(\tau_{\mathrm{Enc}}) \longleftrightarrow t(\tau_{\mathrm{IRIG}}) \, ,
    \label{eq:interpolation}
\end{equation}
where $\tau_{\mathrm{Enc}}$ and $\tau_{\mathrm{IRIG}}$ are the MCU clocks for the encoder and IRIG signals, respectively. Deviations between $\tau_{\mathrm{Enc}}$ and $\tau_{\mathrm{IRIG}}$ on the Arduino lead to additional noise that will not exist when using the BBB in the field. Therefore, we analyze both $\chi(t)$ and $\tau_{\mathrm{IRIG}}(t)$ to distinguish MCU drifts from those of the rotation mechanism and encoder system.

To facilitate the following discussion, we consider two illustrative angle jitter definitions: the residual after subtracting a quadratic fit
\begin{equation}
\delta \chi_{\mathrm{HWP}}^{\mathrm{poly}}(t) = \chi(t) - P_{\mathrm{2}}^{\mathrm{\chi}}(t) \, ,
\label{eq:angle_poly_jitter}
\end{equation}
\noindent
and the residual after subtracting a quadratic fit plus an angle-dependent function $A(\chi)$
\begin{equation}
    \delta \chi_{\mathrm{HWP}}^{\mathrm{temp}}(t) = \chi(t) - \left[ P_{2}^{\chi}(t) + A(\chi) \right] \, .
\label{eq:angle_temp_jitter}
\end{equation}
\noindent
For this test data, we select a simple $2 \pi$-repeating template
\begin{equation}
    A(\chi) = A(\chi - 2 \pi n) \; ; \; n \in \{0, 1, \cdots, N_{\mathrm{rev}} \} \, ,
    \label{eq:template}
\end{equation}
where $N_{\mathrm{rev}}$ is the number of completed revolutions in the data set and where $A(\chi - 2 \pi n)$ is constructed by taking a window-weighted average of the encoder pattern over all $N_{\mathrm{rev}}$. In practice, a more complex function can be used to remove HWP-synchronous signals in the field,\cite{essinger-hileman_systematic_2016} but a uniform template works well for lab characterization. Additionally, we define IRIG clock jitter as
\begin{equation}
    \delta \tau_{\mathrm{IRIG}}^{\mathrm{poly}}(t) = \tau_{\mathrm{IRIG}}(t) - P_{2}^{\tau}(t) \, ,
\end{equation}
which we use to evaluate MCU clock drifts.

\begin{figure}[!t]
    \begin{subfigure}
    \centering
    \includegraphics[width=0.98\linewidth, trim=0cm 1cm 0.1cm 3cm, clip]{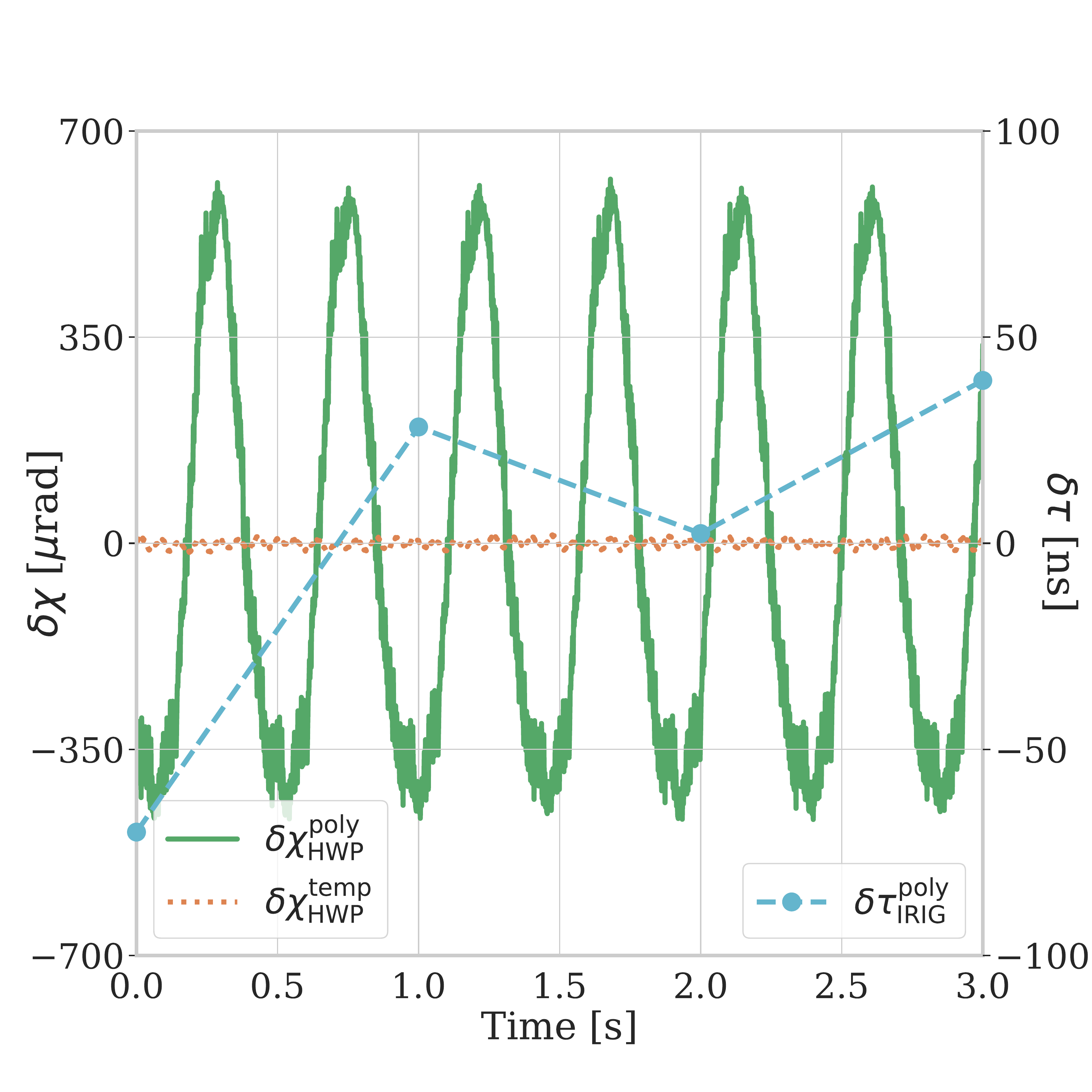}
    \end{subfigure}
    \begin{subfigure}
    \centering
    \includegraphics[width=0.98\linewidth, trim=7cm 4.3cm 7.9cm 5.2cm, clip]{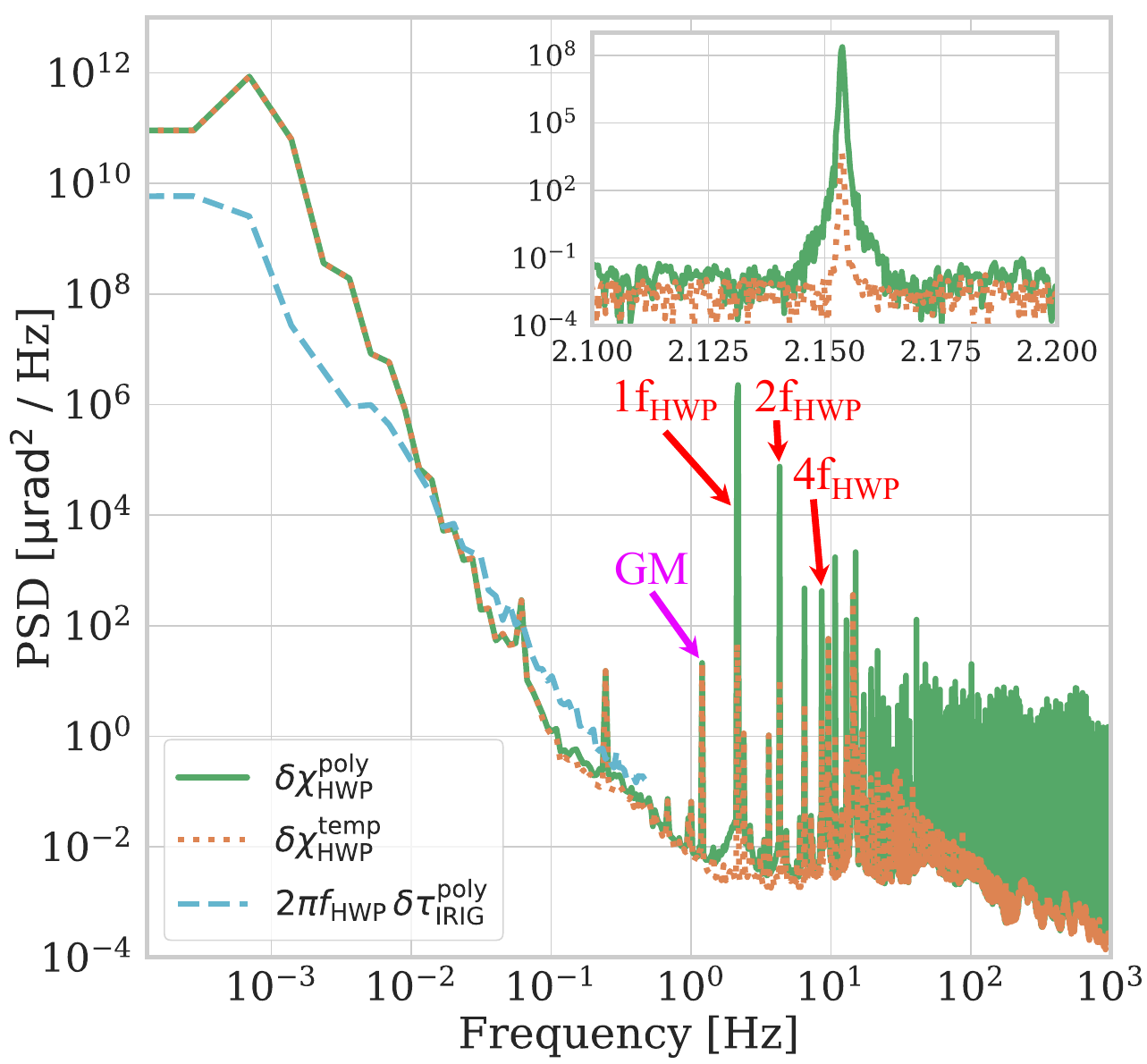}
    \end{subfigure}
    \caption{A measurement of encoder performance during 1 hour of testing in the LBNL setup. \textbf{Top panel:} a three-second sample of the angle and IRIG jitters. The rotor is spinning at $\approx$~2.15~Hz and its angle is sampled 1,140 times per revolution. \textbf{Bottom panel:} PSDs of the angle and IRIG jitters, including a zoomed inset of the $1 f_{\mathrm{HWP}}$ peak with finer binning. We multiply $\delta \tau_{\mathrm{IRIG}}^{\mathrm{poly}}$ by $2 \pi f_{\mathrm{HWP}}$ to covert it from seconds to radians.}
    \label{fig:jitter_psd}
\end{figure}

A three-second segment of angle jitters $\delta \chi_{\mathrm{HWP}}^{\mathrm{poly}}$ and $\delta \chi_{\mathrm{HWP}}^{\mathrm{temp}}$ are shown in the top panel of Fig.~\ref{fig:jitter_psd}. The polynomial-subtracted spectrum $\delta \chi_{\mathrm{HWP}}^{\mathrm{poly}}$ shows a distinct peak-to-peak variation of $\approx$~1~mrad at $f_{\mathrm{HWP}}$ in addition to higher-order structures, and the template-subtracted spectrum $\delta \chi_{\mathrm{HWP}}^{\mathrm{temp}}$ shows that most of this jitter is removed by $A(\chi)$. A unique property of the SMB is that it does not wobble, even if it is gravitationally imbalanced. Therefore the observed $\delta \chi_{\mathrm{HWP}}^{\mathrm{poly}}$ signal is caused by a slight misalignment between the slotted encoder plate and magnet ring, while higher-frequency structures are caused by non-uniform patterning of the encoder slots. Three seconds of IRIG clock jitter $\delta \tau_{\mathrm{IRIG}}^{\mathrm{poly}}$ is also shown in Fig.~\ref{fig:jitter_psd} and is effectively a measurement of Arduino clock noise. Because the IRIG signal is GPS synchronized to $\leq$~40~ns, the observed drift is due to instability of the Arduino's crystal oscillators, which we discuss further below.

Power spectral densities (PSDs) of the angle and clock jitters are shown in the bottom panel of Fig.~\ref{fig:jitter_psd}. The sharpness of the peaks in the $\delta \chi_{\mathrm{HWP}}^{\mathrm{poly}}$ spectrum demonstrates excellent rotational stability, and the suppression of those peaks in the $\delta \chi_{\mathrm{HWP}}^{\mathrm{temp}}$ spectrum confirms that $A(\chi)$ measures all but a few $\delta \chi_{\mathrm{HWP}}$ features with high signal-to-noise. The $A(\chi)$ subtraction also suppresses side-band power in the $f_{\mathrm{HWP}}$ harmonics, better revealing a white-noise level of $\approx \: 0.1 \; \mathrm{\mu rad / \sqrt{Hz}}$.

Both $\delta \chi_{\mathrm{HWP}}$ spectra have a 1/f knee of $\approx$~1~Hz, which is larger than expected from the $\Delta f_{\mathrm{HWP}}$ measurement in Fig.~\ref{fig:rot_stability}. Therefore, we compare the angle and IRIG spectra to analyze the contribution of MCU clock drifts to the observed 1/f noise. The two MCU clocks, $\tau_{\mathrm{IRIG}}$ and $\tau_{\mathrm{Enc}}$ (see Eq.~\ref{eq:interpolation}), fluctuate with $\approx$~90\% coherence, and because common-mode clock drifts are subtracted during angle-time interpolation, $\delta \chi_{\mathrm{HWP}}$ 1/f power dips beneath that of $2 \pi f_{\mathrm{HWP}} \delta \tau_{\mathrm{IRIG}}^{\mathrm{poly}}$ between $\approx$~0.05~Hz and the IRIG's 0.5~Hz Nyquist frequency. Below $\approx$~0.05~Hz, drifts in rotational velocity begin to contribute and the $\delta \chi_{\mathrm{HWP}}$ spectra steepen, but even so, MCU-induced 1/f noise remains dominant down to $\sim$~0.01~Hz. This finding  motivated us to replace the Arduino with the BBB, which uses a single shared clock ($\tau_{\mathrm{Enc}} = \tau_{\mathrm{IRIG}}$) and has a measured 1/f knee of <~0.1~Hz. This configuration allows the 1~Hz IRIG signal to fully subtract MCU clock drifts during angle-time interpolation, and therefore we expect significantly improved low-frequency noise when measuring $\chi(t)$ in the field.

\subsubsection{Magnetic interference}
\label{sec:magnetic_interference}

The CHWP's motor and rotor both introduce magnetic interference that can affect the detectors and their SQUID amplifiers. In particular, interference at 4$f_{\mathrm{HWP}}$ mimics a sky signal and must be especially well controlled. The motor energizes three phases across 114 solenoids at 38$f_{\mathrm{HWP}}$, and its large multipole number causes its $\approx$ 20~G field to decay quickly with distance. The magnet ring, on the other hand, has only 16 segments---resulting in a lower multipole number---and a surface magnetization of $\approx$~5,000~G, posing a greater 4$f_{\mathrm{HWP}}$ interference concern. 

\begin{figure}[!t]
    \begin{subfigure}
        \centering
        \includegraphics[width=0.98\linewidth, trim=0cm 1cm 2cm 2.5cm, clip]{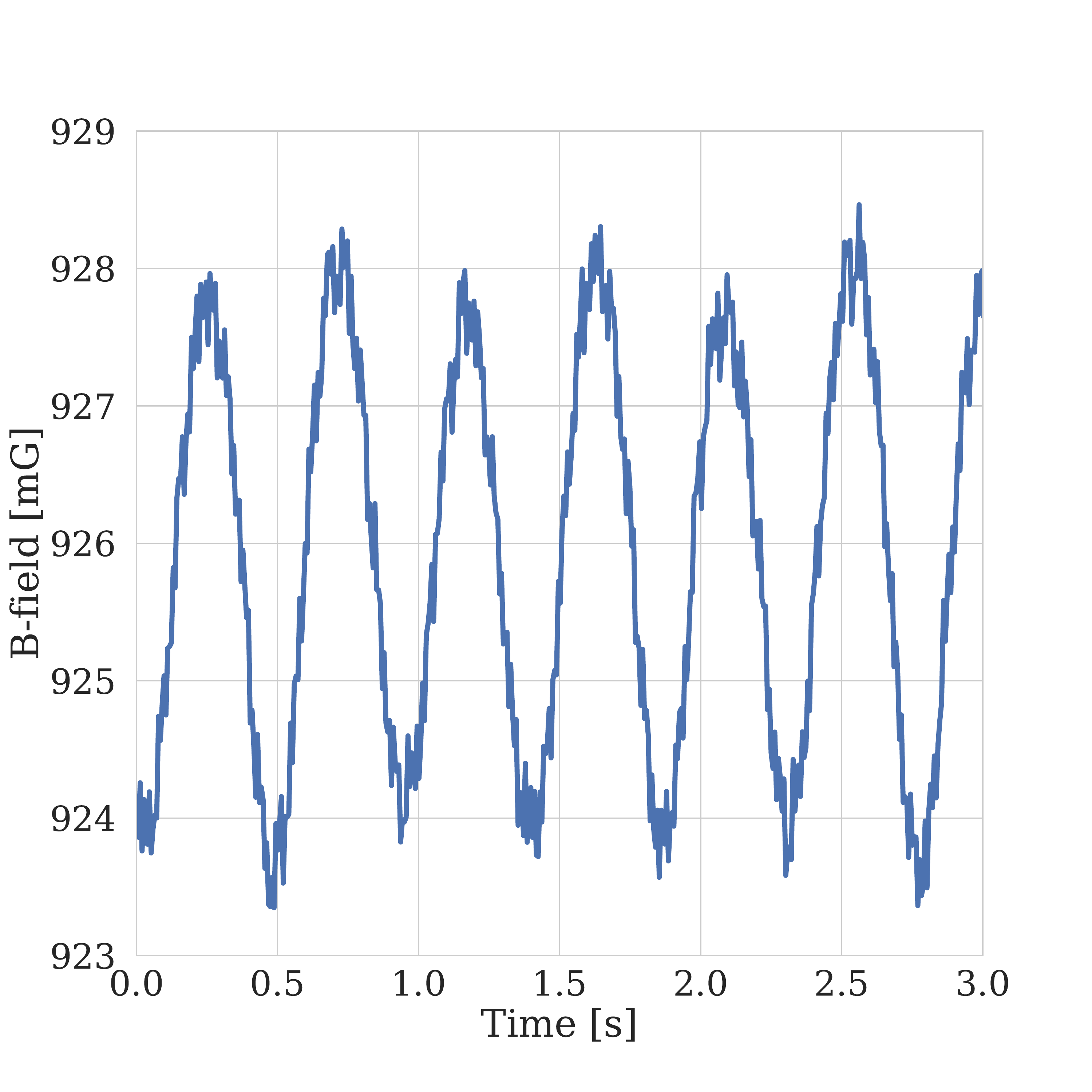}
    \end{subfigure}
    \begin{subfigure}
        \centering
        \includegraphics[width=0.98\linewidth, trim=5.5cm 4cm 6.5cm 5cm, clip]{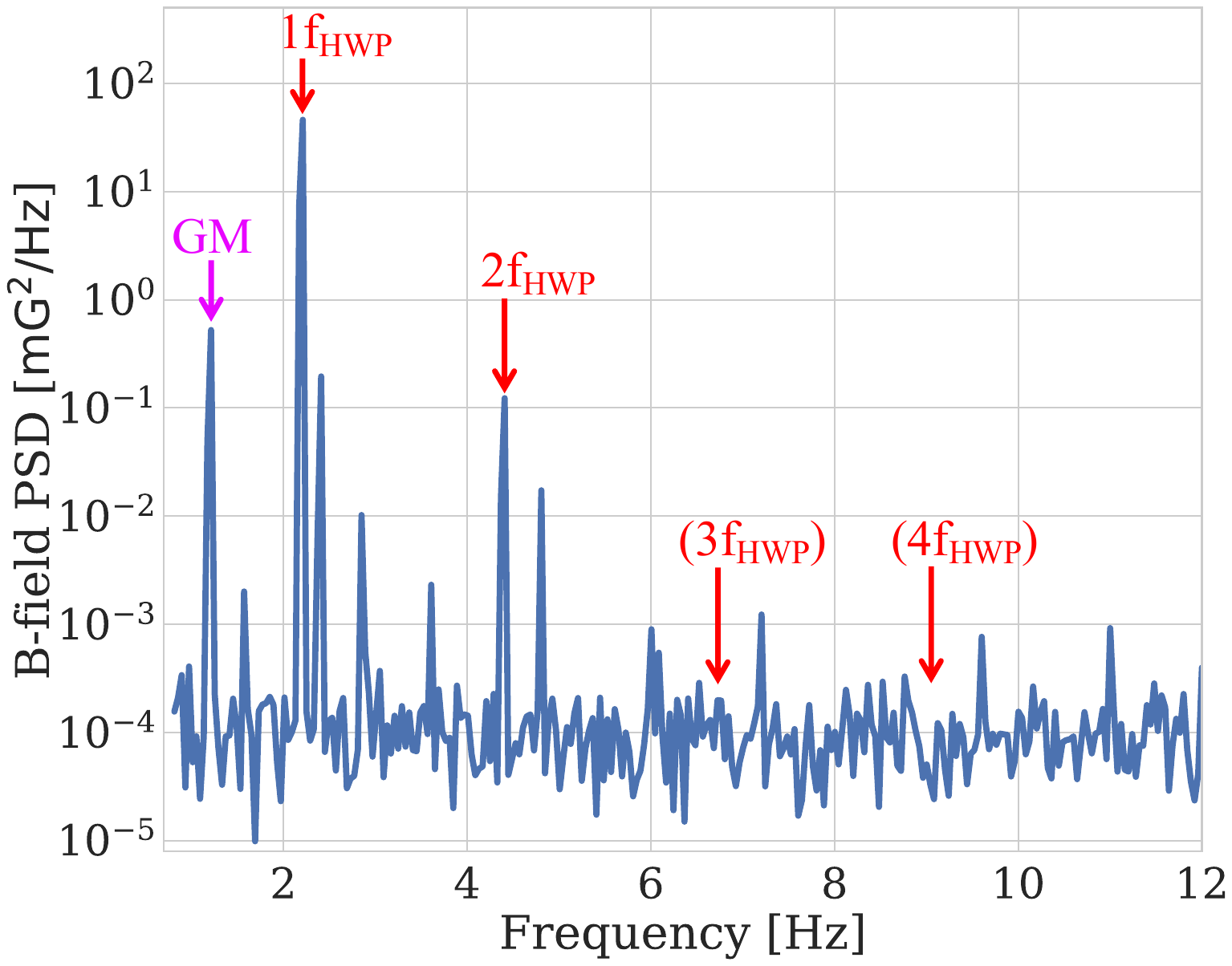}
    \end{subfigure}
    \caption{A measurement of the magnetic field 1.5~m behind the CHWP assembly during 100~s of continuous rotation at 2.15~Hz in the LBNL cryostat. \textbf{Top panel:} 3~s of time-ordered magnetometer data, showing a clear $\approx$~5~mG peak-to-peak variation at 1$f_{\mathrm{HWP}}$. \textbf{Bottom panel:} the B-field's power spectral density. Red arrows mark rotation-synchronous peaks, while the magenta arrow marks that of the GM cooler. The unmarked peaks are environmental and are predominantly GM harmonics.}
    \label{fig:mag_spec}
\end{figure}

Magnetic field testing is performed at LBNL using a room-temperature magnetometer\footnote{Honeywell HMR2300: https://aerospace.honeywell.com/} placed 1.5~m behind the CHWP assembly at the approximate location of a detector near the edge of the focal plane\footnote{We calculate that CHWP-induced magnetic fields are larger near the edge of the focal plane than near the center. Therefore, this measurement represents a worst-case scenario for potential detector interference.} (see Fig.~\ref{fig:pb2_ray_trace}). The CHWP rotates steadily at $\approx$~2.15~Hz, and the ambient magnetic field is measured at 150 samples per second for 100~s. The results are presented in Fig.~\ref{fig:mag_spec}. The time-ordered data show a DC field of 926~mG\footnote{This average field applies a $\approx$~-0.3~mK DC shift to the TES transition temperature $T_{\mathrm{c}}$, which is less than $T_{\mathrm{c}}$ variation across the focal plane\cite{westbrook_polarbear-2_2018} and is calibrated out during detector tuning.} and a $\approx 5$~mG oscillation at $1 f_{\mathrm{HWP}}$, while the spectrum shows peaks at $1 f_{\mathrm{HWP}}$, $2 f_{\mathrm{HWP}}$, and at harmonics of the GM cooler's cycle frequency. However, the $4 f_{\mathrm{HWP}}$ component lies beneath the noise floor, setting a bound on its amplitude of $< \: 10 \; \mathrm{\mu G / \sqrt{Hz}}$.

Full-system interference testing is performed at UCSD by monitoring detector outputs when the CHWP is spinning. During two minutes of rotation at 2~Hz, data is collected from both optical and non-optical bolometers across the full focal plane. While $2 f_{\mathrm{HWP}}$ and $4 f_{\mathrm{HWP}}$ signals are visible in the optical detectors, as expected, no CHWP-induced signals are detected in any of the dark detectors. In addition, because 4$f_{\mathrm{HWP}}$ magnetic interference is $\leq \; 10^{-6}$ that at 1$f_{\mathrm{HWP}}$, as shown in Fig.~\ref{fig:mag_spec}, this test also suggests that no 4$f_{\mathrm{HWP}}$ magnetic features will arise when $\sim \: 5 \times 10^{3}$ detectors are coadded during data analysis.


\subsubsection{Rotor temperature stability}
\label{sec:temperature_stability}

To evaluate thermal stability, we simulate rotor temperature using the thermal model presented in Sec.~\ref{sec:thermal_design} and App.~\ref{app:thermal_sim}. Because it is floating, the rotor's temperature variations are dominated by those of the 50~K stage to which it is radiatively tied. These CHWP fluctuations are expected to be very slow, as the rotor has a $\approx$~900~J/K heat capacity at base temperature and a $\approx$~10~mW/K coupling to its 50~K surroundings (see App.~\ref{app:thermal_sim}). Due to diurnal variations in ambient temperature and changes in telescope elevation---which impact PTR performance---50~K temperature drifts in Chile are substantially larger than those in the lab. Therefore, we use 24 hours of 50~K temperature data from PB-2a, which is operating in the field, to simulate the expected rotor stability. The selected PB-2a data includes telescope slew testing, and therefore the presented 50~K-stage variations represent an upper bound on those expected during science observations.

\begin{figure}[!t]
    \centering
    \includegraphics[width=0.98\linewidth, trim=0.3cm 1cm 2.4cm 2.5cm, clip]{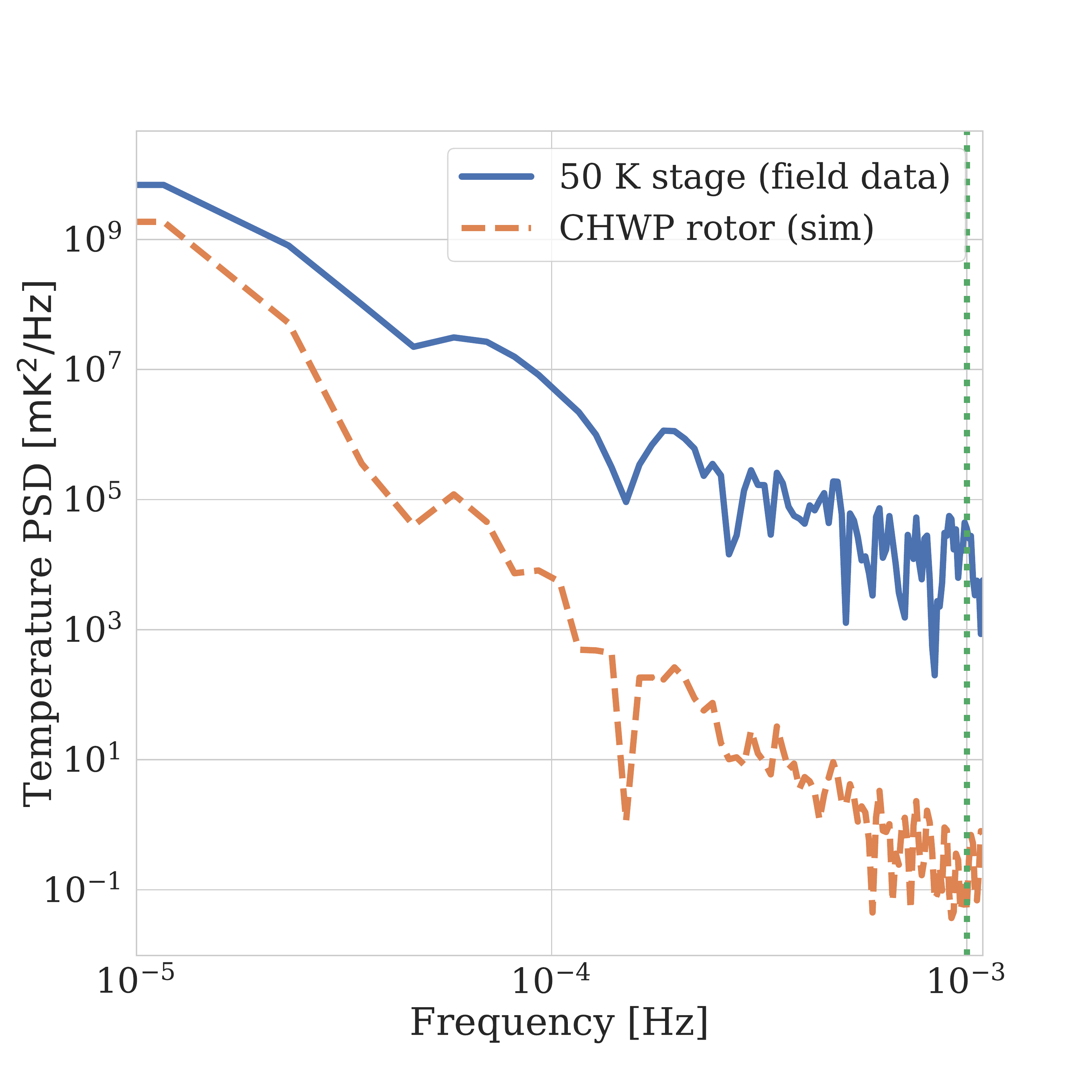}
    \caption{Power spectra of measured 50~K temperature variations from PB-2a in Chile and of the resulting simulated rotor temperature variations. The rotor's large thermal mass and small radiative coupling make it an effective low-pass filter, suppressing IRF fluctuations by >~100$\times$ at 1~mHz.}
    \label{fig:hwp_stability}
\end{figure}

The measured 50~K-stage and simulated rotor PSDs are shown in Fig.~\ref{fig:hwp_stability}. The rotor acts as a low-pass filter, and its simulated drift is $\approx$~0.3~$\mathrm{mK / \sqrt{Hz}}$ at 1~mHz, which is below the 1~$\mathrm{mK / \sqrt{Hz}}$ requirement in Tab.~\ref{tab:requirements}. We note that rotor temperature stability can be substantially improved by regulating the 50~K-stage temperature, which may become necessary for future experiments with tighter noise requirements.

\section{Conclusions}
\label{sec:conclusion}

We have presented the design and evaluation of a cryogenic half-wave plate (CHWP) for the POLARBEAR-2b (PB-2b) CMB receiver. This paper advances the research area of cryogenic polarization modulators for mm-wave and sub-mm astronomical observations, introducing a new motor, encoder, and grip-and-release mechanism while pushing the superconducting magnetic bearing and sapphire diameters to the largest ever deployed. The CHWP has been evaluated in both a non-optical standalone cryostat and an optical test configuration of the PB-2b receiver. Its performance satisfies all PB-2b-specific requirements and includes a 430~mm clear aperture, a rotor temperature of <~53~K, <~1.3~W of dissipation during continuous operation, rotation frequencies up to 2.8~Hz, 0.1~$\mathrm{\mu rad \, / \, \sqrt{Hz}}$ of encoder noise, and < 10~$\mathrm{\mu G \, / \, \sqrt{Hz}}$ of magnetic interference at $4 f_{\mathrm{HWP}}$. Additionally, an emphasis has been placed on system robustness to ensure years of continuous operation in the field. The presented CHWP instrument has deployed to the Chilean observation site and is expected to see first light in 2020 or 2021.

The PB-2b CHWP design is already making its way into other experiments. PB-2c is building an identical CHWP for observation at 220 and 270~GHz, and the Simons Observatory has adopted the presented system, with some system-dependent modifications, for cryogenic polarization modulation on their small aperture telescopes.\cite{galitzki_simons_2018,ali_small_2020} Additionally, several of its core strengths, such as its low-noise encoder and easy-to-operate driver, make the PB-2b CHWP system portable to future CMB experiments including CMB Stage-4.\cite{abitbol_cmb-s4_2017}


\begin{acknowledgments}
The presented CHWP development at LBNL is supported in part by the U.S. Department of Energy, Office of Science, Office of High Energy Physics under contract No. DE-AC02-05CH11231, as well as the LBNL Laboratory Directed Research and Development (LDRD) Program. This work is also supported in part by the U.S. Department of Energy, Office of Science, Office of Workforce Development for Teachers and Scientists (WDTS) under the Science Undergraduate Laboratory Internships Program (SULI). AK acknowledges the support by JSPS Leading Initiative for Excellent Young Researchers (LEADER) and by the JSPS KAKENHI Grant Numbers JP16K21744, JP18H05539, and JP19H00674. This work is additionally supported in part by the World Premier International Research Center Initiative (WPI), MEXT, Japan. PB-2b and Simons Array, whose activities are integrated into much of the presented work, are supported by the Simons Foundation, Gordon and Betty Moore Foundation, Templeton Foundation, and National Science Foundation (NSF) grant numbers AST-0618398 and AST-1212230.

We thank Cory Lee in the LBNL physics division who fabricated and influenced many design choices during early-stage prototyping, and we thank Warner Carlisle, Gordon Long, Tommy Gutierrez, and Abel Gonzalez in the UC Berkeley physics department for machining, welding, and consulting on many of the CHWP's large-diameter, tight-tolerance components. We thank Suzanne Staggs at Princeton University for loaning the CAPMAP dewar, which was used for the presented LBNL testing, and we thank Lewis Hyatt for providing CAD drawings of CAPMAP. We thank Frank Werfel and Uta Flögel-Delor at ATZ in Germany for fruitful collaborative research that led to the presented large-diameter superconducting magnetic bearing, and we thank Carl Johnson, Carl Zhang, and Dr. Yong Ji at GHTOT for pushing the limits of sapphire manufacturing to enable sapphire HWPs for high-throughput CMB instruments. We thank Muhammad Suri at SMC Corporation for many suggestions and ideas around robust gripper operation, and we thank Tijmen de Haan and Darcy Barron for useful discussions about SQUIDs in ambient magnetic fields.

We want to recognize the bright and talented scientists who worked as undergraduates on the CHWP development at LBNL and UC Berkeley: Richard Chen, Hawkins Clay, Alexander Droster, Andrew Fischer, Mael Flament, Chingam Fong, Samantha Gilbert, Grant Hall, Arian Jadbabaie, Alex Madurowicz, Adam Rutkowski, and Danielle Sponseller.

We also want to recognize the broader PB-2b and Simons Array team members who have helped with and consulted on a plethora of CHWP-related areas, including receiver integration, analysis considerations, data acquisition implementation, and detector/readout implications: Darcy Barron, Yuji Chinone, Tucker Elleflot, Neil Goeckner-Wald, John Groh, Logan Howe, Jennifer Ito, Oliver Jeong, Lindsay Lowry, Marty Navaroli, Haruki Nishino, Toki Suzuki, Praween Siritanasak, Nate Stebor, Satoru Takakura, and Calvin Tsai.
\end{acknowledgments}

\appendix


\section{Thermal model}
\label{app:thermal_sim}

\begin{figure}[!t]
    \centering
    \includegraphics[width=0.98\linewidth, trim=8.5cm 4cm 8cm 3cm, clip]{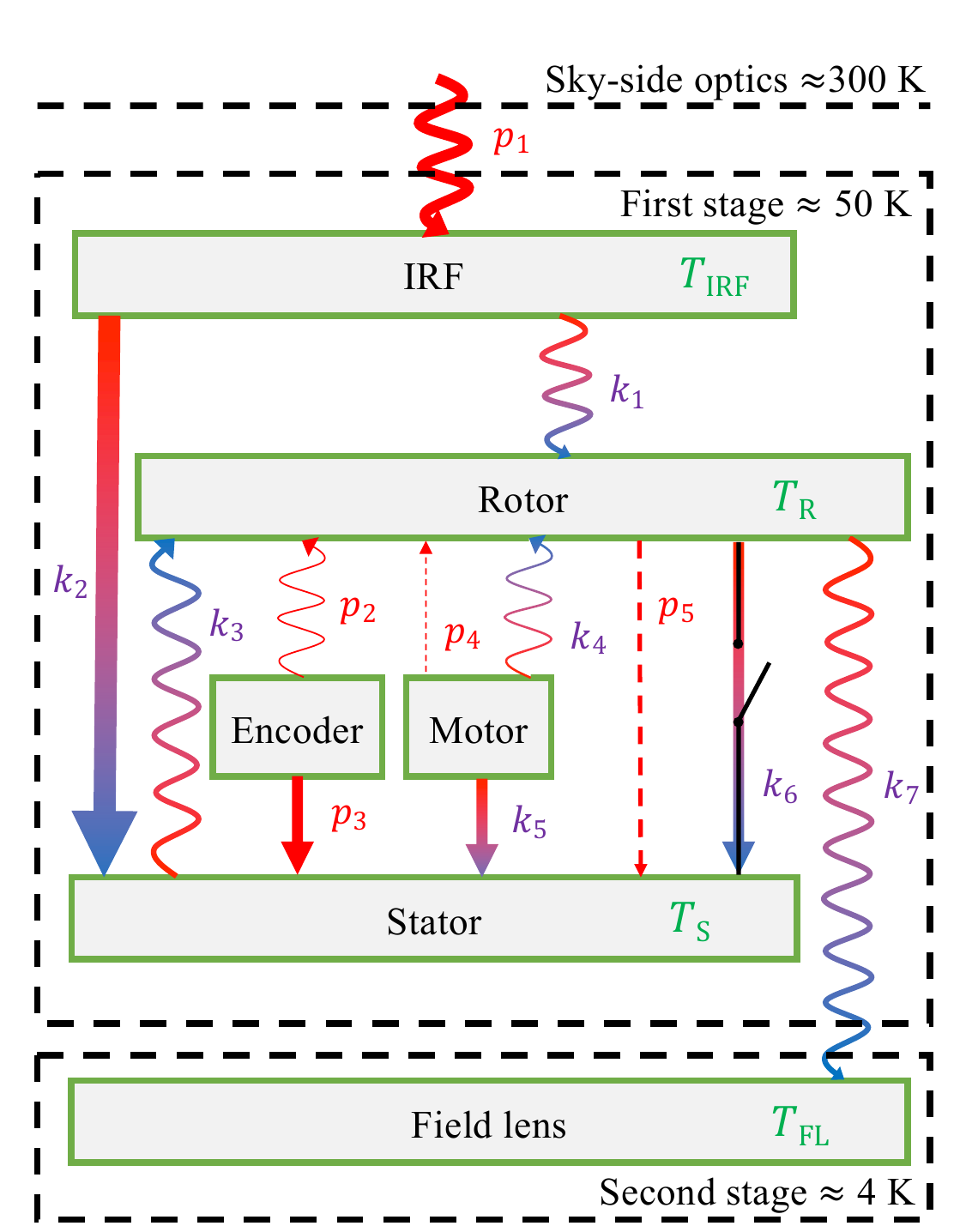}
    \caption{The CHWP thermal circuit during nominal operation. The squiggly lines represent radiative loads, the straight lines conductive loads, and the dotted lines dissipative loads. Loads labeled by $p$ represent constant power, while loads labeled with $k$ represent power that depends on operating temperatures. Each conductivity has a red-to-blue color gradient from hot to cold. The switched conductivity $k_{6}$ represents the gripper connection, which is closed when the rotor is gripped and open when it is not. The lines' thicknesses show the relative magnitudes of the various contributions but are not to scale and are intended only as a visual guide. The IRF ($T_{\mathrm{IRF}}$), rotor ($T_{\mathrm{R}}$), stator ($T_{\mathrm{S}}$), and field lens ($T_{\mathrm{FL}}$) temperatures are the system's merit figures. The measured and calculated values for each symbol are given in Tab.~\ref{tab:thermal_values}.}
    \label{fig:thermal_diagram}
\end{figure}

In this appendix, we review the details of the thermal model used in Secs.~\ref{sec:thermal_design} and~\ref{sec:temperature_stability} to simulate the rotor's equilibrium temperature, the load on the field lens, and the rotor's temperature stability. Fig.~\ref{fig:thermal_diagram} shows a schematic for the model, and Tab.~\ref{tab:thermal_values} shows the measured and calculated values for the schematic. The errors on the measured values contain both measurement and configurational uncertainties, while the errors on the calculated values are driven by uncertainties in the assumptions. We find that the most important contributions to the results shown in Figs.~\ref{fig:rotor_temp} and~\ref{fig:hwp_stability} are the IRF temperature, 50~K stage temperature, and the IR emissivity of the IRF and CHWP sapphire stack.

\begin{table}
\caption{\label{tab:thermal_values} Parameter definitions and values for Fig.~\ref{fig:thermal_diagram}. The bold values are measured, while the non-bold values are estimates. The error bars represent some combination of uncertainties in the calculation or measurement as well as configurational variations.}
\begin{ruledtabular}
\begin{tabular}{l p{4.5cm} r}
    Symbol & Description & Value \\
    \hline \\
    $p_{1}$ & Radiative load from 300~K onto the IRF, including that of the sky, window, and RT-MLI & 17 $\pm$ 5 W \\
    \hline \\
    $p_{2}$ & Radiative load from the encoder LEDs onto the rotor & 10 $\pm$ 3 mW \\
    \hline \\
    $p_{3}$ & Power dissipated by the encoder LEDs onto the stator & 0.8 $\pm$ 0.2 W\\
    \hline \\
    $p_{4}$ & Eddy current and hysteresis dissipation onto the rotor due to the motor's oscillating magnetic field  & < 1 mW \\
    \hline \\
    $p_{5}$ & Rotor frictional dissipation during 2~Hz rotation & \textbf{80 $\pm$ 10 mW} \\
    \hline \\
    $k_{1}$ & Radiative coupling between the IRF and the rotor & 7 $\pm$ 2 mW/K \\
    \hline \\
    $k_{2}$ & Conductivity between the IRF and the stator & \textbf{2.5 $\pm$ 0.3 W/K} \\
    \hline \\
    $k_{3}$ & Radiative coupling between the stator and the rotor & 5 $\pm$ 2 mW/K \\
    \hline \\
    $k_{4}$ & Radiative coupling between the solenoid array and the rotor & < 1 mW/K \\
    \hline \\
    $k_{5}$ & Conductivity between the motor solenoids and the stator & \textbf{0.8 $\pm$ 0.1 W/K} \\
    \hline
    $k_{6}$ & Conductivity between the rotor and the stator via the gripper & \textbf{0.7 $\pm$  0.1 W/K} \\
    \hline
    $k_{7}$ & Radiative coupling between rotor and the field lens & 5 $\pm$ 1 mW/K \\
\end{tabular}
\end{ruledtabular}
\end{table}
 
The CHWP is an 11.4~mm thick stack of three sapphire windows. At 50 K, the sapphire stack has an emissivity of 0.95 at $\sim$~3~THz,\cite{oxford_instruments_windows_nodate} and therefore the CHWP rotor absorbs IR radiation efficiently despite its AR coating. The dominant sources of radiative transfer to and from the rotor are the IRF, stator, and field lens. The IRF is heat-strapped to its fixture using 24 1~mm thick, 50~mm wide, 20~mm long flexible ribbons of 99.9999\% pure aluminum, which have superior conductivity between 100$\sim$50 K,\cite{Woodcraft2005} are malleable enough to engage the alumina surface without thermal grease, and are flexible enough to absorb differential thermal contraction between the IRF and its aluminum fixture. The IRF fixture is machined from aluminum 1100 and is bolted to six 90~mm tall, titanium-helicoiled OFHC copper towers (see Fig.~\ref{fig:chwp_tabletop}), which provide a thermal path to the stator baseplate. The measured conductivity between the IRF and the stator baseplate is $k_{2} = 2.5$~W/K. During laboratory testing with a thick window and no AR coatings (see Sec.~\ref{sec:thermal_performance}), we see an IRF temperature $\approx$~5~K warmer than the 50~K stage, suggesting a $\approx$~13~W sky-side load, which is slightly less than the model's expectation but within its $1 \sigma$ uncertainty.

The sky-side power leaked through the IRF depends on its AR coating. We assume in this paper that the IRF and CHWP sapphire stack are coated with an epoxy-based ARC,\cite{rosen_epoxy-based_2013} which has a non-negligible transmissivity $\lesssim$~2~THz.\cite{halpern_far_1986} Using the effective temperature of a six-layer RT-MLI stack\cite{choi_radio-transparent_2013} and the transmission spectra for Stycast-coated alumina,\cite{inoue_cryogenic_2014} we estimate that $\approx 60 \; \pm \; 10$~mW of sky-side power leaks through the IRF onto the rotor (or, in the heritage PB-2b configuration without the CHWP, onto the field lens).
 
The stator baseplate is made from aluminum 1100, houses the motor and encoder, and includes a 3~mm thick 50~K shield (see Fig.~\ref{fig:gripper_assy}) that encloses the CHWP assembly. The inner wall of the 50~K shield is coated with carbon-loaded Stycast\cite{persky_review_1999} and is intended to absorb any 300~K photons that leak into the CHWP cavity through the gripper ports. The 50 K shield is attached to the stator baseplate by only sixteen bolts, and therefore its temperature is typically $\approx$~5~K higher than that of the baseplate. The radiative coupling between the stator and the rotor is similar to that between the IRF and the rotor.
 
The encoder consists of two read heads, each with a set of five LEDs shining through the slotted encoder plate onto five photodiodes. At 50~K, the LEDs are current biased with 30-50~mA at 1.8~V, and therefore the power dissipated by each read head is 300$\sim$500 mW. We attach each read head to the stator baseplate using four 6~mm diameter aluminum 6061 standoffs with polished ends and interfaced with thermal grease,\footnote{Apiezon N Grease: https://www.apiezon.com/} and the estimated read-head warm-up is <~1~K. The LEDs also shine onto the rotor's slotted encoder plate, whose slit patterns have 50\% duty cycles. The LED emits between 900 and 960 nm, and its beam has a peak intensity of 35 mW/sr at 50 mA bias with a full-width-half-max of $\pm \; 7^{\circ}$. In order to limit stray read-head emission, each LED is collimated by a 1~mm wide, 3~mm deep hole that truncates its beam at $\pm \; 10^{\circ}$. Upon integrating these LED beams over the collimation holes, the estimated loading on the rotor from both encoder read heads is $10 \; \pm \; 3$~mW.
 
During 2~Hz rotation, the motor solenoids carry an RMS current of $\approx$~20~mA, generating a peak-to-peak magnetic field of $\approx$~20~G. The field is generated by a 38$f_{\mathrm{HWP}}$, $V_{\mathrm{D}}\; \approx \; 12 \mathrm{V}$ square-wave (at $f_{\mathrm{HWP}} \approx 2$~Hz), which is low-pass filtered above $\approx$ 300~Hz. To keep the coils cool, we epoxy\footnote{Stycast 2850FT: https://www.henkel-adhesives.com/us/en/product/potting-compounds/loctite\_stycast\_2850ft.html} them to their cores, providing a $7 \; \pm \; 2$~mW/K conductive path to the stator. At 50~K, each coil's resistance is $\approx \; 3 \; \mathrm{\Omega}$, dissipating $\approx \; 1$~mW per coil during continuous rotation. This dissipation warms the coils <~1~K, and their radiative coupling to the rotor is <~1~mW/K. In order to minimize eddy current dissipation on the rotor due to the motor's oscillating magnetic field, the encoder plate is made of G10, limiting eddy losses on the rotor to the sprocket ring. Electromagnetic dissipation within low-carbon steel at $\sim$~10~G and $\sim$~100~Hz frequencies induces <~1~mW of heating on the rotor.

The field lens is $\approx$~50~mm thick and is therefore assumed to have an IR emissivity of 1 for all calculations. Its coupling to the rotor is determined purely by the CHWP aperture diameter and the field-lens view factor and is calculated to be $5 \; \pm \; 1$~mW/K.


\bibliography{mendeley_refs}


\end{document}